\documentclass[12pt]{article}
\usepackage{bbm}
\usepackage{graphics}
\usepackage{color}
\usepackage{authblk}
\usepackage{latexsym}
\usepackage{epsfig,amssymb,euscript, mathrsfs,cite}
\usepackage{amsmath}
\usepackage{slashed}
\usepackage{soul}
\usepackage{cancel}
\usepackage{verbatim}
\usepackage{booktabs}
\usepackage{mathrsfs} 
\usepackage[normalem]{ulem}
\usepackage[dvipsnames]{xcolor}
\usepackage{enumerate}
\definecolor{MyDarkBlue}{rgb}{0.15,0.15,0.45}
\usepackage[linktocpage=true]{hyperref}
\usepackage{bm}
\usepackage[new]{old-arrows}
\usepackage{tikz,tikz-cd}

\newcommand{\ii}{\mathrm{i}}
\newcommand{\ex}{\mathrm{e}}
\newcommand{\dd}{\mathrm{d}}

\newcommand{\R}{\mathbb{R}}
\newcommand{\C}{\mathbb{C}}
\newcommand{\Z}{\mathbb{Z}}
\newcommand{\CP}{\mathbb{CP}}

\newcommand{\mf}[1]{\mathfrak{#1}}

\newcommand{\hook}
{\mathbin{\rule[.2ex]{.4em}{.03em}\rule[.2ex]{.03em}{.9ex}}}
\newcommand{\nn}{\notag}

\hypersetup{
colorlinks=true,
citecolor=MyDarkBlue,
linkcolor=MyDarkBlue,
urlcolor=MyDarkBlue,
pdfauthor={},
pdftitle={},
pdfsubject={hep-th}
}
\newsavebox{\ns}
\newsavebox{\dbrane}
\newsavebox{\dbshort}
\renewcommand{\arraystretch}{1.2}

\newcommand{\vol}{\mathrm{vol}}

\newcommand{\p}{p}

\newlength{\sswidth}

\renewcommand{\Re}{\operatorname{Re}}

\DeclareMathAlphabet{\mathbfsf}{OT1}{cmss}{bx}{n}

\newcommand{\cB}{\mathcal{B}}

\newcommand{\cK}{\mathcal{K}}
\newcommand{\cL}{\mathcal{L}}
\newcommand{\cM}{\mathcal M}
\newcommand{\cN}{\mathcal{N}}
\newcommand{\cO}{\mathcal{O}}
\newcommand{\cP}{\mathcal{P}}

\newcommand{\cS}{\mathcal{S}}

\numberwithin{equation}{section}       

\newcommand{\jpg}[1]{{\color{cyan}\bf [jpg: #1]}}
\newcommand{\jpgred}[1]{{\color{red}\bf [jpg: #1]}}
\let\jpgcyan\jpg
\RenewDocumentCommand{\jpg}{t2 +m}{%
  \IfBooleanTF{#1}{\jpgred{#2}}{\jpgcyan{#2}}}

\definecolor{chatgptteal}{HTML}{007C89}

\begin{document}

\pagestyle{plain}
\setcounter{page}{1}
\newcounter{bean}
\baselineskip18pt

\hypersetup{pageanchor=false}
\begin{titlepage}

\vfill

\begin{center}
   \baselineskip=16pt
   {\Large\bf 
Localizing the M-Theory Path Integral: \\ \vskip 0.4cm 
Membrane Instantons}
  \vskip 1cm
Pietro Benetti Genolini,$^1$ Jerome P. Gauntlett,$^2$ \\Jaeha Park$^2$ 
and James Sparks$^{3}$ \\
     \vskip 1cm     
\begin{small}
\textit{$^1$D\'epartement de Physique Th\'eorique,
	Universit\'e de Gen\`eve,\\
	24 quai Ernest-Ansermet,
	1211 Gen\`eve, Suisse}
\end{small}\\
\begin{small}\vskip .3cm
\textit{$^2$Abdus Salam Centre for Theoretical Physics, 
	Imperial College\\ 
	Prince Consort Rd., London, 
	SW7 2AZ, U.K.}
\end{small}\\
\begin{small}\vskip .3cm
\textit{$^3$Mathematical Institute,
	University of Oxford,\\
	Woodstock Road,
	Oxford OX2 6GG, U.K.}
\end{small}\\
\end{center}

\vfill

\begin{center}
\textbf{Abstract}
\end{center}

\begin{quote}
We propose 
a formula for M2-instanton
contributions to the M-theory path integral, incorporating quantum
corrections to all orders in the Planck length. The M2-brane action
is determined by relative equivariant localization, while an
equivariant Dolbeault character computes the one-loop determinant. 
Continuous families of M2-branes are treated by equivariant localization
on their moduli spaces.  
For M-theory
backgrounds dual to ABJM theory,
we compute single instanton
contributions from linear holomorphic fillings
to the sphere partition function and the
topologically twisted index associated with
supersymmetric $AdS_4$ black holes.
Our results reproduce known field theory
instanton coefficients and give new analytic
predictions for mass and squashing deformations
at all Chern--Simons levels.
For the $V^{5,2}$ theory, we compute new instanton
contributions and establish an exact relation
between its partition function at special
squashing and that of ABJM theory on the round
three-sphere.
\end{quote}

\vfill

\end{titlepage}
\hypersetup{pageanchor=true}

\setcounter{tocdepth}{2}
\tableofcontents

\newpage

\section{Introduction}\label{sec:Intro}

Equivariant localization allows one to compute
supersymmetric M-theory observables to all orders in the Planck length $\ell_p$ \cite{BenettiGenolini:2026kjy}.
Building on earlier results for equivariant localization in supergravity
\cite{BenettiGenolini:2023kxp,BenettiGenolini:2026cdw,
BenettiGenolini:2026cyc},  
reference~\cite{BenettiGenolini:2026kjy} provided a precise prescription for the local perturbative contribution to the M-theory path integral for a class of M2-brane backgrounds. 
Crucially, these computations do not require explicit supergravity solutions, but instead use only global topological and equivariant data. Since the M-theory path integral has no independent first-principles definition, we regard equivariant localization as providing a mathematically well-defined proposal for its exact supersymmetric partition function in this setting. The perturbative construction of \cite{BenettiGenolini:2026kjy} and its extension to M2-instantons in the present work provide successive steps towards making this a complete prescription.

The construction of \cite{BenettiGenolini:2026kjy}
starts from a supersymmetric Killing vector $K$
on the physical eleven-dimensional spacetime $M$,
appearing in the square of a preserved supercharge. 
The localization prescription
then has three main ingredients. First, $K$ defines the
equivariant structure and determines the fixed point set on
which the localization formula is evaluated. Second, the
presence of Chern--Simons couplings requires an auxiliary
twelve-dimensional filling $W$ of the eleven-dimensional
spacetime, with $\partial W=M$ \cite{Witten:1996md}. Extending $K$ and the relevant
geometric and flux data to $W$, 
the Chern--Simons couplings $C\wedge G\wedge G$ and $C\wedge X_8$ on $M$ are equivariantly
extended to give polyforms on $W$ that allow the supersymmetric action
$I_{\rm SUSY}$ to be computed using relative equivariant
localization. Third, an equivariant
Dolbeault character constructed from the holomorphic tangent
and cotangent bundles of $W$ determines the one-loop factor
$Z_{\text{1-loop}}$ after zeta-function regularization. Both
$I_{\rm SUSY}$ and $Z_{\text{1-loop}}$ are computed from
equivariant data at the fixed point set of $K$ in $W$ using
the Berline--Vergne--Atiyah--Bott (BVAB) theorem
\cite{BerlineVergne1982,AtiyahBott1984}. Finally,
$\ex^{-I_{\rm SUSY}}Z_{\text{1-loop}}$ supplies the weight for
an integration over any remaining BPS locus, which is parametrized
by equivariant flux data that is not fixed by the boundary
conditions.

The prescription of \cite{BenettiGenolini:2026kjy} was explored and tested in a
holographic setting for
three-dimensional SCFTs whose M-theory duals are described,
at the two-derivative level, by Freund--Rubin solutions $AdS_4\times Y_7$, where $Y_7$ is a Sasaki--Einstein
seven-manifold. These backgrounds arise as the near-horizon
limits of $N$ M2-branes at the apex of a Calabi--Yau
four-fold cone. The main examples studied in
\cite{BenettiGenolini:2026kjy} include $Y_7=S^7/\Z_k$
dual to ABJM theory \cite{Aharony:2008ug}, and
$Y_7=V^{5,2}$ dual to the $\mathcal{N}=2$ SCFT identified in \cite{Martelli:2009ga}. 
For deformed M-theory backgrounds, describing 
these SCFTs on squashed three-spheres with flavour
chemical potentials that are associated with mass deformations, the M-theory prescription yields Airy functions
for the perturbative partition functions, including their
overall normalization. These partition functions resum the full perturbative
large $N$ expansion and hence provide a gravitational computation
to all perturbative orders in $\ell_p$. The results reproduce
the independent field theory calculations that are available,
obtained from localizing the field theory path integral 
and taking the large $N$ limit,
and also provide analytic predictions for general deformation
parameters. In addition, the framework was applied to supersymmetric
black hole backgrounds dual to the ABJM topologically twisted
index, 
yielding agreement with known field theory results and also 
closed form expressions for the $N$-independent terms in its perturbative large $N$ expansion.

Field theory calculations also reveal non-perturbative corrections,
exponentially suppressed at large $N$, which are not included in
the perturbative expansion \cite{Drukker:2011zy,Marino:2011eh}.
Incorporating these contributions is the next step
in the programme of developing a complete localization
prescription for the M-theory path integral. 
Here we extend the
framework of \cite{BenettiGenolini:2026kjy} to supersymmetric
Euclidean M2-branes wrapping compact three-cycles in the
eleven-dimensional spacetime. Such instantons need not be isolated:
their supersymmetric embeddings can form continuous families. As emphasized
already in \cite{Harvey:1999as}, treating non-isolated
M2-brane instantons requires understanding their
additional zero modes and the interactions that saturate
them. More generally, one must determine the measure
on the moduli space, including the contributions of
bosonic and fermionic fluctuations. Here we propose a
localization prescription that addresses all of these
ingredients together, giving a definite contribution
from an entire supersymmetric M2-brane family.

Let $\cM$ denote a family of supersymmetric M2-brane configurations,
whose generic members admit $K$-invariant holomorphic fillings by
complex surfaces $D\subset W$. 
 Our prescription uses a torus $T$ obtained by taking
the closure of the subgroup generated by the flow
of $K$ 
and, when necessary, additional commuting flavour symmetries chosen so that the action of $T$ on $\mathcal M$ 
has only isolated fixed points (much as in equivariant instanton counting \cite{Nekrasov:2002qd,Nekrasov:2003rj}). We denote the additional equivariant parameters by $\epsilon$ and label the fixed points by $[D_a]$.
At fixed equivariant flux data,\footnote{In the holographic examples that we discuss,
as in \cite{BenettiGenolini:2026kjy}, this corresponds to
the grand canonical ensemble at fixed chemical potential
$\mu$, conjugate to the M2-brane number $N$
\cite{Gautason:2025plx, Bobev:2026gir}. The contribution to the fixed $N$
partition function is obtained by inserting $Z_\cM$ into
the bulk localization integral described above, with weight
$\ex^{-I_{\rm SUSY}}Z_{\text{1-loop}}$, and then performing
the inverse Laplace transform in $\mu$.}
we propose
\begin{align}
\label{eq:Z1M}
 Z_\cM
 =\lim_{\epsilon\to0}\sum_{[D_a]\in\cM^T}
 \frac{e_T(\operatorname{Obs}_{[D_a]})}
      {e_T(T_{[D_a]}\cM)}\,
 Z^{\rm red}_{\text{1-loop},a}(\epsilon)\,
 \ex^{-I_{{\rm M2},a}(\epsilon)}\, .
\end{align}
Here $I_{{\rm M2},a}$ is the localized Euclidean
M2-brane action and $Z^{\rm red}_{\text{1-loop},a}$
is the reduced cohomological one-loop determinant.
Both factors are understood to
incorporate corrections to all orders in $\ell_p$. They are computed from fixed point data of the refined
equivariant generator on the filling $D_a$: the action $I_{{\rm M2},a}$ by relative
equivariant localization, and the determinant by regularizing an
equivariant Dolbeault character. The latter builds on the
cohomological fluctuation analysis of
\cite{Kurlyand:2026yke,Kurlyand:2026gxi} for two-derivative
Freund--Rubin backgrounds.

The remaining factors in \eqref{eq:Z1M} account for the
finite-dimensional zero modes of the supersymmetric fluctuation
complex, which are omitted from the reduced determinant. Bosonic
collective coordinates are identified with tangent directions
to $\cM$, while the relevant fermionic zero modes define
obstruction spaces and are saturated by supersymmetric curvature
couplings. 
The equivariant Euler classes in \eqref{eq:Z1M}
are those of the tangent and obstruction spaces at
each fixed point $[D_a]$, with the induced $T$-action.
In terms of their equivariant weights
$\tau_{a,j}$ and $\phi_{a,\alpha}$, respectively,
these are 
\begin{align}\label{eq:M2_Euler_weights}
 e_T(T_{[D_a]}\cM)=\prod_j\tau_{a,j}\, ,\qquad
 e_T(\operatorname{Obs}_{[D_a]})=\prod_\alpha\phi_{a,\alpha}\, .
\end{align}
When all the modes omitted from the reduced
determinant acquire non-zero weights under refinement,
the Euler ratio restores their contributions,
giving the full cohomological one-loop determinant.
Fermionic obstruction modes can remain invariant
under $T$ even at isolated fixed points.
The vanishing obstruction Euler factors in
\eqref{eq:Z1M} then make these fixed point
contributions zero, provided the reduced
determinants are regular at generic refinement
parameters.
Both cases are explained in
section~\ref{subsec:Quantum_Brane_Families}. 

We expect \eqref{eq:Z1M} to arise from the equivariant
localization of a globally defined, equivariantly closed
integrand on $\cM$, generated by supersymmetric
localization of the M2-brane path integral. In particular,
when the fermionic obstruction spaces form a bundle,
their equivariant Euler class provides a natural global
extension, whose ordinary Euler class we use in some
of our examples. Establishing the corresponding global
equivariant classes for the M2-brane action and reduced
determinant is beyond the scope of this paper. 
Instead,
we take the refined fixed point sum \eqref{eq:Z1M}
as our starting point. 
For the M-theory backgrounds studied here, generic
physical mass deformations give isolated supersymmetric
M2-brane configurations, whose contributions require
no auxiliary refinement.
At values of the deformation parameters for which
continuous families occur, we introduce the auxiliary
refinement described above, so that the torus $T$ has
only isolated fixed points on $\cM$. 
The corresponding refinement parameters can themselves be realized as flavour mass deformations. The limit in \eqref{eq:Z1M} removes these additional deformations after summing the fixed point contributions, while keeping the parameters of the original background fixed.

The choice of $\cM$ is important: compactifications of families
of smooth M2 configurations may contain multiple covers or
multiple membrane components. We do not construct these general
compactifications here, but develop the prescription explicitly
for families of linear holomorphic fillings. Our proposal is intended to capture
the complete contribution associated with such a family,
including perturbative corrections to all orders in $\ell_p$,
without requiring explicit higher-derivative supergravity
solutions or corrected membrane embeddings.

Working directly in eleven dimensions gives a unified
description of instantons with different type IIA
interpretations. 
Upon reduction along the M-theory circle, M2-branes wrapping
this circle give fundamental string worldsheet instantons,
while M2-branes transverse to it give Euclidean D2-brane
instantons \cite{Drukker:2011zy,Beccaria:2023ujc}.
Both arise naturally within our framework of holomorphic
fillings and their moduli spaces. 

Previous studies of M2-instantons include
\cite{Becker:1995kb,Harvey:1999as}. More recently, explicit
holographic calculations have been carried out for
fundamental string worldsheet instantons in type IIA
\cite{Gautason:2023igo} and for M2-instantons in the
two-derivative Freund--Rubin background
$AdS_4\times S^7/\Z_k$
\cite{Beccaria:2023ujc,Kurlyand:2026yke,Tseytlin:2026dhy}. 
For uplifts of minimal four-dimensional gauged supergravity,
\cite{Gautason:2025per} used the $\kappa$-symmetry condition
to localize the external positions of M2-instantons.
The study included backgrounds with squashed sphere boundaries and
one-loop contributions were computed for selected internal
embeddings, with a grand canonical interpretation. 
A different geometric approach to leading WKB
instanton corrections in round sphere partition
functions, based on equivariant disk partition functions
of gauged linear sigma models, was developed in \cite{Hristov:2026nhx}. 

For ABJM theory, non-perturbative corrections can be
organized through the modified grand potential
$J(k,\mu;\ldots)$, with $\mu$ the chemical potential
conjugate to the rank $N$. 
The exponential of
$J$ enters the grand canonical integrand. The M2-brane
contribution $Z_\cM$ in \eqref{eq:Z1M} is an insertion
in this integrand and hence contributes additively to
$J$ at first order in the instanton expansion.
We denote the sum of these contributions from families
of linear holomorphic fillings by $J_{\rm lin}$. 
Our prescription reproduces known round sphere instanton
coefficients in these sectors
\cite{Hatsuda:2012dt,Hatsuda:2013gj}, as well as
the conjectured expressions on a mass-deformed locus
\cite{Nosaka:2024gle}. The agreement with these
independent field theory results provides strong evidence
for the prescription, which also yields new predictions
for more general squashing and flavour deformations.

As a concrete illustration, consider ABJM theory at
Chern--Simons level $k=1$ on the squashed three-sphere
$S^3_b$, with vanishing flavour masses. Introducing
\begin{align}
\label{eq:intro_ABJM_squashing_f}
 f\equiv\frac{b^2-1}{b^2+1}\, ,
\end{align}
we obtain the following closed-form expression for $J_{\rm lin}$:
\begin{align}
\label{eq:Intro_Jlin_1}
J_{\mathrm{lin}}(1,\mu;b)
={}&-\frac{\ex^{-4(1-f^2)\mu}}{4\pi^4}
\Bigl\{
\sin^2(2\pi f)\cP_4(\mu,f)
+\pi f\sin(4\pi f)\cP_3(\mu,f)\nn\\
&+\pi^2 \cP_2(\mu,f)
+\frac{20\pi^3}{3}f(1-4f^2)
\bigl[4(1-f^2)\mu+1\bigr]\cot(2\pi f)
\Bigr\}\, ,
\end{align}
where  $\cP_a(\mu,f)$ are polynomials of degree $a$ in $\mu$ and are given later in \eqref{laterintext}.
For generic $b$, $J_{\mathrm{lin}}(1,\mu;b)$ is an exponential times a quartic polynomial in the chemical potential 
$\mu$. At the special values $b^2=1,3,\frac13$, corresponding to
$f=0,\pm\frac12$, the terms involving $\cP_4$ and $\cP_3$ vanish 
and the formula reduces to
\begin{align}
\label{eq:intro_ABJM_squashed_limits}
 J_{\rm lin}(1,\mu;1)
 &=\frac{4\mu^2+\mu+\frac14}{\pi^2}\ex^{-4\mu}\, ,\nn\\
 J_{\rm lin}(1,\mu;\sqrt3)
 &=J_{\rm lin}(1,\mu;1/\sqrt3)
 =\Big(\frac38-\frac{45\mu^2+18\mu+6}{8\pi^2}\Big)\ex^{-3\mu}\, .
\end{align}
These reproduce the independent field theory results
known for the round sphere
\cite{Hatsuda:2012dt,Hatsuda:2013gj} and for $b^2=3, 1/3$
\cite{Hatsuda:2016uqa}.
The change in polynomial degree has a direct geometric
explanation in our supersymmetric deformation--obstruction
theory. The moduli space of linear M2-instantons has
complex dimension four and persists for all $b$, but
at the special squashing values two complex fermionic
zero modes appear. The virtual complex dimension
therefore drops from four to two, explaining the reduction
from a quartic to a quadratic polynomial. 

We also apply our prescription to supersymmetric,
magnetically charged and rotating $AdS_4$ black holes,
computing M2-instanton corrections to the ABJM
topologically twisted index on $S^1\times S^2$.
These corrections are exponentially suppressed at
large $N$ and probe quantum black hole physics beyond
the entire perturbative large $N$ expansion.
Extending the perturbative construction of
\cite{BenettiGenolini:2026kjy}, we formulate the
calculation using two local gravitational blocks
at general chemical potentials and refinement.

At the unrefined universal twist, our prescription
determines the complete coefficient of the leading
instanton exponential, including the perturbative
corrections about the instanton.
The results reproduce the coefficients reconstructed
from field theory for the index and on-shell twisted
superpotential at $k=1,2,4$
\cite{Hosseini:2026dkj,Hosseini:2026lvj}.
The agreement includes the constant terms in the
instanton prefactors, whose recovery requires the
integration over bulk fluxes and transformation to
fixed $N$. Our construction also provides predictions at
arbitrary Chern--Simons level $k$.
These calculations identify wrapped M2-branes
as the source of exponentially small corrections
to supersymmetric black hole partition functions,
providing precise non-perturbative tests of
M-theory.\footnote{For a localization approach to black
hole quantum entropy in four-dimensional
supergravity, see \cite{Dabholkar:2010uh}.
A calculation reproducing microscopic indices,
including non-perturbative orbifold contributions,
was presented in \cite{Iliesiu:2022kny}.}

For backgrounds with internal manifold
$V^{5,2}/\Z_k$, whose Calabi--Yau cone is singular
and non-toric, 
we obtain analogous analytic
expressions for M2-instanton contributions to
the modified grand potential, at vanishing flavour
masses and arbitrary squashing.
As in the ABJM examples above, these expressions
simplify at special squashing values, which for this model
are at
$b^2=2,\tfrac12$.

We also obtain two new field theory results
for the $U(N)_k\times U(N)_{-k}$
Chern--Simons-matter theory dual to $V^{5,2}$ proposed in
\cite{Martelli:2009ga}.
First, at $b^2=2,\tfrac12$, we establish the exact
partition-function identity
\begin{align}
 Z_{V^{5,2}}(N,k;\sqrt2)
 &=Z_{V^{5,2}}(N,k;1/\sqrt2)
 =Z_{\rm ABJM}(N,2k;1)\, .
\end{align}
This holds at every finite $N$ and relates the
special squashing $V^{5,2}$ partition function
to that of round sphere ABJM.
The known leading ABJM instanton contribution
then reproduces our gravity result for every $k$.

Our second $V^{5,2}$ field theory result is a numerical
calculation that isolates the exponentially small
corrections in the round sphere planar matrix
model.
After converting the large $k$ gravity result
to fixed $N$, the comparison supports the
predicted instanton action and both coefficients
of its prefactor.

The paper is organized as follows.
Section~\ref{sec:m2action} describes the localization
of the M2-brane action.
Section~\ref{sec:Holomorphic_M2_PI} introduces the
holomorphic fillings and fluctuation complex,
and explains how to evaluate \eqref{eq:Z1M}
for isolated membranes and continuous families.
We apply this framework to the M-theory duals of ABJM theory on the
three-sphere in section~\ref{sec:ABJM_Theory},
to its topologically twisted index in
section~\ref{sec:ABJM_TTI}, and to the
$V^{5,2}$ theory in section~\ref{sec:V52_Theory}.
We conclude with a discussion of open questions.
The appendices contain supporting geometric
details, derivations and field theory computations.

\section{Localization of the M2-instanton action}\label{sec:m2action}

\subsection{Two-derivative backgrounds}

We begin with Freund--Rubin $AdS_4\times Y_7$ solutions of $D=11$ supergravity,\footnote{Throughout this paper, by $D=11$
supergravity we mean the two-derivative theory.
Higher-derivative terms are treated as corrections
in the M-theory effective action.
We always work in Euclidean signature.}
where $Y_7$ is a compact Sasaki--Einstein
seven-manifold and the four-form flux $G$ is
proportional to the volume form on $AdS_4$.
These backgrounds arise as near-horizon limits
of M2-branes at the apex of the Calabi--Yau cone
$C(Y_7)$ and are dual to three-dimensional SCFTs.

We also consider supersymmetric deformations
describing these SCFTs on squashed three-spheres,
with flavour chemical potentials corresponding
to mass deformations.
Correspondingly, we take the underlying
topology to be $M\cong M_4\times Y_7$,\footnote{
From section~\ref{sec:Holomorphic_M2_PI}
onwards we also allow $Y_7$ to be fibred over
$M_4$, including backgrounds with
$M_4\cong\R^2\times S^2$ relevant for the
topologically twisted index in
section~\ref{sec:ABJM_TTI}.}
where $M_4\cong\R^4$ is asymptotically locally
Euclidean $AdS_4$.
The product specifies only the topology:
the metric and fluxes need not have a product
form.
Explicit solutions are known in special cases, e.g.
\cite{Martelli:2011fu,Freedman:2013oja}, but not
for general squashing and flavour deformations.
Our localization calculation does not require these explicit
solutions, but instead uses the supersymmetric equivariant
structure and flux data.

A supersymmetric solution admits a Killing vector $K$
constructed as a bilinear in the preserved Killing spinor
\cite{Gauntlett:2002fz}. For the Freund--Rubin backgrounds,
the Sasaki--Einstein structure is determined by a complex
Killing spinor $\chi$ on $Y_7$, whose one-form bilinear
gives the contact form, with metric dual the Reeb Killing vector
$\xi$. 
Together with the $AdS_4$ Killing spinors, $\chi$
determines the preserved eleven-dimensional supersymmetry
and the associated vector $K$. For our choice of
supercharge and complex structure, we normalize $K$
so that, in the undeformed round background, its
action on the two external complex planes has
weights $(1,1)$.
With the standard normalization
of the Reeb vector, for which the holomorphic
$(4,0)$-form on $C(Y_7)$ has charge $4$, the
corresponding internal action is $-\frac12\xi$.
Thus, in Euclidean signature, with
$EAdS_4\cong B^4\subset\R^2_1\oplus\R^2_2$,
we have
\begin{align}
\label{eq:Intro_K_2}
 K=\partial_{\varphi_1}+\partial_{\varphi_2}
 -\frac12\xi\, ,
\end{align}
where $\partial_{\varphi_i}$ generates rotations
of $\R^2_i$ with weight $+1$. 

For the deformed backgrounds, we assume that the
projection of $K$ onto $M_4$ has a unique isolated fixed
point, which we call the origin, and that its action on
$T_0M_4\cong\R^2_1\oplus\R^2_2$ has weights $(b_1,b_2)$.
The internal component of $K$ is likewise determined by
the supersymmetric background, and incorporates the
appropriate flavour deformations.

The other ingredient is the equivariantly closed extension
of the $C$-field curvature $G=\dd C$. On the physical
eleven-dimensional spacetime, the Killing spinor bilinears
determine a two-form $\Omega$ such that
\begin{align}
 \Phi^G=G+\Omega\, ,
 \qquad
 \dd_K\Phi^G=0\, ,
 \qquad
 \dd_K\equiv\dd-K\hook\, ,
\end{align}
as follows from the two-derivative supersymmetry equations
\cite{Gauntlett:2002fz}.

To formulate the Chern--Simons couplings, in particular
$C\wedge G\wedge G$, we introduce an auxiliary
twelve-dimensional filling $W$, with $\partial W=M$
\cite{Witten:1996md}. We take
\begin{align}
 W\cong M_4\times Z_8\, ,
 \qquad \partial Z_8=Y_7\, ,
\end{align}
and extend $K$ and the equivariant flux data to $W$.
On this filling, the equivariantly closed polyform
$\Phi^G$ also has a zero-form component $\Phi^G_0$,
which vanishes at the boundary physical spacetime $M$ but can be non-zero at
interior fixed points. This is the equivariant flux
data entering the localized M2-brane action below.

\subsection{M2-brane action and equivariant localization}
\label{subsec:CalibratedMembranes}

We consider a supersymmetric Euclidean M2-brane located at
the origin of $M_4$ and wrapping a compact three-cycle
$\Sigma\subset Y_7$. Such cycles are generalized calibrated \cite{Gutowski:1999tu}.
With the M2-brane orientation chosen compatibly with the
preserved supersymmetry, the calibration condition takes the
form\footnote{This relation follows from the $\kappa$-symmetry
projection for a supersymmetric M2-brane. The Euclidean
M2-brane projector in Freund--Rubin backgrounds is
discussed explicitly in section~2.1 of
\cite{Kurlyand:2026yke}.}
\begin{align}
 \Omega|_\Sigma=-\ii\mskip1mu K\hook\vol_\Sigma\, .
\end{align}

We choose a $K$-invariant four-dimensional filling
$D\subset Z_8$ in the internal fibre over the origin
of $M_4$, with $\partial D=\Sigma$. For the Freund--Rubin backgrounds, let $r\geq 0$ denote the radial
coordinate on the Calabi--Yau cone $C(Y_7)$, with $Y_7$
identified with the hypersurface $\{r=1\}$. We then take
$Z_8$ to be the radially truncated region $0\leq r\leq 1$, and $D$ to be
the cone over $\Sigma\subset Y_7$ in this region,
resolving the apex if necessary. 
More generally, we may use any filling
$D\subset W$ satisfying the same boundary
and equivariance conditions.

Defining
$\varphi_3\equiv-\ii\,\vol_\Sigma$, the calibration
condition implies
\begin{align}
\label{eq:Relative_equivariant_PhiG}
 \dd_K\Phi^G=0\, ,
 \qquad
 \dd_K\varphi_3=-\Phi^G|_\Sigma\, .
\end{align}
Here we use the fact that the zero-form component of
$\Phi^G$ vanishes on the physical spacetime $M$.
These relations define a relative equivariant class \cite{BenettiGenolini:2026cdw}
on the pair $(D,\Sigma)$. 
At leading order in the M-theory expansion, the Euclidean M2-brane action 
\cite{Bergshoeff:1987cm} is
\begin{align}
\label{eq:M2_Rel_1}
 I_{\rm M2}
 &=T_2\int_\Sigma
   \left(\vol_\Sigma+\ii C\right)=\ii T_2\int_D
   \left(G+\dd(f\varphi_3)\right)\, ,
\end{align}
where $T_2=1/[(2\pi)^2\ell_p^3]$ and $f$ is a
$K$-invariant smooth function on $D$ which equals one
near $\Sigma$ and vanishes away from the boundary.
The second equality follows locally from $G=\dd C$ and Stokes'
theorem;  its global interpretation is discussed in the next subsection.

The advantage of the relative formulation is that
\begin{align}
 \widehat\Phi^G
 \equiv\Phi^G+\dd_K(f\varphi_3)
\end{align}
is equivariantly closed and vanishes on restriction to
$\Sigma$. Its four-form component is precisely the
integrand in the right-hand expression of
\eqref{eq:M2_Rel_1}. We may therefore apply equivariant
localization on the filling $D$, without boundary
contributions, provided $K$ has no zeroes on $\Sigma$.
The result is
\begin{align}
\label{eq:M2_Rel_Eq_Loc}
 I_{\rm M2}
 =2\pi\ii\,\frac{1}{(2\pi\ell_p)^3}
 \int_{\mathscr F}
 \frac{\Phi^G}{e_K(N\mathscr F)}\, ,
\end{align}
where $\mathscr F\subset D$ is the fixed point set of $K$,
$N\mathscr F$ is its normal bundle, and $e_K$ is the
equivariant Euler class. This is the relative BVAB
localization formula discussed in
\cite{BenettiGenolini:2026cdw}.

When $\mathscr F$ consists of isolated fixed points
labelled by $a$, let $u_{1,a},u_{2,a}$ be the weights
of $K$ on $T_aD$. The formula \eqref{eq:M2_Rel_Eq_Loc} reduces to
\begin{align}
\label{eq:SM2_Generic}
 I_{\rm M2}
 =\sum_a
 \frac{b_1b_2}{h_a u_{1,a}u_{2,a}}\,\mu_a\, ,
\end{align}
where $h_a=1$ for a smooth filling, while a finite
stabilizer of order $h_a$ contributes the corresponding
orbifold factor. Here we have defined, as in \cite{BenettiGenolini:2026cyc, BenettiGenolini:2026kjy}, 
\begin{align}
\label{eq:mu_a_Def}
 \mu_a
 \equiv\frac{\ii}{b_1b_2\ell_p^3}\Phi^G_0|_a
 =\frac{\ii}{(2\pi)^2\ell_p^3}
 \left.\int_{M_4}^{\rm eq}\Phi^G\right|_a\, .
\end{align}
The second expression follows from equivariant
localization of the flux on $M_4$, using a compatible
$K$-invariant four-dimensional slice through $a$.

Thus, at this order, the action of a supersymmetric M2-brane
is determined entirely by fixed point weights and
equivariant flux data, the latter being the same
variables that enter the perturbative M-theory
localization formula of
\cite{BenettiGenolini:2026cyc,BenettiGenolini:2026kjy}.\footnote{An analogous computation using equivariant localization has been carried out for D3-brane probes on asymptotically $AdS_5 \times SE_5$ IIB backgrounds \cite{BenettiGenolini:2026hmz}.}
The independence of the complete M2-brane contribution
from the choice of filling $D$ is an important
consistency condition, which we next discuss.

\subsection{Filling independence and higher-derivative corrections}
\label{subsec:HD_Corrections}

The localized expression for the M2-brane action in
\eqref{eq:M2_Rel_Eq_Loc} uses an auxiliary filling $D$ of
the M2-brane worldvolume $\Sigma$. The physical M2-brane
contribution must be independent of this choice. As explained
in \cite{Witten:1996md}, this independence holds for the
complete M2-brane amplitude, including its fermion Pfaffian,
rather than for the bosonic action alone.

Let $D'$ be another admissible filling of $\Sigma$, such
that $B\equiv D'\cup_\Sigma(-D)$ defines a compact integral
four-cycle in the spin manifold $W$, with $D'$ and $-D$ glued along their common boundary $\Sigma$. We assume compatible
extensions of the $C$-field and the worldvolume fermions.
The $G$-flux obeys the shifted quantization condition
\cite{Witten:1996md}
\begin{align}
\label{eq:Witten_flux_quantization}
 \left[\frac{G}{(2\pi\ell_p)^3}\right]
 -\frac14 p_1(TW)\in H^4(W;\mathbb Z)\, .
\end{align}
Writing $\lambda\equiv\tfrac12 p_1(TW)$, which is integral
since $W$ is spin, the change in the M2-brane action is
\begin{align}
\label{eq:Witten_Change}
 I_{\rm M2}[D']-I_{\rm M2}[D]
 =2\pi\ii\int_B\frac{G}{(2\pi\ell_p)^3}
 =2\pi\ii\left(n+\frac12\int_B\lambda\right)\, ,
 \qquad n\in\mathbb Z\, .
\end{align}
The integer ambiguity $2\pi \ii n$ disappears upon exponentiation,
while the shifted quantization condition can produce a
sign $(-1)^{\int_B\lambda}$. The M2-brane fermion Pfaffian
transforms by the same sign \cite{Witten:1996md},
so that the complete M2-brane amplitude is independent
of the choice of filling $D$. This requires the appropriate global $C$-field
data; in particular, the bosonic action need not be
individually well-defined.

We now consider higher-derivative corrections. The
M-theory effective action includes corrections to the
two-derivative $D=11$ supergravity action, including
the $C\wedge X_8$ coupling, which generically modify
the supersymmetric background. The bulk localization
prescription of \cite{BenettiGenolini:2026kjy}
incorporates these corrections through a relative
equivariant class on $W$, constructed from the
Chern--Simons couplings and the gauge-invariant terms
in the action. This determines the supersymmetric
bulk action from fixed point data without requiring
the explicit higher-derivative supergravity solution.
The complete bulk M-theory phase likewise requires
the appropriate global flux quantization and fermionic
contributions for independence of the choice of $W$.

The M2-brane action also receives higher-derivative
corrections, both through the background and through
the worldvolume effective action. 
Motivated by the bulk proposal of
\cite{BenettiGenolini:2026kjy}, we propose that
these corrections preserve the relative
equivariant localization structure. 
More precisely, on the
corrected supersymmetric background and embedding,
we take the M2-brane action to admit a boundary
representative $\varphi_3^{\rm corr}$ satisfying
\begin{align}
\label{eq:Membrane_Conjecture}
 I_{\rm M2}
 &=\ii T_2\left(
 \int_\Sigma\varphi_3^{\rm corr}+\int_D G
 \right)\, , \qquad
 \dd_K\varphi_3^{\rm corr}
 =-\Phi^G\big|_\Sigma\, ,
\end{align}
where $G$ and $\Phi^G$ are understood as the corrected
flux and its equivariant extension.
At fixed equivariant flux and global $C$-field data,
the corrected relative class is taken to represent
the same equivariant cohomology class as in the
two-derivative construction. The localized action is
then given to all orders by the same fixed locus
formula \eqref{eq:M2_Rel_Eq_Loc}, with the exact
equivariant data. For isolated fixed points this
reduces to \eqref{eq:SM2_Generic}, where $\mu_a$ in
\eqref{eq:mu_a_Def} is defined by the corresponding
exact equivariant flux $\Phi^G_0|_a$.

\section{Holomorphic fillings and M2-brane fluctuations}
\label{sec:Holomorphic_M2_PI}

Having described the localized M2-brane action in
section~\ref{sec:m2action}, we now explain how to compute
the remaining factors in \eqref{eq:Z1M}.
We describe the holomorphic fillings and their
fluctuation complex, identify the finite-dimensional
bosonic and fermionic zero mode sectors associated with the
Euler factors, and construct the reduced one-loop
determinant.
We then explain when these factors can  be
combined into a full one-loop determinant.

\subsection{Holomorphic fillings}
\label{subsec:M2_complex_geometry}

We first recall the geometric setting of
\cite{BenettiGenolini:2026kjy}, which we also
adopt here. We take the physical eleven-dimensional
spacetime $M$ and its auxiliary twelve-dimensional filling $W$
to be fibrations of the form
\begin{align}
\label{eq:W_Productfibred}
Y_7\hookrightarrow M\rightarrow M_4\, ,
\qquad
Z_8\hookrightarrow W\rightarrow M_4\, ,
\qquad \partial Z_8=Y_7\, . 
\end{align}
Taking the boundary of each internal fibre gives
$\partial W=M$. For the sphere backgrounds,
$M_4\cong\mathbb R^4$ and the fibrations are
topologically trivial, giving
$M\cong M_4\times Y_7$ and $W\cong M_4\times Z_8$.
For the topologically twisted index considered in
section~\ref{sec:ABJM_TTI},
$M_4\cong\mathbb R^2\times S^2$, and the internal
spaces are non-trivially fibred over $M_4$.

Let $p\in M_4$ denote an isolated fixed point of the
external action of $K$. The supersymmetric M2-branes
considered here wrap three-cycles $\Sigma\subset Y_{7,p}$
in the internal fibre over $p$. Their four-dimensional
fillings $D\subset Z_{8,p}$ satisfy $\partial D=\Sigma$
and lie in the corresponding fibre of $W$. Thus $W$
is the auxiliary filling of the spacetime, while $D$
is the filling of the M2-brane worldvolume used to
localize its action. 
Recall that $\cM$ denotes a family of
supersymmetric M2-brane configurations, as
introduced in \eqref{eq:Z1M}. An M2-brane
configuration is specified by a map from its
worldvolume into $M$, up to worldvolume
reparametrizations, rather than by its image alone.
In particular, maps with different covering
degrees can have the same image, and this
information must be retained in specifying
$\cM$.
The construction below applies
near each such fibre, using a compatible local
holomorphic trivialization and the induced equivariant
weights. 

For the two-derivative Freund--Rubin solutions, we take
$Z_8$ to be the radially truncated region $0\leq r\leq 1$ of the
Calabi--Yau cone $C(Y_7)$, as in section \ref{subsec:CalibratedMembranes}.
The Sasaki--Einstein structure on $Y_7$ determines a
Calabi--Yau structure $(J,\omega,\Omega_4)$ on this
cone. For the sector of supersymmetric M2-branes
preserving the two internal Killing spinors defining
this structure, $\Sigma$ is a Sasakian link and its
cone $D$ is a complex surface
\cite{Kurlyand:2026yke}. Thus the holomorphic filling
arises naturally in the undeformed geometry.

We take this holomorphic description to extend
to supersymmetric backgrounds with squashing
and flavour deformations, including
higher-derivative corrections to all orders
in $\ell_p$. We take $Z_8$ to admit a complex structure
$J$ for which the relevant M2-brane worldvolumes have
holomorphic fillings $D$, preserved by $K$. In particular,
we require
\begin{align}
\label{eq:BPS_Condition_Invariance}
\mathcal L_K J=0\, ,
\qquad \text{$K$ preserves $D$}\, .
\end{align}
We further take this complex structure to extend to a
$K$-invariant complex structure on $W$, compatible with
a local holomorphic trivialization near each relevant
internal fibre. This provides the complex geometry used
in the equivariant fluctuation complex below.

For the backgrounds studied here, we choose the
extension $W$ so that its canonical bundle $K_W$
is equivariantly trivial. This means that $W$ admits
a nowhere-vanishing holomorphic volume form with
zero weight under $K$.
This is additional structure of our chosen fillings.
The relevant supersymmetry condition is the invariance
of the preserved spinor under $K$, which does not
in general require a globally defined holomorphic
volume form on $W$. 

\subsection{The fluctuation complex and M2-brane moduli problem}
\label{subsec:Quantum_Brane_Families}

We now use the holomorphic fillings to compute
the fluctuation factors entering \eqref{eq:Z1M}.
We encode the unpaired supersymmetric
M2-brane fluctuations in an equivariant Dolbeault
complex on $D\subset W$.
For a smooth filling in the internal fibre over
$p\in M_4$, the complex structure introduced in
section~\ref{subsec:M2_complex_geometry} gives
\begin{align}
\label{eq:VD_def}
 N^{1,0}_{D/W}
 &\cong
 \bigl(\mathcal O_D\otimes T^{1,0}_pM_4\bigr)
 \oplus N^{1,0}_{D/Z_8}
 \equiv V_D\, .
\end{align}
Here $N^{1,0}_{D/W}$ denotes the holomorphic normal
bundle of $D$ in $W$, describing complex directions
transverse to $D$, and similarly for
$N^{1,0}_{D/Z_8}$.
The trivial holomorphic line bundle $\mathcal O_D$
extends the external tangent space over $D$.
The two summands describe the external and internal
holomorphic normal directions, respectively.
Each has complex rank two, so $V_D$ has complex
rank four. 

The fluctuation complex is taken to be the graded
difference of the Dolbeault complexes with
coefficients in $V_D$ and $\Lambda^3V_D$.
We encode it in the virtual bundle
\begin{align}
\label{eq:ED_def}
 E_D
 &\equiv \Lambda^1V_D-\Lambda^3V_D\, ,
\end{align}
where $\Lambda^1V_D=V_D$.
The positive and negative terms encode the
bosonic and fermionic fibres of the unpaired
complex, respectively, with their difference
understood in equivariant $K$-theory.
With this choice of extension, adjunction gives
$\det V_D\cong K_D$.\footnote{For a general extension,
adjunction gives
$\det V_D\cong K_D\otimes K_W^{-1}|_D$.
\label{footnote:Equivariant_Triviality}}
Using
$\Lambda^3V_D\cong(\det V_D)\otimes V_D^*$,
we may therefore write
\begin{align}
\label{eq:ED_def_2}
 E_D
 &=V_D-K_D\otimes V_D^*\, ,
\end{align}
where $K_D$ is the canonical bundle of $D$ and
$V_D^*$ is the dual bundle.

The virtual bundle \eqref{eq:ED_def_2} follows from the same prescription that gives the
tangent--cotangent virtual bundle $TW-T^*W$, which governs the
perturbative one-loop factor in \cite{BenettiGenolini:2026kjy}.
In each case, we can think of the one-loop factor as accounting for
the holomorphic fluctuations of a geometric object in $W$. These are
described by a Dolbeault complex with values in a holomorphic bundle
$F$, and we encode their fermionic partners using the Serre dual $K_W\otimes F^*$.\footnote{On a compact complex manifold $Y$ of dimension $n$, with holomorphic bundle $F$, Serre duality is the statement that $H^q(Y,F) \cong H^{n-q}(Y,K_Y \otimes F^*)^*$ for any $q$.
The pairing is $\int_Y \beta \wedge \alpha$ with $\alpha \in \Omega^{0,q}(F)$ and $\beta \in \Omega^{n,n-q}(F^*) = \Omega^{0,n-q}(K_Y \otimes F^*)$.
The integrand is an $(n,n)$-form, so $K_Y\otimes F^*$ is the bundle whose sections can be paired with those of $F$ and integrated over $Y$. }
Our prescription is therefore a single rule on the twelve-dimensional
filling: in both cases, the one-loop factor is determined by the
virtual bundle
\begin{align}
\label{eq:XW_def_main}
 X_W(F) \equiv F-K_W\otimes F^* \, ,
\end{align}
where the dual is taken on $W$.
For the bulk, the geometric object is the complex
structure of $W$. Its infinitesimal deformations
are described by the Dolbeault complex with
coefficients in $TW$, so we take $F=TW$.
Since $K_W$ is equivariantly trivial,
$X_W(TW)=TW-T^*W$.
For an M2-brane, the object is the embedding of its filling,
$i\colon D\hookrightarrow W$. Its deformations are sections of the
normal bundle \eqref{eq:VD_def}, which we regard as fluctuations on
$W$ supported on $D$, $F = i_*N_{D/W}$, that is, the normal bundle
extended by zero.
To pair such a fluctuation with a field on $W$, we integrate over
$W$. The delta function supported on $D$ absorbs the normal
directions, and the top form on $W$ becomes a top form on $D$:
$K_W|_D\otimes\det N_{D/W} = K_D$, by adjunction.
The Serre dual on $W$ of $i_*N_{D/W}$ is therefore
$i_*\bigl(K_D\otimes N^*_{D/W}\bigr)$. Hence
$ X_W(i_*N_{D/W}) = i_*\bigl(N_{D/W}-K_D\otimes N_{D/W}^*\bigr) = i_*E_D $,
and, since the index of $i_*E_D$ on $W$ equals that of $E_D$ on $D$,
the same rule reproduces \eqref{eq:ED_def_2}.

We now evaluate the one-loop factor from the
Dolbeault complex on $D$ with coefficients in the
virtual bundle $E_D$ of \eqref{eq:ED_def_2}.
Its full equivariant character is
\begin{align}
\label{eq:M2_full_character}
 \cK^{\rm full}_D(t)
 &\equiv
 \operatorname{Ind}_{-2\pi tK}
 \bigl(\overline{\partial}_D\otimes E_D\bigr)
 \nonumber\\
 &=
 \int_{\mathscr F_D}
 \frac{
 \operatorname{ch}_{-2\pi tK}(E_D)\,
 \operatorname{Td}_{-2\pi tK}(T^{1,0}D)}
 {e_{-2\pi tK}(N\mathscr F_D)}\, .
\end{align}
Here $\mathscr F_D$ is the fixed locus on $D$,
$N\mathscr F_D$ is its normal bundle in $D$, and
$\operatorname{ch}$, $\operatorname{Td}$ and $e$
denote the equivariant Chern character, Todd class
and Euler class.
For a compact manifold, the fixed point expression
follows from equivariant Hirzebruch--Riemann--Roch
and BVAB localization.
For fillings with boundary, we use this expression
to define the localized character, as in
\cite{BenettiGenolini:2026kjy}.

When the fixed points are isolated, this expression
can be written directly in terms of the tangent and fibre weights.
At $x\in\mathscr F_D$, let
$\lambda_{x,1},\lambda_{x,2}$ be the weights of
local holomorphic coordinate functions, so that
the tangent weights are
$-\lambda_{x,1},-\lambda_{x,2}$.
Let $v_{x,A}$, $A=1,\ldots,4$, be the fibre
weights of $V_D$, with our convention that a
weight $v$ contributes $\ex^{-tv}$.
The canonical bundle $K_D$ has weight
$\kappa_{D,x}=\lambda_{x,1}+\lambda_{x,2}$.
In the exterior power description \eqref{eq:ED_def},
the four components of the fermionic term
$\Lambda^3V_D$ are obtained by omitting one normal
basis vector. The component omitting the direction
of weight $v_{x,A}$ has weight
$\sum_{B\ne A}v_{x,B}=\kappa_{D,x}-v_{x,A}$,
since $\det V_D\cong K_D$.
This is the fermionic fibre weight corresponding
to the bosonic fibre weight $v_{x,A}$, equivalently
the weight in $K_D\otimes V_D^*$ in
\eqref{eq:ED_def_2}.
The full character therefore becomes
\begin{align}
\label{eq:M2_character_weights}
 \cK^{\rm full}_D(t)
 &=
 \sum_{x\in\mathscr F_D}
 \frac{
 \sum_{A=1}^{4}
 \left(
 \ex^{-tv_{x,A}}
 -
 \ex^{-t(\kappa_{D,x}-v_{x,A})}
 \right)}
 {(1-\ex^{-t\lambda_{x,1}})
  (1-\ex^{-t\lambda_{x,2}})}\, .
\end{align}
This is the M2-brane analogue of the explicit
bulk character in equation~(2.11) of
\cite{BenettiGenolini:2026kjy}.
The numerator records the fibre weights of
$V_D$ and $K_D\otimes V_D^*$, the bosonic
and fermionic terms in $E_D$, respectively,
while the denominator generates holomorphic
functions on the filling $D$.
For a linear plane, suppressing the fixed point
label, the monomial expansion gives the bosonic
and fermionic mode weights
\begin{align}
\label{eq:M2_mode_weights}
 \lambda^{\rm B}_{A,m,n}
 &=v_A+m\lambda_1+n\lambda_2\, ,
 \qquad\lambda^{\rm F}_{A,m,n}
 =\kappa_D-v_A+m\lambda_1+n\lambda_2\, ,
\end{align}
where $m,n\geq0$.
The superscripts denote the bosonic and fermionic
grading, rather than the sign of the weights.
For a finite quotient, we retain the invariant
modes, including the quotient action on both the
bundle fibres and the holomorphic functions.

If neither the bosonic nor the fermionic spectrum
contains zero weights, the full one-loop determinant
is obtained directly by zeta-function regularization:
\begin{align}
\label{eq:1loop_def}
 \zeta_{\cK}(s)
 &=
 \frac{1}{\Gamma(s)}
 \int_0^\infty \dd t\,t^{s-1}\cK(t)\, ,
 \qquad \log Z^{\rm full}_{\text{1-loop},D}
 =
 \zeta'_{\cK^{\rm full}_D}(0)\, .
\end{align}
The Mellin transform is understood by analytic
continuation, with consistent choices of the
logarithms of the spectral weights.
Equivalently, the determinant is the
zeta-function regularized product of fermionic weights
divided by the product of bosonic weights.
The choice of logarithms fixes the phases of
the local regularized determinants.
A globally defined M2-brane amplitude also requires
a consistent combination of the fermion Pfaffian
and the $C$-field coupling, as discussed in
section~\ref{subsec:HD_Corrections}.
Our fixed point sums assume compatible phase
choices for their components, with undetermined
relative phases between membrane sectors
kept explicit.

For two-derivative Freund--Rubin backgrounds,
membrane fluctuations were studied by direct
quantization of the quadratic worldvolume action
\cite{Beccaria:2023ujc,Gautason:2025per} and through
equivariant Dolbeault-type operators transverse
to the Reeb foliation
\cite{Kurlyand:2026yke,Kurlyand:2026gxi}.
In Appendix~\ref{app:fluctuation_complex}, we show
that the complex proposed here reproduces the graded cohomological
fluctuation spectrum of a membrane in the round
ABJM Freund--Rubin background, including its
$\Z_k$ projection and zero modes.

We next consider configurations for which the
fluctuation complex has zero modes.
By ``zero modes'', without further qualification,
we will always mean\footnote{These should be distinguished from
zero modes of the quadratic kinetic operators.
Some kinetic zero modes belong to supersymmetric
doublets with non-zero equivariant weights and
are treated by supersymmetric pairing and
regularization.
For an explicit discussion in Freund--Rubin
backgrounds, see section~4.1 of
\cite{Kurlyand:2026yke}.} cohomological zero modes:
unpaired modes with vanishing equivariant weight
for the chosen supercharge $\mathcal Q$.
In \eqref{eq:Z1M}, the bosonic 
zero modes are identified with infinitesimal
supersymmetric deformations tangent to the
family $\cM$.
The residual fermionic zero modes define the
obstruction spaces.
For a smooth family with constant fermionic
rank, the fermionic zero modes form a bundle
$\operatorname{Obs}\to\cM$; more generally,
the tangent and obstruction data have a
virtual interpretation.
Fermionic zero modes can also occur at isolated
configurations, where there are no bosonic
collective coordinates.
 The fluctuation complex must be compatible
with the geometry of the supersymmetric M2-brane
family $\cM$. In particular, its bosonic cohomological
zero modes must reproduce the tangent directions
to $\cM$, including their weights under equivariant
refinement. This agreement provides a non-trivial
consistency check of \eqref{eq:Z1M}.
The residual fermionic zero modes determine the
obstruction data entering the same expression.
Their interpretation as supplying the fermionic
Euler factor is part of our localization proposal;
these data need not coincide with the ordinary
obstruction theory of holomorphic submanifolds.

The zero modes must be treated separately.
The remaining fluctuations in the complex define the reduced
character and one-loop determinant,
\begin{align}
\label{eq:Kred_D_def}
 \cK^{\rm red}_D(t)
 &\equiv
 \bigl[\cK^{\rm full}_D(t)\bigr]'\, ,
 \qquad \log Z^{\rm red}_{\text{1-loop},D}
 =
 \zeta'_{\cK^{\rm red}_D}(0)\, .
\end{align}
Here the prime denotes omission of the bosonic and
fermionic 
zero modes assigned
to the chosen family. 
Bosonic and fermionic zero modes are counted
separately, even when their contributions
cancel in the character.
The bosonic collective coordinates and residual
fermionic zero modes supply the
contribution represented by the Euler factors
in \eqref{eq:Z1M}, whose expressions in terms
of fixed point weights are given in
\eqref{eq:M2_Euler_weights}.
The fermionic Euler class represents saturation
of the zero modes by supersymmetric curvature
couplings \cite{Witten:1988ze,Cordes:1994fc}.
It records their bundle over the family and
can be non-zero even though the corresponding
modes have zero equivariant weight under $K$.

In principle, the family contribution could
be evaluated without auxiliary refinement,
by combining the reduced determinant and
M2-brane action with a finite-dimensional
integration over $\cM$.
This would require identifying suitable global classes
for the action and determinant, together with
the family measure.
We do not construct such global data here.
Instead, we evaluate the fixed point expression
\eqref{eq:Z1M}, using the auxiliary torus $T$
to isolate the configurations in $\cM$.

Let $T$ be generated by $K$ and additional commuting
flavour generators $H_\ell$, chosen so that its action
on $\cM$ has only isolated fixed points;
we denote the fixed point set by $\cM^T$.
We use the induced action on the fillings and
their fluctuation complexes, with refined generator
$K_\epsilon=K+\sum_\ell\epsilon_\ell H_\ell$.
At each fixed point $[D_a]\in\cM^T$, denote the refined
weights of the bosonic family modes by
$\tau_{a,j}(\epsilon)$ and those of the fermionic
obstruction modes by $\phi_{a,\alpha}(\epsilon)$.
For a smooth family, isolation of the fixed points
ensures that the tangent weights $\tau_{a,j}$
are non-zero at generic refinement parameters.
Some fermionic weights $\phi_{a,\alpha}$ may
nevertheless remain zero.

The reduced character continues to exclude the
modes associated with the original family,
including those that acquire non-zero weights
after refinement.
The full and reduced characters are therefore
related by
\begin{align}
\label{eq:M2_character_recombination}
 \cK^{\rm full}_{D_a}(t;\epsilon)
 &=
 \cK^{\rm red}_{D_a}(t;\epsilon)
 +\sum_j\ex^{-t\tau_{a,j}(\epsilon)}
 -\sum_\alpha\ex^{-t\phi_{a,\alpha}(\epsilon)}\, .
\end{align}
Thus, to compute the reduced determinant, we
evaluate the character with $K_\epsilon$ and
subtract the finite contribution of the family
modes.
The same quotient projection is used throughout.
Together with their tangent and obstruction
Euler factors and the localized action evaluated
with $K_\epsilon$, this determines the fixed point
contribution in \eqref{eq:Z1M}.

Suppose first that the refinement also lifts
all the fermionic obstruction modes.
The omitted modes then give a finite correction
to the zeta function,
\begin{align}
 \zeta_{\cK^{\rm full}_{D_a}}(s)
 -
 \zeta_{\cK^{\rm red}_{D_a}}(s)
 &=
 \sum_j\tau_{a,j}^{-s}
 -
 \sum_\alpha\phi_{a,\alpha}^{-s}\, .
\end{align}
Using the same choices of logarithms and
differentiating at $s=0$, each restored bosonic
mode contributes $\tau_{a,j}^{-1}$ and each
restored fermionic mode contributes
$\phi_{a,\alpha}$.
Consequently,
\begin{align}
\label{eq:M2_family_recombination}
 Z^{\rm full}_{\text{1-loop},a}(\epsilon)
 &=
 \frac{e_T(\operatorname{Obs}_{[D_a]})}
      {e_T(T_{[D_a]}\cM)}
 Z^{\rm red}_{\text{1-loop},a}(\epsilon)\, .
\end{align}
The full determinant therefore includes the
Euler ratio and can be used directly in the
fixed point sum, without additional tangent
or obstruction Euler factors.
Here ``full'' refers to the unreduced
cohomological determinant.

If instead a fermionic obstruction mode remains
invariant under $T$, its refined weight
$\phi_{a,\alpha}(\epsilon)$ vanishes identically.
This mode must remain excluded from the zeta
function, so we retain the reduced determinant
and explicit Euler factors in \eqref{eq:Z1M}.
The mode contributes a zero factor to
$e_T(\operatorname{Obs}_{[D_a]})$, while the tangent
Euler factor $e_T(T_{[D_a]}\cM)$ is non-zero at generic auxiliary
parameters. The corresponding fixed point
contribution therefore vanishes, provided the
reduced determinant is regular at those parameters.

To recover the contribution of the original family,
we sum the fixed point contributions at generic
auxiliary parameters and only then take
$\epsilon\to0$.
The parameters of the original background and
the equivariant flux data are held fixed.
Individual terms can have poles even when
their sum has a finite limit.
The refinement dependence of both the action
and the determinant must therefore be retained
when extracting that limit.
Appendix~\ref{app:M2_family_localization}
explains the cancellation of these poles and
the resulting Taylor coefficient calculation. 
In the examples below, the bosonic zero mode weights
extracted from the fluctuation complex agree with the
geometric tangent weights of the family.
The cancellation of refinement poles and independence
of the limit from the generic direction in which the
auxiliary parameters are removed provide further
consistency checks of~\eqref{eq:Z1M}. 

Finally, in our holographic examples, physical mass
deformations are parametrized by flavour chemical
potentials. These enter the M2-brane action and
fluctuation complex through the same equivariant
weights used for the auxiliary refinement.
When the physical mass deformations isolate the
supersymmetric M2-brane configurations, no auxiliary
refinement is needed. If continuous families remain,
we introduce an additional refinement as described
above. Only the auxiliary parameters are then
removed, while the physical masses are held fixed.

\section{ABJM theory}
\label{sec:ABJM_Theory}

We now apply the localization formula \eqref{eq:Z1M} to
the M-theory dual of ABJM theory.
ABJM theory is a $3d$ $\cN=6$ $U(N)_k \times U(N)_{-k}$ Chern--Simons-matter theory with 
global symmetry algebra $\mf{so}(6)_R \oplus \mf{u}(1)_\text{top}$ for $k>2$, enhanced to
 $\mf{so}(8)_R$ for $k=1,2$ \cite{Aharony:2008ug}.
The theory can be placed on a three-sphere with a $U(1)\times U(1)$-preserving squashing parameter $b$, while turning on real masses for the flavour symmetries commuting with a chosen $\cN=2$ subalgebra. These deformations generically preserve $\cN=2$ supersymmetry, but on special loci can preserve more.

For backgrounds dual to ABJM theory on a squashed
three-sphere, we specialize the fibrations in \eqref{eq:W_Productfibred}
to the
topologically trivial case
$W\cong M_4\times Z_8$, with
$M_4\cong\mathbb R^4$ and $Z_8=\mathbb C^4/\mathbb Z_k$.
The Euclidean M2-branes are localized at the fixed point
of $K$ in $M_4$ and wrap three-cycles
$\Sigma\subset S^7/\mathbb Z_k$.
We restrict to holomorphic fillings represented on the covering space by complex two-planes $D\cong\C^2\subset\C^4$. We first classify such linear fillings and their stabilizers, impose $K$-invariance, and compute their localized actions. We then evaluate the fluctuation determinants and fixed point sums for M2-brane families, and compare the resulting instanton contributions with field theory. Throughout this section, $J_{\rm full}$, $J_{\rm gen}$ and $J_{\rm lin}$ refer only to contributions obtained from these linear fillings.

\subsection{Linear fillings and their stabilizers}
\label{subsec:Geometry_Semiclassical_Action}

We consider linear holomorphic fillings in the internal factor
$Z_8\cong\C^4/\Z_k$. In the complex structure used for M2-brane
localization, the generator of $\Z_k$ acts with weights $(1,1,-1,-1)$:
\begin{align}
\label{eq:CG_ABJM_quotient}
 (z_1,z_2,z_3,z_4)
 \longmapsto
 (\omega z_1,\omega z_2,\omega^{-1}z_3,\omega^{-1}z_4) \, , 
 \qquad \omega=\ex^{2\pi\ii/k} \, .
\end{align}
The holomorphic four-form is invariant under this action and
hence descends to the quotient. 
In these coordinates the Reeb  vector instead generates
the common phase rotation of all four $z_I$,
with weights $(1,1,1,1)$, and is distinct
from the circle action defining the quotient~\eqref{eq:CG_ABJM_quotient}.
At the round superconformal point, the internal
component of $K$ is $-\frac12$ times this
Reeb vector, consistently with
\eqref{eq:Intro_K_2}.

\begin{figure}
\begin{center}
\begin{tikzpicture}[scale=.8,
  every node/.style={font=\small},
  vertex/.style={circle,fill=black,inner sep=1.4pt}]

\coordinate (A) at (0,2.2);      
\coordinate (Ap) at (0,-2.2);   

\coordinate (B) at (-2,0);       
\coordinate (Bp) at (2,0);       

\coordinate (C) at (0.8,0.75);   
\coordinate (Cp) at (-0.8,-0.75);

\draw[dashed,gray]
  (Cp)--(A)
  (Cp)--(B)
  (Cp)--(Ap)
  (Cp)--(Bp);

\draw[thick]
  (C)--(A)
  (C)--(B)
  (C)--(Ap)
  (C)--(Bp);

\draw[thick]
  (A)--(B)
  (B)--(Ap)
  (Ap)--(Bp)
  (Bp)--(A);

\node[vertex,label=above:$W_{12}$] at (A) {};
\node[vertex,label=below:$W_{34}$] at (Ap) {};

\node[vertex,label=left:$W_{13}$] at (B) {};
\node[vertex,label=right:$W_{24}$] at (Bp) {};

\node[vertex,label=above right:$W_{14}$] at (C) {};
\node[vertex,label=below left:$W_{23}$] at (Cp) {};

\end{tikzpicture}
\end{center}
\caption{The moment polytope of $\operatorname{Gr}(2,4)$. It is the moment map image of the family of holomorphic two-planes in the $\C^4$ geometry of ABJM at level $k=1$.}
\label{fig:Delta24}
\end{figure}
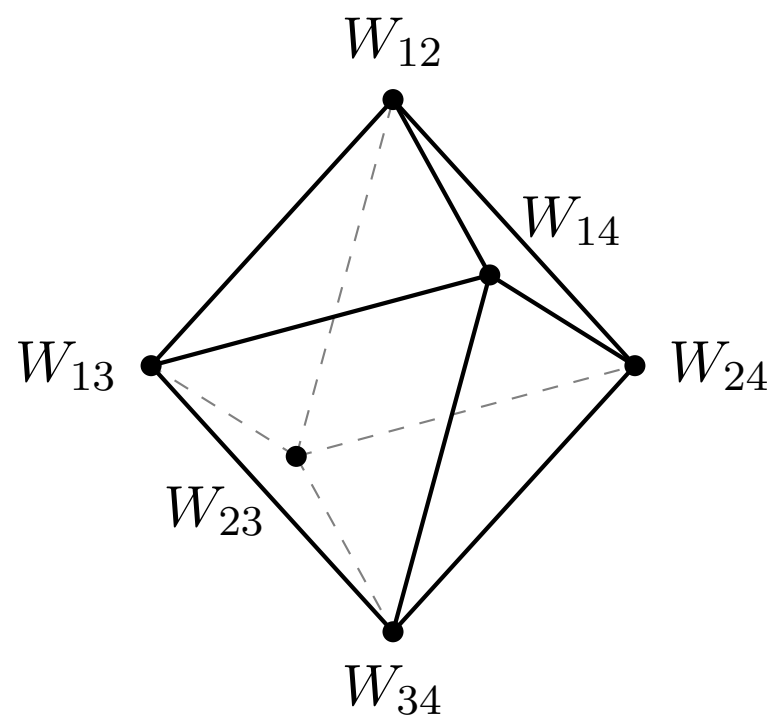

A linear filling is represented on the covering space $\C^4$ by a
complex two-plane $D\cong\C^2\subset\C^4$ through the origin.
These planes are parametrized by
\begin{align}
\label{eq:Mhol}
 \cM_{\rm hol}=\operatorname{Gr}(2,4)\, ,
\end{align}
which has complex dimension four. 
At the round superconformal point, a holomorphic filling on
$\C^4$ preserved by the common phase Reeb action is a cone over
a projective curve $C\subset\CP^3_r$, where $\CP^3_r$ is the
Reeb quotient of $S^7$ (distinct from the projective space
used in the type IIA reduction; see Appendix~\ref{app:ABJM_IIA}).
The linear fillings are cones over degree one curves, namely
projective lines, and hence form $\operatorname{Gr}(2,4)$.
Higher-degree curves, with $d=\int_C c_1(\mathcal O(1))$,
have 
localized
actions $I_{\rm M2}=4d\mu$ in the unquotiented
round background ($k=1$). With generic flavour parameters,
the equivariant action depends on the invariant curve, rather
than on $d$ alone. Multiple covers and reducible or non-reduced
limits can occur in compactifications of the higher-degree
moduli problem. We restrict our explicit calculations to
linear fillings, and do not attempt to construct those more
general moduli spaces here.
We write
$W_{IJ}=\operatorname{span}_{\C}(\ex_I,\ex_J)$
for the six coordinate planes, corresponding to the vertices
of the moment polytope in Figure~\ref{fig:Delta24}.

The quotient geometry of a filling depends on the setwise
stabilizer of its covering plane,
\begin{align}\label{staborderGammaD}
 \Gamma_D=\{g\in\Z_k\mid gD=D\} \, , 
 \qquad s=|\Gamma_D| \, .
\end{align}
The link of $D$ upstairs is a three-sphere, and $\Gamma_D$
acts freely on it. The parametrized M2-brane worldvolume
on the quotient is therefore
\begin{align}
\label{eq:M2-worldvolumes}
 \Sigma[D]\cong(D\cap S^7)/\Gamma_D
 \cong S^3/\Z_s\, ,
\end{align}
where $\Gamma_D$ acts on $D\cap S^7$ by restriction of
\eqref{eq:CG_ABJM_quotient}.
Planes in the same $\Z_k$ orbit describe the same quotient
configuration.
We denote the locus of planes preserved by the entire
quotient group by
$\cM_k\equiv\operatorname{Gr}(2,4)^{\Z_k}$.

The quotient geometry depends on $k$ as follows:
\begin{enumerate}
\item For $k=1$, there is no quotient. Thus
$\cM_1=\operatorname{Gr}(2,4)$, every plane has
trivial stabilizer and its worldvolume is $S^3$.

\item For $k=2$, the generator of $\Z_k$ acts as $-1$
on $\C^4$ and preserves every plane. Thus
$\cM_2=\operatorname{Gr}(2,4)$, every plane has stabilizer
$\Z_2$, and its worldvolume is $\mathbb{RP}^3$.

\item For $k>2$, planes divide into two classes:
those preserved by the full $\Z_k$, forming $\cM_k$,
and the remaining planes, which we call generic planes.
The latter have stabilizer $\Z_h$, where $h=\gcd(k,2)$.
\end{enumerate}

For the remainder of this subsection we take $k>2$
and describe these two classes, starting with $D\in\cM_k$.
The generator of $\Z_k$ has two eigenspaces,
$W_{12}$ and $W_{34}$, with eigenvalues $\omega$ and
$\omega^{-1}$, respectively, and
$\C^4=W_{12}\oplus W_{34}$.
A plane preserved by the generator of $\Z_k$ splits as
$D=(D\cap W_{12})\oplus(D\cap W_{34})$.
An invariant two-plane is therefore either $W_{12}$,
$W_{34}$, or the direct sum of one complex line in
$W_{12}$ and one in $W_{34}$. The latter choices are
parametrized by $\CP^1\times\CP^1$, so
\begin{align}
\label{eq:ABJM_full_stabilizer_locus}
 \cM_k=
 \{W_{12}\}\sqcup(\CP^1\times\CP^1)\sqcup\{W_{34}\} \, , 
 \qquad k>2 \, .
\end{align}
Every plane in $\cM_k$ has stabilizer $\Z_k$ and M2-brane
worldvolume $S^3/\Z_k$.
The four coordinate planes $W_{13},W_{14},W_{23},W_{24}$
lie in the $\CP^1\times\CP^1$ component. Thus, in particular, all six
coordinate planes have the full stabilizer. 

For a generic plane $D\notin\cM_k$, the stabilizer contains the identity and, for even $k$, the antipodal element. It cannot contain any other element: such an element has distinct eigenvalues on $W_{12}$ and $W_{34}$, so invariance would force $D$ to split as above and hence belong to $\cM_k$. Its stabilizer is therefore precisely $\Z_h$, with $h=\gcd(k,2)$. For odd $k$, the stabilizer is trivial and the M2-brane worldvolume is $S^3$; for even $k$, the stabilizer is the antipodal $\Z_2$ and the worldvolume is $\mathbb{RP}^3$.

The orbit of a generic plane contains $k/h$ distinct planes.
For example, near $W_{12}$ a plane can be written as
\begin{align}
\label{eq:Holomorphic_Plane_z12}
 \begin{pmatrix}z_3\\z_4\end{pmatrix}
 =
 A\begin{pmatrix}z_1\\z_2\end{pmatrix}\, ,
 \qquad A\in\operatorname{Hom}(W_{12},W_{34}) \, , 
\end{align}
and the quotient acts by $A\mapsto\omega^{-2}A$. 
As $A\to0$, the $k/h$ orbit representatives
approach the same coordinate plane $W_{12}$.
However, the limiting map of the M2-brane
worldvolume onto the quotient link of $W_{12}$
is a covering of degree $k/h$: 
\begin{align}
 S^3/\Z_h\longrightarrow S^3/\Z_k\, ,
 \qquad \deg=\frac{k}{h}\, ,\qquad k>2\, .
\end{align}
Thus a generic orbit M2-brane can degenerate
to a multiple cover of the quotient link of
a coordinate plane. This limiting configuration
must be distinguished from an M2-brane
wrapping the same cycle once.

The relation of these M2-brane configurations to worldsheet
and D2-brane instantons in type IIA is discussed in
Appendix~\ref{app:ABJM_IIA}.

\subsection{The action of \texorpdfstring{$K$}{K} and the M2-brane action}
\label{subsec:ABJM_K_Action}
We now 
determine which of the
linear fillings described above are preserved by $K$,
as required by \eqref{eq:BPS_Condition_Invariance}.
 The M2-brane lies over the fixed
point of $K$ in $M_4$, and its internal filling must be
preserved by the action of $K$ on $\C^4/\Z_k$.

The external weights of $K$ are labelled $b_1,b_2$ and we normalize $K$ so that 
$b_1+b_2=2$. The ratio $b_1/b_2=b^2$ determines the squashing of the
three-sphere at the conformal boundary of $M_4$, where it is also convenient to introduce $Q$ via
\begin{align}
\label{eq:ABJM_ExternalWeights}
 b_1=\frac{2b}{Q} \, , 
 \qquad b_2=\frac{2}{bQ}  \, , 
 \qquad Q\equiv b+b^{-1} \, .
\end{align}
In section \ref{subsec:ABJM_Squashed_Sphere} it will be convenient to use $ f \equiv \frac{b^2-1}{b^2+1}$. 
We write the internal weights as
\begin{align}
\label{eq:ABJM_InternalWeights}
 u_A=-\Delta_A \, , 
 \qquad \sum_{A=1}^4\Delta_A=2 \, .
\end{align}
Together these satisfy
$b_1+b_2+\sum_Au_A=0$. This expresses the
vanishing $K$-charge of the preserved complex
Killing spinor: the contributions from the external
and internal spinor representations cancel.
Equivalently, the local holomorphic volume element
on $W$ has zero $K$-weight, in accordance with the
choice of extension in
section~\ref{subsec:M2_complex_geometry}. 
The $\Delta_A$ encode chemical potentials for the global
flavour symmetries of the dual field theory, with physical real
masses obtained by analytic continuation. 
At the round superconformal point, where
$b=1$ and $\Delta_A=\frac12$, the internal
component of $K$ is
$-\frac12\sum_A\partial_{\arg z_A}
=-\frac12\xi$, in agreement with
\eqref{eq:Intro_K_2}.

The planes preserved by $K$ are the fixed points of its
induced action on $\operatorname{Gr}(2,4)$. For distinct
internal weights $\Delta_A$, these are precisely the six
coordinate planes $W_{IJ}$ (the six vertices in Figure~\ref{fig:Delta24}).
For every $k$, each coordinate
plane is also preserved by the entire quotient group
$\Z_k$ and therefore has full stabilizer.
Varying the squashing parameter $b$ at fixed $\Delta_A$
changes the external weights but leaves the invariant
planes unchanged.

To evaluate the M2-brane action, we use the fixed
locus of $K$ within each filling. For a $K$-invariant plane
$D$ with non-zero tangent weights, this consists only of
the origin of the plane. 
All linear fillings pass through the origin of
$\C^4/\Z_k$ and hence share the same value of
$\Phi^G_0$, the zero-form component of the
ambient equivariant flux class. There is therefore a single flux variable $\mu$, defined
by \eqref{eq:mu_a_Def}, which is conjugate to the M2-brane
number $N$.

Let $u_{D,1}$ and $u_{D,2}$ be the two tangent weights of $K$
on $D$, and let $s=|\Gamma_D|$ be its stabilizer order defined in \eqref{staborderGammaD}.
Using \eqref{eq:SM2_Generic} and $b_1b_2=4/Q^2$, we obtain
\begin{align}
 I_{\rm M2}[D]
 =\frac{4\mu}{sQ^2u_{D,1}u_{D,2}}\, .
\end{align}
We take the M2-brane orientation compatible with the preserved
supersymmetry, as in section~\ref{subsec:CalibratedMembranes}. 
For an M2-brane wrapping the quotient link of the coordinate plane $W_{IJ}$ once, the
tangent weights are $-\Delta_I$ and $-\Delta_J$, and $s=k$.
Its action is therefore
\begin{align}
\label{eq:ClassicalS_IJ_h}
 I_{\rm M2}[W_{IJ}]
 =\frac{4\mu}{kQ^2\Delta_I\Delta_J}\, .
\end{align}
For $k>2$, the same coordinate support can also represent
the limiting configuration inherited from a generic orbit.
As explained in the preceding subsection, this configuration
retains covering degree $k/h$, with $h=\gcd(k,2)$, and hence
has action
\begin{align}
\label{eq:ABJM_generic_orbit_action}
 I^{\rm gen}_{IJ}
 =\frac{k}{h}I_{\rm M2}[W_{IJ}]
 =\frac{4\mu}{hQ^2\Delta_I\Delta_J}\, . 
\end{align}
The covering multiplicity thus distinguishes the 
exponential
weights even when the coordinate supports coincide.
The treatment of fluctuations about these limiting
configurations will be discussed in the next subsection.

When internal weights coincide, additional planes can
become $K$-invariant. In particular, near $W_{IJ}$ the
linear graph deformations have weights
$\Delta_i-\Delta_A$, with $i\in\{I,J\}$ and
$A\notin\{I,J\}$. Vanishing differences allow continuous
families of invariant planes. Their quotient stabilizers
remain those classified in the preceding subsection.

When all four internal weights coincide, 
$\Delta_A=\tfrac12$, the internal action of $K$ then
preserves every plane in $\operatorname{Gr}(2,4)$.
This holds for arbitrary squashing $b$, and the action
of an M2-brane on a linear filling with stabilizer order $s$ is
\begin{align}
 I_{\rm M2}[D]=\frac{16\mu}{sQ^2}\, .
\end{align}
At the round superconformal point, $b=1$ and $Q=2$.
We find
\begin{align}
\label{eq:ABJM_Round_M2_Action}
 I_{\rm M2}[D]=\frac{4\mu}{s}\, ,
 \qquad
 \ex^{-I_{\rm M2}[D]}=\ex^{-4\mu/s}\, .
\end{align}
For $k=1,2$, every plane has $s=k$, giving a common action
$4\mu/k$ throughout the Grassmannian family.
For $k>2$, planes in $\cM_k$ have action $4\mu/k$, while
generic planes have action $4\mu/h$, in agreement with
the action of their covered limiting representatives.

\subsection[Fluctuation determinants and fixed point sums]{\texorpdfstring{Fluctuation determinants and fixed point sums}{Fluctuation determinants and fixed point sums}}
\label{subsec:ABJM_Fluctuations}

We now compute the fluctuation determinants and family
contributions entering  \eqref{eq:Z1M} for the linear fillings
discussed above, 
using the construction in
section~\ref{subsec:Quantum_Brane_Families} and the localized
actions obtained in section~\ref{subsec:ABJM_K_Action}.
We first work at generic values of $b$ and $\Delta_A$,
where the six coordinate planes $W_{IJ}$ are isolated
$K$-invariant fillings with no bosonic or fermionic zero modes.
We then discuss special loci with continuous families,
residual fermionic zero modes, or both. 

\subsubsection{The projected fluctuation determinant}

Let $D=W_{IJ}$ and denote the complementary indices by $K,L$.
The normal bundle $V_D$ in \eqref{eq:VD_def} has fibre weights
\begin{align}
\label{eq:ABJM_Weights_VD}
 (v_1,v_2,v_3,v_4)
 =(b_1,b_2,-\Delta_K,-\Delta_L)\, ,
 \qquad
 \sum_{a=1}^4v_a=\Delta_I+\Delta_J\, ,
\end{align}
where $b_1,b_2$ are given in \eqref{eq:ABJM_ExternalWeights}.
Using the convention of section~\ref{subsec:Quantum_Brane_Families},
in which a weight $\lambda$ contributes $\ex^{-t\lambda}$,
the full character on the covering plane is
\begin{align}
\label{eq:ABJM_Unreduced_KD}
 \cK^{\rm full}_{IJ}(t)
 =\frac{\displaystyle\sum_{a=1}^4
    \left[\ex^{-tv_a}-\ex^{-t(\Delta_I+\Delta_J-v_a)}\right]}
   {(1-\ex^{-t\Delta_I})(1-\ex^{-t\Delta_J})}\, .
\end{align}
The numerator follows from $E_D=V_D-K_D\otimes V_D^*$,
while the denominator accounts for holomorphic functions on $D$.
The superscript ``full'' means that no bosonic tangent or fermionic obstruction modes
have been omitted. Quotient projection is indicated
by an additional stabilizer subscript $s$.
Expanding the denominator gives
\begin{align}
 \cK^{\rm full}_{IJ}(t)
 =\sum_{a=1}^{4}\sum_{m,n\geq0}
 \left[
 \ex^{-t\lambda^{\rm B}_{a,m,n}}
 -\ex^{-t\lambda^{\rm F}_{a,m,n}}
 \right]\, ,
\end{align}
where the bosonic and fermionic weights are, respectively,
\begin{align}
\label{eq:ABJM_linear_mode_weights}
 \lambda^{\rm B}_{a,m,n}=v_a+m\Delta_I+n\Delta_J\, ,
 \qquad
 \lambda^{\rm F}_{a,m,n}
 =\Delta_I+\Delta_J-v_a+m\Delta_I+n\Delta_J\, ,
\end{align}
for $m,n\geq0$.

For an M2-brane with stabilizer $\Z_s$, we retain
the $\Z_s$-invariant part of this spectrum.
Write
$(\gamma_1,\gamma_2,\gamma_3,\gamma_4)=(1,1,-1,-1)$
for the charges in \eqref{eq:CG_ABJM_quotient}.
In addition to its $K$-weight $v_a$, each component of the bosonic term $V_D$
carries a charge $c_a$ under the quotient action. In the
same ordering as the weights in \eqref{eq:ABJM_Weights_VD},
these charges are
\begin{align}
 (c_1,c_2,c_3,c_4)=(0,0,\gamma_K,\gamma_L)\, .
\end{align}
The charge of the canonical bundle $K_D$ is
\begin{align}\label{ceedeedef}
 c_D=-\gamma_I-\gamma_J\, ,
\end{align} so 
the fermionic term $K_D\otimes V_D^*$
has fibre charges $c_D-c_a$. 
With the action on functions defined by
$(g\cdot f)(z)=f(g^{-1}z)$, a monomial $z_I^mz_J^n$
has charge $-m\gamma_I-n\gamma_J$.
Thus the Molien projection
\begin{align}
\label{eq:ABJM_Molien_character}
 \cK^{\rm full}_{IJ,s}(t)
 =\frac1s\sum_{g\in\Z_s}\cK^{\rm full}_{IJ}(t;g)
\end{align}
retains a mode with fibre charge $c$ precisely when
$c-m\gamma_I-n\gamma_J=0\pmod s$.
Here $c=c_a$ for a bosonic component and $c=c_D-c_a$
for its fermionic partner.
In particular, the projection acts on the bundle fibres as well
as on the holomorphic functions.

For a fixed plane, define the projected double zeta function
\begin{align}
\label{eq:ABJM_projected_double_zeta}
 \zeta^{IJ}_{2,s}(z,x;c)
 =\sum_{\substack{m,n\geq0\\
             c-m\gamma_I-n\gamma_J=0\ ({\rm mod}\ s)}}
       (x+m\Delta_I+n\Delta_J)^{-z}\, .
\end{align}
For $s=1$ this is the Barnes double zeta function with periods
$\Delta_I,\Delta_J$. For general $s$, splitting $m,n$ into
residue classes modulo $s$ expresses it as a finite sum of
Barnes functions with periods $s\Delta_I,s\Delta_J$.
The congruence condition in the sum already implements the group average;
there is no additional factor $1/s$.
 We write $\zeta^{\rm full}_{IJ,s}=\zeta_{\cK^{\rm full}_{IJ,s}}$
for the spectral zeta function of the projected character
and denote the full and reduced one-loop determinants
by $Z^{\rm full}_{IJ,s}$ and $Z^{\rm red}_{IJ,s}$, respectively.
The full determinant on the quotient worldvolume
$S^3/\Z_s$ is therefore given by
\begin{align}
\label{eq:ABJM_generic_Barnes_determinant}
 \zeta^{\rm full}_{IJ,s}(z)
 &=\sum_{a=1}^4\left[
   \zeta^{IJ}_{2,s}(z,v_a;c_a)
   -\zeta^{IJ}_{2,s}(z,\Delta_I+\Delta_J-v_a;c_D-c_a)
   \right]\, ,\nonumber\\
 \log Z^{\rm full}_{IJ,s}(b,\Delta)
 &=\left.\partial_z\zeta^{\rm full}_{IJ,s}(z)\right|_{z=0}\, .
\end{align}
These expressions are understood by analytic continuation,
with the fixed spectral branch and phase convention used in
\eqref{eq:1loop_def}. 

In particular, for $s=1$ there is no quotient projection and the determinant
can be written as a product of double sine functions:
\begin{align}
  Z^{\rm full}_{IJ,1}(b,\Delta) &= \prod_{a=1}^4 S_2(v_a | \Delta_I, \Delta_J)^{-1} \, .
\end{align}
The zeroes and poles of $S_2(x|\Delta_I, \Delta_J)$ occur at 
 $-m\Delta_I-n\Delta_J$ and $(m+1)\Delta_I+(n+1)\Delta_J$, respectively, for $m,n\geq0$ (note the correspondence with \eqref{eq:ABJM_linear_mode_weights}).
Using the properties of the double sine function and analytic continuation, we can find some additional structure in this result. Rescaling the argument and periods by the same factor gives
\begin{align}
  S_2(v_a | \Delta_I, \Delta_J) & = S_2 \left( \frac{Q' r_a}{2} \bigg| b', b'^{-1} \right) \, , \nonumber\\ b' & = \sqrt{\frac{\Delta_I}{\Delta_J}} \, , \quad Q' = b' + b'^{-1} \, , \quad r_a = \frac{2v_a}{\Delta_I + \Delta_J} \, .
\end{align}
Thus, we find that $Z_{IJ,1}^{\rm full}(b,\Delta)$ is the product of the $S^3_{b'}$ partition functions of four $\cN=2$ chiral multiplets with R-charges $r_a$ \cite{Hama:2011ea}.\footnote{These are more commonly written in terms of $s_b(z) \equiv S_2(Q/2 - \ii z|b,b^{-1})$, so a chiral multiplet of R-charge $r$ contributes $s_b( \ii Q(1-r)/2) = S_2(Qr/2|b,b^{-1})^{-1}$.}
Their appearance can be structurally traced back to the four complex normal directions of $V_D$ (or, equivalently, the eight real transverse scalars to the M2-brane), with the charge relation $\sum_a r_a = 2$ due to the relevant bundle being $E_D$ rather than $V_D$ \eqref{eq:ED_def_2}.\footnote{The appearance of $3d$ $\cN=2$ chiral determinants was noted in \cite{Gautason:2025per}. Our proposed fluctuation complex makes their organization explicit.}

\subsubsection{Isolated saddles and brane families}

For generic values of $b$ and $\Delta_A$, we first compute
the contribution of M2-branes wrapping once the quotient
links of the six coordinate planes $W_{IJ}$. These planes are
isolated $K$-invariant fillings with stabilizer $\Z_k$, and
there are no unpaired bosonic or fermionic zero modes.
The Euler factors in \eqref{eq:Z1M} are therefore
trivial, and the reduced and full determinants
coincide.  
Combining them with the M2-brane actions
\eqref{eq:ClassicalS_IJ_h} gives the connected contribution
to the modified grand potential:
\begin{align}
\label{eq:ABJM_generic_fixed_point_sum}
 J_{\rm full}(k,\mu;b,\Delta)
 =\sum_{I<J}Z^{\rm full}_{IJ,k}(b,\Delta)
       \exp\!\left[-\frac{4\mu}{kQ^2\Delta_I\Delta_J}\right]\, .
\end{align}
We count each singly wrapped quotient configuration once, with
no additional overall factor $1/k$; its action and local
spectrum already incorporate the worldvolume quotient.
The subscript ``full'' in $J_{\rm full}$ labels
the contribution from the full-stabilizer locus $\cM_k$.
The superscript ``full'' in $Z^{\rm full}_{IJ,k}$
denotes the unreduced cohomological one-loop determinant
defined in section~\ref{subsec:Quantum_Brane_Families}.

Depending on $b$ and $\Delta_A$, the projected spectrum may
contain bosonic zero modes, residual fermionic zero modes,
both, or neither. The degree one internal bosonic modes have
weights $\Delta_i-\Delta_A$ and charges $\gamma_A-\gamma_i$,
where $i\in\{I,J\}$ and $A\notin\{I,J\}$; when their weights
vanish, they describe infinitesimal deformations of the plane
within a continuous family $\cM$. The two lowest external
fermions have weights $\Delta_I+\Delta_J-b_\alpha$,
$\alpha=1,2$, and charge $c_D$.
Only modes invariant under the stabilizer $\Gamma_D=\Z_s$
survive: $s=k$ in $\cM_k$, while $s=h=\gcd(k,2)$ on the
generic orbit stratum for $k>2$. The complete zero mode sector
must be determined from \eqref{eq:ABJM_linear_mode_weights}
and its stabilizer projection, since additional modes can
become massless at special values of $b$ and $\Delta_A$.

When zero modes occur, we evaluate the fixed point expression
\eqref{eq:Z1M} using the reduced determinant and Euler factors,
as explained in section~\ref{subsec:Quantum_Brane_Families}. 
For a family whose auxiliary fixed fillings are coordinate
planes, we assume that compatible lifts of the fluctuation
complex are represented by the corresponding weight shifts
in \eqref{eq:ABJM_Unreduced_KD}, with the stabilizer projection
retained. 
When the auxiliary torus action lifts all omitted modes, the
projected zeta functions compute the full determinants in
\eqref{eq:M2_family_recombination}, which already include
the Euler ratio. If fermionic modes remain unlifted, we
instead use the reduced determinant and Euler ratio in
\eqref{eq:Z1M}. Both cases occur in the
round sphere examples below, where the fixed point
contributions are summed before removing the auxiliary
parameters.
For $k=1,2$, every covering plane has full
stabilizer $\Z_k$. There is therefore a single
linear sector, described by
\eqref{eq:ABJM_generic_fixed_point_sum} and its
family limits. 

For $k>2$, the generic orbit configurations have covered
coordinate limits described in
sections~\ref{subsec:Geometry_Semiclassical_Action}
and~\ref{subsec:ABJM_K_Action}.
Their covering multiplicity fixes the action
\eqref{eq:ABJM_generic_orbit_action}, but does not determine
the fluctuation determinant. We propose extending the
$h$-projected complex and residual quotient measure to
these limiting configurations. This gives
\begin{align}
\label{eq:ABJM_generic_orbit_refined_allk}
 J_{\rm gen}(k,\mu;b,\Delta)
 =\ex^{\ii\vartheta_k}\frac{h}{k}
   \sum_{I<J}Z^{\rm full}_{IJ,h}(b,\Delta)
       \exp\!\left[-\frac{4\mu}{hQ^2\Delta_I\Delta_J}\right]\, ,
 \qquad k>2\, .
\end{align}
The factor $h/k$ reflects the identification of
the $k/h$ representatives of a generic orbit.
We assume that this factor also applies to their
covered limits, and that the corresponding
fluctuations are described by the continued
$\Z_h$-projected determinant. Neither assumption
follows from the local calculation.
The common spectral convention of
section~\ref{subsec:Quantum_Brane_Families}
fixes the local determinant phases.
The factor $\ex^{\ii\vartheta_k}$ parametrizes
the remaining undetermined global phase of
$J_{\rm gen}$ relative to $J_{\rm full}$.
Its relation to the fermion Pfaffian and
$C$-field data discussed in
section~\ref{subsec:HD_Corrections}
remains to be established.
We assume that this phase is unchanged under
the continuous deformations considered,
with the global data held fixed.

For generic physical masses, the only $K$-invariant
linear planes are the six coordinate planes.
However, their quotient links also admit
$k/h$-fold covered M2-brane maps, which remain
$K$-invariant. These are distinct from the
singly wrapped configurations contributing to
$J_{\rm full}$, even though their images coincide.
At the superconformal point, the covered maps
arise as limits of the generic orbit family,
as described in
section~\ref{subsec:Geometry_Semiclassical_Action}.
This suggests interpreting $J_{\rm gen}$ through
localization on a suitable compactification of
the moduli space of M2-brane maps, rather than
as a contribution from non-invariant generic
planes at generic masses. 
We have not constructed this compactification,
and the continuation
\eqref{eq:ABJM_generic_orbit_refined_allk}
remains conjectural. It treats the covered limits
of the linear families; a general treatment of
multiply covered M2-branes remains beyond the
present analysis.

These two sector sums provide the starting point for the
specializations below. We first evaluate the round sphere
two-mass locus and its superconformal limit, before turning
to squashing deformations.

\subsection{Round sphere: masses and the superconformal limit}
\label{subsec:ABJM_Round}

We now evaluate the sector sums of the preceding subsection
on the round sphere, $b=1$. A two-mass locus first reduces the
projected determinants to elementary functions. Its
superconformal limit then exhibits the contributions of the
continuous M2-brane families.

\subsubsection{The Nosaka two-mass locus}
\label{subsubsec:ABJM_Nosaka}

Set $b=1$, so that $b_1=b_2=1$, and take
\begin{align}
\label{eq:ABJM_round_Nosaka_weights}
 (\Delta_1,\Delta_2,\Delta_3,\Delta_4)
 =\frac12(1+\xi,1+\eta,1-\xi,1-\eta)\, .
\end{align}
This is the two-mass locus studied in
\cite{Nosaka:2015iiw,Nosaka:2024gle}.
For generic $\xi,\eta$, the four internal weights are
distinct, so the six coordinate planes $W_{IJ}$ are the
only $K$-invariant linear fillings.
All six planes belong to $\cM_k$ and have stabilizer $\Z_k$.
We first compute M2-branes wrapping their quotient
links once.
At $W_{13}$ and $W_{24}$, however,
$\Delta_I+\Delta_J=1$, so the two lowest external fermions
have zero weight. Their quotient charge $c_D$ in \eqref{ceedeedef} also vanishes,
so they survive the $\Z_k$ projection. 
At generic $\xi,\eta$, the reduced determinants are finite.
The vanishing fermionic Euler factors in
\eqref{eq:Z1M} therefore imply that the
$W_{13}$ and $W_{24}$ instanton contributions vanish.

For each of the four remaining planes, choose the complementary
indices so that
$\Delta_K=1-\Delta_I$, $\Delta_L=1-\Delta_J$ and
$\gamma_K=-\gamma_I$, $\gamma_L=-\gamma_J$.
The covering character \eqref{eq:ABJM_Unreduced_KD} then splits
into two one-dimensional characters\footnote{Formally, this represents the fact that $E_D$ in (equivariant) K-theory is the sum of two bundles supported on the coordinate axes. Concretely, each of the summands represents the index of the Dolbeault operator on the line $\ell_\Delta \cong \C$ valued in the rank two bundle $\cL \oplus (K_{\ell_\Delta} \otimes \cL^{-1})$, where $\cL$ is an external normal bundle of equivariant weight $1$, and $K_{\ell_\Delta}$ has equivariant weight $\Delta$.}:
\begin{align}
\label{eq:ABJM_round_Nosaka_character}
 \cK^{\rm full}_{IJ}(t)
 =\frac{\ex^{-t}+\ex^{-t(\Delta_I-1)}}{1-\ex^{-t\Delta_I}}
  +\frac{\ex^{-t}+\ex^{-t(\Delta_J-1)}}{1-\ex^{-t\Delta_J}}\, .
\end{align}
For the term associated with $\Delta_i$, the two towers have
charges $-m\gamma_i$ and $-(m+1)\gamma_i$, respectively.
Since $\gamma_i=\pm1$, the $\Z_k$ projection retains $m=0$
modulo $k$ in the first tower and $m+1=0$ modulo $k$ in the
second. Thus
\begin{align}
\label{eq:ABJM_round_Nosaka_projected_character}
 \cK^{\rm full}_{IJ,k}(t)
 =\sum_{i\in\{I,J\}}\sum_{n\geq0}
 \left[\ex^{-t(1+kn\Delta_i)}
       +\ex^{-t(k\Delta_i-1+kn\Delta_i)}\right]\, .
\end{align}
The double zeta functions of section~\ref{subsec:ABJM_Fluctuations}
therefore reduce to 
\begin{align}
\label{eq:ABJM_round_Nosaka_Hurwitz}
 \zeta_{\Delta,k}(z)
 =(k\Delta)^{-z}\left[
 \zeta_H\!\left(z,\frac1{k\Delta}\right)
 +\zeta_H\!\left(z,1-\frac1{k\Delta}\right)\right] \, , 
\end{align}
where the Hurwitz zeta function is initially defined by 
\begin{align}
	\zeta_H\left( z, x \right) = \sum_{n \geq 0} \frac{1}{(x + n)^z} \, , \qquad \Re z > 1 \, , \ \Re x > 0 \, ,
\end{align}
and then analytically continued in $z$. We use its value and derivative at $z=0$, together with
the reflection formula for the Gamma function:
\begin{align}
\label{eq:ABJM_Hurwitz_Cool}
	\zeta_H(0,x) = \frac{1}{2} - x \, , \quad \zeta'_H(0,x) = \log \Gamma(x) - \frac{1}{2} \log (2\pi) \, , \quad \Gamma(x) \Gamma(1-x) = \frac{\pi}{\sin(\pi x)} \, .
\end{align}
Using these identities, with continuation in the weights
and the logarithm convention specified in
\eqref{eq:1loop_def}, we obtain the simple result
\begin{align}
\label{eq:ABJM_round_Nosaka_determinant}
 Z^{\rm full}_{IJ,k}
 =\exp\!\left[\zeta'_{\Delta_I,k}(0)
             +\zeta'_{\Delta_J,k}(0)\right]
 =\frac1{4\sin\!\left(\frac{\pi}{k\Delta_I}\right)
           \sin\!\left(\frac{\pi}{k\Delta_J}\right)}\, .
\end{align}
These full determinants apply away from zero weight loci.

For each of these four planes, one coordinate
index lies in $\{1,3\}$ and the other in $\{2,4\}$.
We label these choices by $\sigma,\tau=\pm1$.
Combining \eqref{eq:ABJM_round_Nosaka_determinant} with the
actions of section~\ref{subsec:ABJM_K_Action} gives
\begin{align}
\label{eq:ABJM_round_Nosaka_block}
 \cB^{\sigma,\tau}(k,\mu;\xi,\eta)
 =\frac{
    \exp\!\left[-\frac{4\mu}
      {k(1+\sigma\xi)(1+\tau\eta)}\right]}
   {
    4\sin\!\left(\frac{2\pi}{k(1+\sigma\xi)}\right)
     \sin\!\left(\frac{2\pi}{k(1+\tau\eta)}\right)}\, .
\end{align}
The full-stabilizer sum over linear fillings
\eqref{eq:ABJM_generic_fixed_point_sum} becomes
\begin{align}
\label{eq:ABJM_round_Nosaka_sectors}
 J_{\rm full}(k,\mu;\xi,\eta)
 =F_k(\mu;\xi,\eta)\equiv \sum_{\sigma,\tau=\pm1}\cB^{\sigma,\tau}(k,\mu;\xi,\eta)\, .
\end{align}
This gives the complete contribution from singly wrapped full-stabilizer linear fillings on
the Nosaka locus
for generic $\xi,\eta$.

After writing $\xi=\ii\mathfrak m_1/\pi$ and
$\eta=\ii\mathfrak m_2/\pi$, $F_k$ reproduces the four
first worldsheet contributions conjectured and tested
against high-precision evaluations of the finite $N$
field theory partition function in \cite{Nosaka:2024gle}.
For equal masses, the opposite-sign terms differ from the
leading 't Hooft calculation in \cite{Gautason:2023igo}, as
already noted in \cite{Nosaka:2024gle}.

The preceding result applies for all $k$ at generic
$\xi,\eta$. On special mass loci where internal weights
coincide, continuous families of $K$-invariant planes
can appear. For $k>2$, these families may include planes
outside $\cM_k$, so the generic orbit sector must also
be considered. We examine these possibilities at the
round point $\xi=\eta=0$ below.

\subsubsection{The superconformal point at \texorpdfstring{$k=1,2$}{k=1,2}}
\label{subsubsec:ABJM_Round_k12}

At the superconformal point $\xi=\eta=0$, individual blocks
in \eqref{eq:ABJM_round_Nosaka_block} are singular for $k=1,2$.
However, their sum has the finite limit
\begin{align}
\label{eq:ABJM_round_F_collision}
 \lim_{\xi,\eta\to0}F_k(\mu;\xi,\eta)
 =\frac{4\mu^2+k\mu+k^2/4}{\pi^2}\ex^{-4\mu/k}\, ,
 \qquad k=1,2\, .
\end{align}
For $k=1,2$, this gives the complete M2-brane
contribution from linear fillings at the round
point. It agrees with the
complete leading coefficients of the modified grand potentials
computed from field theory at these levels \cite{Hatsuda:2012dt,Hatsuda:2013gj}.

We now interpret this result directly at the round
superconformal point, where the physical mass
parameters satisfy $\xi=\eta=0$, using the
continuous M2-brane family
$\cM_k=\operatorname{Gr}(2,4)$ for $k=1,2$.
Here the background is held fixed at $b=1$, $\Delta_A=\tfrac12$.
The projected spectrum has four complex bosonic collective
coordinates and two complex residual external fermions.
Introduce an auxiliary torus with weights $\epsilon_A$,
satisfying $\sum_A\epsilon_A=0$, and write the combined
internal weights entering the lifted character as
$\widehat\Delta_A=\tfrac12+\epsilon_A$.
The choice
\begin{align}
\label{eq:ABJM_round_auxiliary_weights}
 (\epsilon_1,\epsilon_2,\epsilon_3,\epsilon_4)
 =\frac12(\xi,\eta,-\xi,-\eta)
\end{align}
reproduces the fixed point functions above, using the matching
of the lifted action and fluctuation data assumed in
section~\ref{subsec:ABJM_Fluctuations}. In this interpretation
$\xi,\eta$ are auxiliary parameters for the fixed round family.

Let $\cS$ be the tautological rank two bundle on
$\operatorname{Gr}(2,4)$, with auxiliary fibre weights
$-\epsilon_I,-\epsilon_J$ at $W_{IJ}$.
Each residual fermion has weight $\epsilon_I+\epsilon_J$,
consistent with the candidate bundle and its equivariant
Euler factor
\begin{align}
\label{eq:ABJM_round_Gr_Obs}
 \operatorname{Obs}\cong(\det\cS^*){\oplus}(\det\cS^*)\, ,
 \qquad
 e_T(\operatorname{Obs}|_{W_{IJ}})
 =(\epsilon_I+\epsilon_J)^2\, .
\end{align}
This bundle captures the residual fermions over the smooth
Grassmannian. Its equivariant Euler factor in \eqref{eq:Z1M}
represents the saturation of these zero modes, as explained
in section~\ref{subsec:Quantum_Brane_Families}.

For use also at $k>2$, we record the tangent Euler factor
of the round full-stabilizer family. The graph modes surviving
the stabilizer projection give
\begin{align}
\label{eq:ABJM_round_tangent_Euler}
 e_T(T_{W_{IJ}}\cM_k)
 =\prod_{\substack{i\in\{I,J\},\ A\notin\{I,J\}\\
                   \gamma_A-\gamma_i=0\ ({\rm mod}\ k)}}
       (\epsilon_i-\epsilon_A)\, .
\end{align}
For $k=1,2$, all four factors survive, giving
\begin{align}
\label{eq:ABJM_round_Gr_Euler_ratio}
 \frac{e_T(\operatorname{Obs}_{W_{IJ}})}
      {e_T(T_{W_{IJ}}\operatorname{Gr}(2,4))}
 =\frac{(\epsilon_I+\epsilon_J)^2}
       {\displaystyle\prod_{i\in\{I,J\}}
        \prod_{A\notin\{I,J\}}(\epsilon_i-\epsilon_A)}\, .
\end{align}
For generic, sufficiently small $\xi,\eta$ in
\eqref{eq:ABJM_round_auxiliary_weights}, the tangent weights
are non-zero at all six fixed points. The reduced determinants
are finite: after removing the four bosonic and two fermionic
zero modes, the remaining projected spectrum has no zero
weights at the round point. 
At $W_{13}$ and $W_{24}$, the two omitted fermions retain
zero auxiliary weight. Their Euler factors in
\eqref{eq:Z1M} therefore make these
two contributions vanish. 
The ordinary Euler class
$e(\operatorname{Obs})=c_1(\det\cS^*)^2$
is nevertheless non-zero as a cohomology class on the
Grassmannian.
At the other four fixed points all omitted modes are lifted;
\eqref{eq:M2_family_recombination} then expresses their
contributions as the full blocks already computed.
The refined fixed point sum is therefore $F_k(\mu;\xi,\eta)$.
Removing the auxiliary parameters after this sum gives
\eqref{eq:ABJM_round_F_collision}.
Thus the fixed point calculation of the round family
contribution agrees with taking the superconformal limit
of the mass-deformed result, as expected.
An explicit evaluation of this round limit using the general collision formula is given in Appendix~\ref{app:ABJM_round_collision_example}.

\subsubsection{The superconformal point at \texorpdfstring{$k>2$}{k>2}}
\label{subsubsec:ABJM_Round_kgt2}

For $k>2$, we consider the full-stabilizer locus
$\cM_k$ and the generic stratum in turn.
The four blocks in
\eqref{eq:ABJM_round_Nosaka_block} are regular
as $\xi,\eta\to0$. Their sum gives the
full-stabilizer contribution
\begin{align}
\label{eq:ABJM_round_full_result}
 J_{\rm full}(k,\mu)
 &=\lim_{\xi,\eta\to0}F_k(\mu;\xi,\eta)
 =\frac{\ex^{-4\mu/k}}{\sin^2(2\pi/k)}\, .
\end{align}

We now interpret this result directly at the round
superconformal point $b=1$, $\Delta_A=\tfrac12$.
Here $K$ preserves every covering plane in
$\operatorname{Gr}(2,4)$, so the invariant locus
includes both $\cM_k$ and the generic stratum
$\operatorname{Gr}(2,4)\setminus\cM_k$.
As in section~\ref{subsubsec:ABJM_Round_k12},
we keep this background fixed and introduce the
auxiliary torus weights
\eqref{eq:ABJM_round_auxiliary_weights}.
Thus $\xi,\eta$ now parametrize the auxiliary
action and are removed after summing the
fixed point contributions.

We first resolve $J_{\rm full}$ into the three
components of $\cM_k$.
At $W_{12}$ and $W_{34}$, the graph modes and
the two lowest external fermions are projected
out for $k>2$. Each isolated component has local
determinant $1/[4\sin^2(2\pi/k)]$.
On $\CP^1\times\CP^1$, two tangent modes and
two residual fermions survive.
With the auxiliary weights
\eqref{eq:ABJM_round_auxiliary_weights},
$W_{13}$ and $W_{24}$ give zero, while
$W_{14}$ and $W_{23}$ have Euler ratio one
and the same limiting local determinant as
the isolated components.
The mixed family therefore contributes twice
the isolated value. Thus the two isolated
components and the mixed family supply equal
halves of~\eqref{eq:ABJM_round_full_result}.

We now turn to the generic stratum, whose stabilizer
is $\Z_h$, with $h=\gcd(k,2)$.
Let $F_h$ denote the four block function in
\eqref{eq:ABJM_round_Nosaka_sectors}, with $k$
replaced by $h$ in both the determinants and the
instanton actions.
Since $h=1,2$, the limit
\eqref{eq:ABJM_round_F_collision} and the proposed
continuation \eqref{eq:ABJM_generic_orbit_refined_allk}
give
\begin{align}
\label{eq:ABJM_round_orbit_result}
 J_{\rm gen}(k,\mu)
 &=\ex^{\ii\vartheta_k}\frac hk
   \lim_{\xi,\eta\to0}F_h(\mu;\xi,\eta)
 =\ex^{\ii\vartheta_k}\frac hk
   \frac{4\mu^2+h\mu+h^2/4}{\pi^2}
   \ex^{-4\mu/h}\, ,
\end{align}
with the global phase $\vartheta_k$ undetermined by the local calculation.

We now interpret this result directly at the round
superconformal point using the generic orbit family.
The auxiliary torus introduced above has no fixed
planes in this generic stratum.
We use the family's limiting covered configurations:
 their supports are
coordinate planes in $\cM_k$, while the maps
retain covering degree $k/h$.
The continuation assigns these configurations
the $h$-projected fluctuation spectrum and the
residual quotient factor $h/k$.
Together with the covered actions, these are
the data entering $F_h$.
Removing the auxiliary parameters after summing
therefore reproduces
\eqref{eq:ABJM_round_orbit_result}.

Table~\ref{tab:Comparison_Round} compares the sum of
\eqref{eq:ABJM_round_full_result} and
\eqref{eq:ABJM_round_orbit_result} with the field theory
instanton expansion.\footnote{For $k>2$, the M2-branes contributing to
$J_{\rm full}$ wrap the M-theory circle and reduce
to unit worldsheet instantons. Transverse representatives
of the generic orbit family reduce to singly wrapped
D2 instantons for even $k$, and doubly wrapped D2 instantons
for odd $k$. However, this family also has covered coordinate
limits, so this geometric interpretation does not by itself
identify $J_{\rm gen}$ with a separately defined
D2-instanton coefficient. See Appendix~\ref{app:ABJM_IIA}
for the reduction geometry.} The $\cM_k$ contribution reproduces
the leading term in each row. Its prefactor also agrees
with the direct finite $k$ M2 one-loop calculation
\cite{Beccaria:2023ujc}.\footnote{
A comparison with the cohomological calculation of
\cite{Kurlyand:2026yke}, including the different
organization of the total coefficient into saddle
multiplicities and family contributions, is given
in Appendix~\ref{app:fluctuation_complex}.}
If we choose $\ex^{\ii \vartheta_k} = (-1)^{k/h-1}$, then remarkably we find \eqref{eq:ABJM_round_orbit_result}
 reproduces the entire
quadratic polynomial proportional to $1/\pi^2$,
including its constant coefficient and its dependence
on $k$. It would be interesting to have a geometric determination of $\vartheta_k$
rather than fixing the sign using field theory input.

To reproduce the complete field theory coefficients,
we must go beyond the linear holomorphic fillings
considered here. At the generic orbit action, the
remaining $\mu$-independent differences, namely
$20/9$ at $k=3,6$ and $(204-60\sqrt{5})/5$ at
$k=5$, arise from genus one curves rather than
the genus zero curves considered in this section,
as will be shown elsewhere. At $k=4$, our result
already reproduces the complete coefficient.

\begin{table}[tbph!]
\centering
\small
\setlength{\tabcolsep}{7pt}
\renewcommand{\arraystretch}{1.5}
\newcommand{\padcell}[1]{%
  \raisebox{0pt}[\dimexpr\height+6pt\relax]%
               [\dimexpr\depth+6pt\relax]{#1}}
\begin{tabular}{|c|l|l|}
\hline
\padcell{$k$} & \padcell{$J_{\rm full}+J_{\rm gen}$}
    & \padcell{$J^{\rm np}(k,\mu)$ from field theory} \\
\hline
3 & \padcell{$\begin{aligned}[t]
 &\frac43\ex^{-4\mu/3}\\
 &+\frac{4\mu^2+\mu+\tfrac14}{3\pi^2}
       \ex^{-4\mu+\ii\vartheta_3}
\end{aligned}$}
& \padcell{$\begin{aligned}[t]
 &\frac43\ex^{-4\mu/3}-2\ex^{-8\mu/3}\\
 &+\left[\frac{4\mu^2+\mu+\tfrac14}{3\pi^2}
           +\frac{20}{9}\right]\ex^{-4\mu}
  +\cO(\ex^{-16\mu/3})
\end{aligned}$} \\
\hline
4 & \padcell{$\begin{aligned}[t]
 &\ex^{-\mu}\\
 &+\frac{4\mu^2+2\mu+1}{2\pi^2}
       \ex^{-2\mu+\ii\vartheta_4}
\end{aligned}$}
& \padcell{$\begin{aligned}[t]
 &\ex^{-\mu}\\
 &-\frac{4\mu^2+2\mu+1}{2\pi^2}\ex^{-2\mu}
  +\cO(\ex^{-3\mu})
\end{aligned}$} \\
\hline
5 & \padcell{$\begin{aligned}[t]
 &\frac{2\sqrt5-2}{\sqrt5}\ex^{-4\mu/5}\\
 &+\frac{4\mu^2+\mu+\tfrac14}{5\pi^2}
       \ex^{-4\mu+\ii\vartheta_5}
\end{aligned}$}
& \padcell{$\begin{aligned}[t]
 &\frac{2\sqrt5-2}{\sqrt5}\ex^{-4\mu/5}
  +\frac{-3\sqrt5+1}{\sqrt5}\ex^{-8\mu/5}\\
 &+\frac{20\sqrt5-16}{3\sqrt5}\ex^{-12\mu/5}
 +\frac{-33\sqrt5+47}{2\sqrt5}\ex^{-16\mu/5}\\
 &+\left[\frac{4\mu^2+\mu+\tfrac14}{5\pi^2}
           +\frac{204-60\sqrt5}{5}\right]\ex^{-4\mu}
  +\cO(\ex^{-24\mu/5})
\end{aligned}$} \\
\hline
6 & \padcell{$\begin{aligned}[t]
 &\frac43\ex^{-2\mu/3}\\
 &+\frac{4\mu^2+2\mu+1}{3\pi^2}
       \ex^{-2\mu+\ii\vartheta_6}
\end{aligned}$}
& \padcell{$\begin{aligned}[t]
 &\frac43\ex^{-2\mu/3}-2\ex^{-4\mu/3}\\
 &+\left[\frac{4\mu^2+2\mu+1}{3\pi^2}
           +\frac{20}{9}\right]\ex^{-2\mu}
  +\cO(\ex^{-8\mu/3})
\end{aligned}$} \\
\hline
\end{tabular}\caption{Comparison of the two contributions
\eqref{eq:ABJM_round_full_result} and
\eqref{eq:ABJM_round_orbit_result} with the field theory
modified grand potential, through the generic orbit action.
The phases $\vartheta_k$ are left explicit;
choosing $\ex^{\ii\vartheta_k}=(-1)^{k/h-1}$,
with $h=\gcd(k,2)$,
reproduces the displayed field theory polynomials
proportional to $1/\pi^2$ at the generic orbit action. The $k=3,4,6$
expansions are tabulated in (A.1) of \cite{Hatsuda:2013gj};
the $k=5$ expansion follows from (2.12), Table~1 and (B.2)
of that reference after pole cancellation. 
The field theory column gives the complete coefficients
at each displayed action; our calculation includes only
the linear filling M2-brane contributions.
}
\label{tab:Comparison_Round}
\end{table}

\subsection{Squashed sphere}
\label{subsec:ABJM_Squashed_Sphere}

We now evaluate the fluctuation determinants and family
contributions on the squashed sphere using the same fixed point
construction.
The fixed point formulae have already been derived in \eqref{eq:ABJM_generic_fixed_point_sum} and \eqref{eq:ABJM_generic_orbit_refined_allk}.  
We first consider physical mass deformations on
the Nosaka locus
\eqref{eq:ABJM_round_Nosaka_weights}
at arbitrary squashing.
We then evaluate the pure-squashing family
contributions, using the same Nosaka parameters
as auxiliary refinements and removing them
after the fixed point sum.
Finally, we consider a different mass assignment
adapted to the external weights, which restores
the elementary determinants of the round sphere
calculation.

Before specializing to these mass assignments,
we record how squashing enters the fluctuation
spectrum for general internal weights $\Delta_A$.
Using the normalization of
\eqref{eq:ABJM_ExternalWeights}, we introduce $f$
and write the external weights as
\begin{align}
\label{eq:ABJM_squashing_f}
  b_1=1+f \, ,  \qquad b_2=1-f \, , \qquad f\equiv \frac{b^2-1}{b^2+1} \, , \qquad |f|<1 \, .
\end{align}
The transformation $b\leftrightarrow b^{-1}$ exchanges the external directions and sends $f\leftrightarrow-f$.
From \eqref{eq:ABJM_linear_mode_weights}, we find that the weights of modes proportional to $z_I^m z_J^n$ are
\begin{align}
\label{eq:ABJM_sq_weight_dependence}
  \lambda^{\rm B}_{{\rm int},A} & = -\Delta_A+m\Delta_I+n\Delta_J \, , \nn \\
  \lambda^{\rm F}_{{\rm int},A} & = \Delta_I+\Delta_J+\Delta_A+m\Delta_I+n\Delta_J \, , \nn \\
  \lambda^{\rm B}_{{\rm ext},\alpha} & = b_\alpha+m\Delta_I+n\Delta_J \, ,\nn \\
  \lambda^{\rm F}_{{\rm ext},\alpha} & = -b_\alpha + (m+1)\Delta_I + (n+1) \Delta_J \, .
\end{align}
Here $A \notin \{ I,J \}$ labels the normal directions, and we vary $b$ at fixed $\Delta_A$, so only the external weights change.
The quotient charges are also unchanged, so the same stabilizer projection as in the round case applies. 
In particular, squashing can change which external fermions have zero $K$-weight while preserving the internal bosonic graph modes.

\subsubsection{The Nosaka locus}
\label{subsubsec:ABJM_squashed_ordinary_Nosaka}

We keep the physical mass assignment
\eqref{eq:ABJM_round_Nosaka_weights}, but allow
arbitrary squashing $b>0$.
For generic $\xi,\eta$, the six coordinate planes
$W_{IJ}$ remain the only $K$-invariant linear
fillings, each with full stabilizer $\Z_k$,
independently of $b$.
The change in the character is entirely due to
the external weights:
\begin{align}
\label{eq:ABJM_ordinary_Nosaka_character_continuation}
 \cK^{\rm full}_{IJ,s}(t;b)
 &=\cK^{\rm full}_{IJ,s}(t;1)\nonumber\\
 &\quad+\frac{2[\cosh(ft)-1]}{s}
 \sum_{j=0}^{s-1}
 \frac{\ex^{-t}
 -\omega^{-j(\gamma_I+\gamma_J)}
  \ex^{-t(\Delta_I+\Delta_J-1)}}
 {(1-\omega^{-j\gamma_I}\ex^{-t\Delta_I})
  (1-\omega^{-j\gamma_J}\ex^{-t\Delta_J})}\, ,
\end{align}
where $\omega=\ex^{2\pi\ii/s}$.
This is an identity of the full characters;
the terms on the right are combined before
identifying and removing any zero modes.

At $W_{13}$ and $W_{24}$, the two lowest external
fermions now have weights $-f$ and $f$, so these
modes are lifted when $b\neq1$.
The four-block simplification of
section~\ref{subsubsec:ABJM_Nosaka} therefore no
longer applies at generic squashing.
All six determinants are specified by the projected
double zeta functions in
\eqref{eq:ABJM_projected_double_zeta} and
\eqref{eq:ABJM_generic_Barnes_determinant}.

For use also in the family calculation below,
we retain a general stabilizer order $s$ and define
\begin{align}
\label{eq:ABJM_ordinary_Nosaka_squashed_family_sum}
 F_s^{(b)}(\mu;\xi,\eta)
 &=\sum_{I<J}Z^{\rm full}_{IJ,s}
       (b,\Delta(\xi,\eta))
   \exp\!\left[
   -\frac{4\mu}{sQ^2\Delta_I\Delta_J}
   \right]\, .
\end{align}
The singly wrapped quotient configurations considered
here have $s=k$, giving
\begin{align}
\label{eq:ABJM_ordinary_Nosaka_squashed_sectors}
 J_{\rm full}(k,\mu;b,\xi,\eta)
 &=F_k^{(b)}(\mu;\xi,\eta)\, ,
\end{align}
for generic $b,\xi,\eta$ and every $k$.
At $b=1$, the fermionic Euler factors eliminate
$W_{13}$ and $W_{24}$, recovering $F_k$ in
\eqref{eq:ABJM_round_Nosaka_sectors}.

We next take the massless limit, in which
continuous families reappear. For $k>2$, these
include the generic orbit family described in
section~\ref{subsec:Geometry_Semiclassical_Action},
with its covered coordinate limits.

\subsubsection{Pure squashing}
\label{subsubsec:ABJM_pure_squashing}

We first take the pure squashing limit, that is, we take the masses to zero, $\Delta_A \to \frac{1}{2}$, while keeping $b$ arbitrary.
We can obtain the full-stabilizer contribution directly from \eqref{eq:ABJM_ordinary_Nosaka_squashed_sectors} by taking the limit $\xi,\eta\to0$ after summing the fixed point contributions.
At generic squashing, the tangent Euler factors make individual blocks singular as the masses vanish, but the negative powers cancel in the sum, leaving a polynomial.
In this calculation, the masses serve as auxiliary equivariant refinements of the massless problem.

For generic squashing, the evaluated sums take the form
\begin{align}
\label{eq:ABJM_pure_squashing_Limit_s}
  \lim_{\xi,\eta\to0} F_s^{(b)}(\mu;\xi,\eta) = \exp \left[-\frac{4\mu(1-f^2)}{s}\right] P_s(\mu;f) \, ,
\end{align}
where the exponent is the action of a membrane on a linear filling with stabilizer $\Z_s$
\begin{align}
\label{eq:ABJM_pure_squashing_action}
  I_{\rm M2} = \frac{16\mu}{sQ^2} =\frac{4\mu(1-f^2)}{s} \, ,
\end{align}
and $P_s(\mu;f)$ is the complete polynomial
prefactor for stabilizer order $s$, with degree
\begin{align}
\label{eq:ABJM_pure_squashing_degree}
  \deg_\mu P_s = \begin{cases}
    4 & s = 1,2 \, ,\\
    2 & s>2 \, .
 \end{cases} 
\end{align}
The $s=k$ limit gives the full-stabilizer contribution. 
As in the round case, for $k>2$, the massless theory also admits the generic orbit family, whose contribution is obtained from the $s=h$ limit, multiplied by $\ex^{\ii\vartheta_k}h/k$, where $h=\gcd(k,2)$. 

The massless sector sums are therefore
\begin{align}
\label{eq:ABJM_pure_squashing_family_sums}
 J_{\rm full}(k,\mu;b)
 &=\lim_{\xi,\eta\to0}F_k^{(b)}(\mu;\xi,\eta)\, ,\nonumber\\
 J_{\rm gen}(k,\mu;b)
 &=\ex^{\ii\vartheta_k}\frac{h}{k}
   \lim_{\xi,\eta\to0}F_h^{(b)}(\mu;\xi,\eta)\, ,
 \qquad k>2\, .
\end{align}
Recall that for $k=1,2$, every linear filling has full
stabilizer $s=k$, so $J_{\rm lin}=J_{\rm full}$.
For $k>2$, the total contribution from linear
fillings is $J_{\rm lin}=J_{\rm full}+J_{\rm gen}$.
In particular, the $k=1$ result is
\eqref{eq:Intro_Jlin_1}, quoted in the introduction.
A common compact expression for $s=1,2$ is given in
\eqref{eq:ABJM_sq_k1_closed_form}, with the
polynomials $\cP_a^{(s)}$ defined in
\eqref{laterintext} and $\cP_a\equiv\cP_a^{(1)}$.
For $s>2$, the complete quadratic prefactor,
including the isolated plane contributions,
is given in \eqref{eq:ABJM_sq_higher_k_decomposition}
and \eqref{eq:app_sq_higher_k}, with the
coefficients $B_k(f)$ and $C_k(f)$ given in
\eqref{eq:ABJM_sq_mixed_coefficients}.

The degree of the polynomial can decrease at special values of $b$. For instance, at $s = 1,2$ the leading coefficient is
\begin{align}
  \label{eq:ABJM_pure_squashing_Ps4}
 P_s(\mu;f) = -\frac{16(1-f^2)^4}{3\pi^4} \sin^2 \left(\frac{2\pi f}{s}\right) \mu^4 +\mathcal O(\mu^3) \, , \qquad s = 1,2 \, ,
\end{align}
which for $s=1$ vanishes when the squashing is $b^2 = 1, 3, \frac{1}{3}$, and for $s=2$ only on the round sphere.

This reduction of the degree has a geometric interpretation. Pure squashing leaves the internal bosonic deformation modes unchanged but can change the residual external fermionic zero modes, explaining the degree of $P_s(\mu;f)$.
Concretely, a connected family of complex dimension $d$,
with an obstruction bundle of complex rank $r$, has virtual
complex dimension $n=d-r$.
For $n\geq0$, under the regularity and $\mu$-independence
assumptions stated in
Appendix~\ref{app:M2_family_collisions}, a finite refinement
limit has a polynomial prefactor of degree at most $n$. 

The bosonic families and their stabilizers are the same as at the round superconformal point. For the full-stabilizer problem $k=1,2$ 
the family is $\cM_k=\operatorname{Gr}(2,4)$. 
At generic squashing there are no residual fermionic zero modes, and the non-zero quartic coefficient in \eqref{eq:ABJM_pure_squashing_Ps4} saturates the degree bound.
To understand the exceptional squashing values, we examine when external fermions acquire zero weight and survive the stabilizer projection.

For a monomial of total degree $\ell = m+n$, the external fermionic modes have weights
\begin{align}
  \lambda^{\rm F}_{{\rm ext},1} = - f + \frac{\ell}{2} \, , \qquad \lambda^{\rm F}_{{\rm ext},2} =  f + \frac{\ell}{2} \, .
\end{align}
The only possible zero weights are the following.
At $f=0$, equivalent to $b=1$, there are two degree zero fermions, which are those identified in the round computation of the previous section. 
At $f= \pm \frac{1}{2}$, corresponding to $b^2 = 3$, $\frac{1}{3}$, one external weight reaches $\frac{3}{2}$ and its two degree one monomials have zero weight. 
In keeping with the symmetry of the problem, the two non-round values are exchanged by $b \leftrightarrow b^{-1}$.

The quotient group average preserves a fermionic weight zero mode only if its $\Z_s$ charge is
\begin{align}
	q_{\rm F} = - (m+1)\gamma_I - (n+1)\gamma_J = 0 \mod s \, .
\end{align}
Recall that the M-theory circle acts with charges
$(\gamma_1,\gamma_2,\gamma_3,\gamma_4)=(1,1,-1,-1)$.
The quotient therefore acts differently on the
planes $W_{12}$ and $W_{34}$, whose two coordinates
have equal charges, and on the four mixed planes,
whose coordinates have opposite charges. 
Concretely, the degree zero charges (relevant to the $b=1$ case) are $\pm 2$ on the $\{ W_{12}, W_{34}\}$ planes, and zero on the mixed planes. The degree one charges (relevant to the $b^2=3, 1/3$ cases) are $\pm 3$ on the $\{ W_{12}, W_{34}\}$ planes, and $\pm 1$ on the mixed planes.
The numbers of fermionic zero modes surviving the projection, and the virtual dimension of the moduli space, are summarized in Table~\ref{tab:ABJM_pure_squashing_dimensions}.

\begin{table}
\begin{center}
\small
\begin{tabular}{@{}lccccc@{}}
\toprule
 & & & \multicolumn{3}{c}{Virtual dimension $n = d-r$} \\
\cmidrule(lr){4-6}
Family/component & $s$ & $d$ & Generic $b$ & $b = 1$ & $b^2 = 3,\tfrac13$ \\
\midrule
$\operatorname{Gr}(2,4)$ & $1$ & $4$ & $4$ & $2$ & $2$ \\
$\operatorname{Gr}(2,4)$ & $2$ & $4$ & $4$ & $2$ & $4$ \\
$\CP^1\times\CP^1$ & $>2$ & $2$ & $2$ & $0$ & $2$ \\
Isolated planes & $3$ & $0$ & $0$ & $0$ & $-2$ \\
Isolated planes & $>3$ & $0$ & $0$ & $0$ & $0$ \\
\bottomrule
\end{tabular}
\par\smallskip
\caption{Virtual dimensions in the proposed pure-squashing localization problem. For $n\geq0$, the entries give polynomial degree bounds. The exceptional $n = -2$ denotes a vanishing isolated contribution. For $k>2$, the full-stabilizer sector uses $s = k$, while the proposed generic orbit calculation uses $s = h = \gcd(k,2)$ and hence one of the first two rows.}
\label{tab:ABJM_pure_squashing_dimensions}
\end{center}
\end{table}

For $k=1$, the first row of
Table~\ref{tab:ABJM_pure_squashing_dimensions}
shows that all three special squashing values
reduce the virtual dimension from four to two,
but the corresponding fermionic bundles differ.
At $b = 1$, the two degree zero fermions give the obstruction bundle \eqref{eq:ABJM_round_Gr_Obs}.
At $b^2 = 3$, instead, the weights of the fermionic modes are $2\epsilon_I+\epsilon_J$ and $\epsilon_I+2\epsilon_J$, which match the fibre weights of $\cS^*\otimes\det\cS^*$ and give the candidate bundle
\begin{align}
\label{eq:ABJM_b3_fermion_bundle}
 \operatorname{Obs}_{b^2=3}
 &\cong \cS^*\otimes\det\cS^* \, , \nn\\
 e_T(\operatorname{Obs}_{b^2=3}|_{W_{IJ}})
 &=(2\epsilon_I+\epsilon_J)(\epsilon_I+2\epsilon_J) \, .
\end{align}

For generic auxiliary parameters, all four bosonic and two fermionic zero modes are lifted at each coordinate plane. We therefore use the full determinants through \eqref{eq:M2_family_recombination},
\begin{align}
  \label{ObsIJ}
    Z^{\rm full}_{IJ,1}(\epsilon) = \frac{(2\epsilon_I+\epsilon_J)(\epsilon_I+2\epsilon_J)}
{ \prod_{i\in\{I,J\}} \prod_{A\notin\{I,J\}} (\epsilon_i-\epsilon_A)} Z^{\rm red}_{IJ,1}(\epsilon) \, ,
\end{align}
with the Euler ratio
already included.  
Summing the refined fixed point contributions
and then removing the auxiliary parameters gives
\begin{align}
\label{eq:ABJM_squashed_b3_result}
 J_{\rm lin}(1,\mu;\sqrt3) &=\lim_{\xi,\eta\to0}F_1^{(\sqrt3)}(\mu;\xi,\eta) \nn\\
 &=\left[-\frac{45\mu^2}{8\pi^2}
         -\frac{9\mu}{4\pi^2}
         +\frac38-\frac{3}{4\pi^2}\right]\ex^{-3\mu}\, .
\end{align}
Details of the fixed point calculation are
given in Appendix~\ref{app:ABJM_b3_fixed_point_data},
where we also express the sum as an integral
of characteristic classes over
$\operatorname{Gr}(2,4)$.
The result \eqref{eq:ABJM_squashed_b3_result} reproduces the complete leading coefficient of the modified grand potential computed from field theory in \eqref{eq:ABJM_squashed_Jnp_b3} \cite{Hatsuda:2016uqa}.
The result at $b^2=\tfrac13$ is identical after exchanging the two external directions.

For $k=2$, the second row of
Table~\ref{tab:ABJM_pure_squashing_dimensions}
shows that the virtual dimension decreases only
on the round sphere, where the two degree zero
fermions survive the $\Z_2$ projection.
The prefactor is consequently quadratic at $b=1$,
as in \eqref{eq:ABJM_round_F_collision},
and quartic otherwise.

For $k>2$, the final three rows of
Table~\ref{tab:ABJM_pure_squashing_dimensions}
describe the mixed and isolated components of
the full-stabilizer locus $\cM_k$ in
\eqref{eq:ABJM_full_stabilizer_locus}.
We consider the isolated planes first,
followed by the mixed family.

On the isolated planes $\{ W_{12}, W_{34}\}$, there are no tangent modes, so $d=0$, but there are two fermionic zero modes at $b^2 = 3, 1/3$ for $k=3$. This leads to a quadratic zero at $f=\pm \frac{1}{2}$ for their contribution when $k=3$.
Each isolated component contributes the
$\mu$-independent full determinant
$Z^{\rm full}_{{\rm iso},k}(f)$,
multiplying the common instanton exponential.
Writing $x_\pm = 2(1\pm f)/k$, its Hurwitz zeta evaluation is
\begin{align}
\label{eq:ABJM_sq_full_isolated}
 Z^{\rm full}_{{\rm iso},k}(f) = \frac{1}{4\sin\left(\frac{2\pi(1+f)}{k}\right)
                       \sin\left(\frac{2\pi(1-f)}{k}\right)}
 \left[\frac{G(1+2/k)^2G(1-x_+)G(1-x_-)}
 {G(1-2/k)^2G(1+x_+)G(1+x_-)}\right]^k \, ,
\end{align}
where $G(z)$ is the Barnes $G$-function, defined for $z>0$ by
\begin{align}
	\log G(z) \equiv (z-1) \log \Gamma(z) + \zeta'(-1) - \partial_s \zeta_H(s,z) \rvert_{s=-1} \, ,
\end{align}
and extended elsewhere by analytic continuation, which satisfies $G(1) = 1$ and $G(z+1) = \Gamma(z)G(z)$. Note that, as anticipated, $Z^{\rm full}_{{\rm iso},3}(f)$ vanishes quadratically as $f\to \pm \frac{1}{2}$, owing to the two surviving fermionic zero modes, and is non-zero for integer $k$ greater than $3$.

On the family $\CP^1\times \CP^1$ of mixed planes, instead, there are two tangent modes but it is only in the round case that there are fermionic zero modes, leading to a vanishing virtual dimension consistent with the $\mu$-independent prefactor in \eqref{eq:ABJM_round_full_result}.

At generic squashing, the mixed component gives a quadratic prefactor, while each isolated component gives a constant in $\mu$. The full-stabilizer polynomial consequently decomposes as
\begin{align}
\label{eq:ABJM_Squashed_FullStabilizer}
 P_k(\mu;f) & = 2 \, Z^{\rm full}_{{\rm iso},k}(f)+P_k^{\rm mix}(\mu;f) \, , \nn \\
 P_k^{\rm mix}(\mu;f)
 & = -\frac{4(1-f^2)^2}{\pi^2}
       \frac{\sin^2(2\pi f/k)}{\sin^2(2\pi/k)}\mu^2
       +\mathcal O(\mu) \, .
\end{align}
The complete quadratic prefactor is given in
\eqref{eq:ABJM_sq_higher_k_decomposition}, \eqref{eq:app_sq_higher_k}, with the coefficients
$B_k(f)$ and $C_k(f)$ given explicitly in
\eqref{eq:ABJM_sq_mixed_coefficients}. 
In the round limit its positive powers of $\mu$ disappear, with $P_k^{\rm mix}\to[2\sin^2(2\pi/k)]^{-1}$ and $Z^{\rm full}_{{\rm iso},k}(f)\to[4\sin^2(2\pi/k)]^{-1}$: the mixed family and the two isolated planes supply equal halves of \eqref{eq:ABJM_round_full_result}.

The quotient relevant for the generic orbit sector, instead, is $\Z_h$ with $h=\gcd(2,k)$, so odd $k$ uses the $s=1$ covering calculation, and even $k$ uses $s=2$, with the previously specified residual quotient and phase.

We can compare our results with the pure-squashing
expression in equation~(24) of \cite{Gautason:2025per},
restricting to $k>2$ as in their analysis.
The instanton exponents agree at leading large $N$:
using $N\sim2(1-f^2)^2\mu^2/(\pi^2k)$ gives
$I_{\rm M2}=4(1-f^2)\mu/k
\sim2\pi\sqrt{2N/k}$.
However, the local determinants differ.
Let $D_{\rm GvM}(f,k)$ denote the one-loop determinant
in their expression, with the overall factor of two
and the instanton exponential removed. Comparison
with our isolated-plane determinant gives
\begin{align}
\label{IDGvMcomp}
 Z^{\rm full}_{{\rm iso},k}(f)
 =\frac{D_{\rm GvM}(f,k)}{1-f^2}\, .
\end{align}
In our complex, the additional factor comes from
the two lowest external bosonic displacement modes.
These are constant in the filling coordinates,
neutral under $\Z_k$, and have non-zero weights
$1+f$ and $1-f$ for $b>0$.
They therefore contribute $1/(1-f^2)$.
Retaining these modes is also required for our
agreement with the exact mass--squashing identity
\eqref{eq:ABJM_mass_squashing_identity}, discussed
in the next subsection.

The counting of configurations also differs.
Reference~\cite{Gautason:2025per} attributes its
overall factor of two to two inequivalent internal
embeddings, both with worldvolume $S^3/\Z_k$ and
identical fluctuation spectra.
Our full-stabilizer contribution instead includes
the two isolated planes together with the continuous
mixed family $\CP^1\times\CP^1$.
At generic squashing, this family produces a quadratic
polynomial in $\mu$, absent from equation~(24) of
\cite{Gautason:2025per}.
The difference therefore cannot be resolved by
an overall multiplicity.

\subsubsection{An adapted mass-squashing locus}
\label{subsubsec:ABJM_squashed_Nosaka}

We now consider a mass assignment adapted to the
squashing, for which the local M2-brane contributions
reduce to the four elementary blocks of the
round sphere Nosaka calculation.
This allows a further test against an exact field
theory identity relating mass-deformed partition
functions on the squashed and round spheres.
We take
\begin{align}
\label{eq:ABJM_squashed_Nosaka_locus}
  (\Delta_1,\Delta_2,\Delta_3,\Delta_4) = \frac{1}{2} \left( b_1(1+\xi),\ b_2(1+\eta),\ b_1(1-\xi),\ b_2(1-\eta) \right) \, .
\end{align}
Thus $\Delta_1+\Delta_3 = b_1$ and $\Delta_2+\Delta_4 = b_2$, with $b_\alpha$ as in \eqref{eq:ABJM_ExternalWeights}. At $b = 1$ this is the Nosaka locus already considered. However, the resulting parameters cannot be tuned to obtain the pure squashing behaviour of the previous section.

We evaluate the same six fixed point terms as in \eqref{eq:ABJM_generic_fixed_point_sum}. On $W_{13}$ the lowest external fermion of weight $\Delta_1+\Delta_3-b_1$ vanishes, and on $W_{24}$ the fermion of weight $\Delta_2+\Delta_4-b_2$ vanishes. Both are neutral under $\Z_k$. For generic parameters on this locus, the corresponding Euler numerators vanish and these two terms drop out, 
since the reduced determinants are regular, as in the round calculation. For the four remaining planes, choose $I\in\{1,3\}$ and $J\in\{2,4\}$, with complementary indices satisfying $\Delta_K = b_1-\Delta_I$ and $\Delta_L = b_2-\Delta_J$.

The non-trivial simplification is again a cancellation in the numerator of \eqref{eq:ABJM_Unreduced_KD}:
\begin{align}
\label{eq:ABJM_squashed_axis_character}
 \cK^{\rm full}_{IJ}(t)& = \kappa_{\Delta_I;b_1}(t)+\kappa_{\Delta_J;b_2}(t) \, ,\nn\\
 \kappa_{\Delta;r}(t)& = \frac{\ex^{-rt}+\ex^{-(\Delta-r)t}}{1-\ex^{-\Delta t}} \, .
\end{align}
This identity also holds with the finite group phases of \eqref{eq:ABJM_Molien_character}. 
As in the round case, the two towers of $\Z_s$-invariant modes have $K$-weights $r+sn\Delta$ and $s\Delta-r+sn\Delta$, $n\geq0$. 
Applying the same Hurwitz identities \eqref{eq:ABJM_Hurwitz_Cool}, with $1/(s\Delta)$ replaced by $r/(s\Delta)$, gives
\begin{align}
\label{eq:ABJM_squashed_Nosaka_determinant}
 Z^{\rm full}_{IJ,s}
 & = \frac{1}{4\sin\!\left(\frac{\pi b_1}{s\Delta_I}\right)
                  \sin\!\left(\frac{\pi b_2}{s\Delta_J}\right)} \, ,\nn\\
 Z^{\rm full}_{IJ,s}\exp\!\left[-\frac{4\mu}{sQ^2\Delta_I\Delta_J}\right]
 & = \cB^{\sigma,\tau}(s,\mu;\xi,\eta) \, .
\end{align}
Here $\Delta_I = b_1(1+\sigma\xi)/2$, $\Delta_J = b_2(1+\tau\eta)/2$, and the second line follows from $b_1b_2 = 4/Q^2$. The spectral continuation and phase convention are inherited from the general determinant. Consequently, the full-stabilizer contribution is exactly the four-block sum \eqref{eq:ABJM_round_Nosaka_sectors} with $s = k$, now on a squashed background. 
For $k=1,2$, it gives the contribution from
singly wrapped linear fillings.
The candidate generic orbit sum \eqref{eq:ABJM_generic_orbit_refined_allk} likewise reduces to the blocks with $s = h$, retaining its proposed quotient measure and global phase.

This reduction has an exact field theory
counterpart. The identity
\eqref{eq:ABJM_mass_squashing_identity}, reviewed
in Appendix~\ref{app:ABJM_squashed_field_theory},
relates mass-deformed partition functions on the
squashed and round spheres for arbitrary $N,k$.
It holds at finite $N$ and therefore also constrains
the non-perturbative contributions.
In the mass conventions of \cite{Chester:2021gdw},
our locus corresponds to
\begin{align}
\label{eq:ABJM_squashed_Nosaka_mass_dictionary}
 m_3 = \frac{\ii}{2}(b-b^{-1})\, ,\qquad
 \xi = -\ii b^{-1}m_+\, ,\qquad \eta = \ii b m_-\, ,\qquad m_\pm = m_2\pm m_1 \, .
\end{align}
These masses are distinct from the $\mathfrak m_i$ used in the round Nosaka comparison.\footnote{In the chiral ordering of \cite{Chester:2021gdw}, let $M_A$ denote the four chiral masses, with
\begin{align}
 (M_1,M_2,M_3,M_4) = \tfrac12(m_1+m_2+m_3,\ m_1-m_2-m_3,\ -m_1-m_2+m_3,\ -m_1+m_2-m_3) \, .
\notag
\end{align}
Comparing the double-sine factors $s_b(\ii Q/4-M_A)$ with $s_b(\ii Q(1-\Delta_A)/2)$ gives $\Delta_A = \frac12-2\ii M_A/Q$, and hence \eqref{eq:ABJM_squashed_Nosaka_locus} with \eqref{eq:ABJM_squashed_Nosaka_mass_dictionary}. Gauge-dependent arguments are held fixed in this comparison.}
The identity
\eqref{eq:ABJM_mass_squashing_identity}
absorbs the squashing into the mass combinations
$b^{-1}m_+$ and $bm_-$.
At fixed $\xi,\eta$, the explicit $b$-dependence
cancels from both the local determinants and
instanton actions.
With the mass dictionary above, this reproduces
the relation for the linear filling M2-brane
contributions computed here.

\section{The ABJM topologically twisted index}
\label{sec:ABJM_TTI}

We now extend the ABJM calculation to the topologically twisted
index (TTI) on $S^1\times S^2$. In \cite{BenettiGenolini:2026kjy}, the perturbative
M-theory contribution was obtained by gluing two local gravitational
blocks associated with the fixed points of a supersymmetric, magnetically charged and rotating
black hole
background. Each block is determined by the same local data as the
squashed sphere partition function. We extend this construction by
including the single M2-brane instanton contributions computed in section~\ref{sec:ABJM_Theory}
in the local grand potentials. 
We first compute the local grand potentials at
general chemical potentials and refinement, then perform the inverse Laplace transform
to obtain the partition function for the fixed $N$ ensemble and extract the unrefined limit, associated with static black holes. 
At the universal twist, we reproduce field theory single instanton coefficients at $k=1,2,4$
\cite{Hosseini:2026dkj,Hosseini:2026lvj} and obtain
predictions for other $k$.\footnote{
Related non-perturbative contributions to the
topologically twisted index of ADHM theory
have been studied using the Bethe Ansatz
formulation in \cite{Peng:2026udd}.
}

\subsection{The refined index and the local grand potentials}
\label{subsec:TTI_grand_potentials}

The refined TTI is a supersymmetric partition function on
$S^1\times S^2$, with flavour chemical potentials $\Delta_I$,
magnetic fluxes $\mathfrak n_I$ through $S^2$, and a refinement parameter
$\epsilon$ associated with rotations of $S^2$
\cite{Benini:2015noa,Benini:2015eyy,Bobev:2022jte}.
In our conventions we have
\begin{align}
\sum_{I=1}^4\Delta_I&=2\, ,
 \qquad
 \sum_{I=1}^4\mathfrak n_I=2\, .
\label{eq:TTI_constraints}
\end{align}
The corresponding Euclidean black hole geometry has topology
$M_4\simeq\R^2\times S^2$, with conformal boundary
$S^1\times S^2$. The boundary circle is filled by the $\R^2$
factor and contracts at its origin, where the $S^2$ black hole horizon lies.
As in \cite{BenettiGenolini:2026kjy}, the auxiliary
twelve-dimensional filling is the total space of a fibration
\begin{align}
Z_8&\longhookrightarrow W\longrightarrow M_4\, ,
 \qquad \partial Z_8=S^7/\Z_k\, ,
\label{eq:TTI_twelve_dimensional_filling}
\end{align}
where $Z_8$ is the radially truncated Calabi--Yau cone
$\C^4/\Z_k$. Taking the boundary of each internal fibre gives
the physical eleven-dimensional spacetime, an $S^7/\Z_k$
fibration over $M_4$. The magnetic charges $\mathfrak n_I$
specify the twisting of the four internal complex directions
over the horizon $S^2$, and the restriction on their sum in 
\eqref{eq:TTI_constraints} implements the topological twist required by supersymmetry.

Thus the filling $W=M_4\times Z_8$ used for the $S^3$ backgrounds in section \ref{sec:ABJM_Theory},
where $M_4\simeq\R^4$, is replaced by
\eqref{eq:TTI_twelve_dimensional_filling}. Locally the filling
still has product form, and 
we use the same localized M2-brane action and fluctuation
complex in the internal fibre over each external fixed point
as described in section~\ref{subsec:M2_complex_geometry}.

For non-zero $\epsilon$, the supersymmetric Killing vector $K$ has two
isolated external fixed points, at the poles of $S^2$, labelled
by $\sigma=\pm$. 
With the weight of $K$ on $\R^2$ normalized to one,
the six local holomorphic tangent weights on $W$ are given by
\cite{BenettiGenolini:2026kjy}
\begin{align}
w_\sigma&=
 \left(1,\sigma\epsilon,
 -\frac{\Delta_1+\sigma\epsilon\mathfrak n_1}{2},\ldots,
 -\frac{\Delta_4+\sigma\epsilon\mathfrak n_4}{2}\right)\, .
\label{eq:TTI_local_weights}
\end{align}
Their sum vanishes by \eqref{eq:TTI_constraints},
so the local holomorphic volume element on $W$
has zero $K$-weight, as in
section~\ref{subsec:ABJM_K_Action}. 
After exchanging the
two external directions and making a common rescaling, these are
the sphere weights of section~\ref{subsec:ABJM_K_Action}, with
\begin{align}
b_\sigma^2&=\sigma\epsilon\, ,
 \qquad
 \Delta_{\sigma,I}
 =\frac{\Delta_I+\sigma\epsilon\mathfrak n_I}
        {1+\sigma\epsilon}\, ,
\label{eq:TTI_local_sphere_parameters}
\end{align}
such that the unrefined limit $\epsilon \rightarrow 0$ of the TTI corresponds to the $b \rightarrow 0$ limit of the individual gravitational blocks.
The flux variable $\mu_\sigma$, defined as in
\eqref{eq:mu_a_Def} using the local external weights, is invariant
under the common rescaling.

The quantity we compute at each pole is the local grand potential,
including perturbative and non-perturbative terms:
\begin{align}
J_\sigma(\mu_\sigma)
 &=J_\sigma^{\rm pert}(\mu_\sigma)
   +J_\sigma^{\rm np}(\mu_\sigma)\, ,\nonumber\\
 J_\sigma^{\rm pert}(\mu_\sigma)
 &=\frac{\mathcal C_\sigma}{3}\mu_\sigma^3
   +\mathcal B_\sigma\mu_\sigma+\mathcal A_\sigma\, .
\label{eq:TTI_local_grand_potential}
\end{align}
Here the perturbative coefficients are the sphere functions evaluated
at \eqref{eq:TTI_local_sphere_parameters}. Explicitly,
\begin{align}
\mathcal C(k,b,\Delta)
 &=\frac{2}{\pi^2kQ^4\prod_I\Delta_I}\, ,\nonumber\\
 \mathcal B(k,b,\Delta)
 &=\frac{1}{24k}\left[
 1-2\sum_I\frac{1}{\Delta_I}
 +\frac{4}{Q^2}\sum_{I<J}\frac{1}{\Delta_I\Delta_J}
 \right]+\frac{k^2-1}{24k}\, ,
\label{eq:TTI_sphere_Airy_coefficients}
\end{align}
with $Q=b+b^{-1}$, while
$\mathcal A(k,b,\Delta)=\mathcal A_{\rm ABJM}(k,b,\Delta)$
is the bulk one-loop function obtained in
\cite{BenettiGenolini:2026kjy}.

At generic local weights, the single M2-brane contribution from full-stabilizer linear fillings
to
$J_\sigma^{\rm np}(\mu_\sigma)$ in
\eqref{eq:TTI_local_grand_potential} is obtained by
applying \eqref{eq:ABJM_generic_fixed_point_sum}
at the pole $\sigma$:
\begin{align}
 J_\sigma^{(1)}(\mu_\sigma)
 &=\sum_{I<J}
 Z^{\rm full}_{IJ,k}(b_\sigma,\Delta_\sigma)
 \exp\!\left[
 -\frac{4\sigma\epsilon\mu_\sigma}
 {k(\Delta_I+\sigma\epsilon\mathfrak n_I)
    (\Delta_J+\sigma\epsilon\mathfrak n_J)}
 \right]\, .
 \label{eq:TTI_local_membrane_grand_potential}
\end{align}
The determinants are specified analytically by the projected double
zeta functions in \eqref{eq:ABJM_projected_double_zeta} and
\eqref{eq:ABJM_generic_Barnes_determinant}. When zero modes occur,
we evaluate the corresponding family contribution using the fixed
point formula \eqref{eq:Z1M}, as explained in
section~\ref{subsec:Quantum_Brane_Families}. 
Thus the same projected spectra specify the M2-brane contributions at arbitrary
chemical potentials $\Delta_I$ and refinement $\epsilon$; the magnetic fluxes $\mathfrak{n}_I$ enter through
the local parameters in \eqref{eq:TTI_local_sphere_parameters}.

The two local grand potentials must still be glued. As in
\cite{BenettiGenolini:2026kjy}, the boundary conditions fix only
$\mu=\mu_++\mu_-$, leaving a relative bulk flux to be integrated.
Writing $\mu_+=\nu$ and $\mu_-=\mu-\nu$, we extend the perturbative
flux measure to include the M2-brane insertions and define the
grand canonical contribution of this bulk saddle by
\begin{align}
\mathcal Z_{\rm TTI}(\mu)
 &=\int\frac{\dd\nu}{2\pi\ii}
   \exp\!\left[J_+(\nu)+J_-(\mu-\nu)\right]
 \equiv\ex^{J_{\rm TTI}(\mu)}\, .
\label{eq:TTI_grand_convolution}
\end{align}
Writing
$J_{\rm TTI}=J_{\rm TTI}^{\rm pert}
+J_{\rm TTI}^{(1)}+\cdots$,
the first order expansion of
\eqref{eq:TTI_grand_convolution} gives
\begin{align}
J_{\rm TTI}^{(1)}(\mu)
 &=
 \frac{
 \int\frac{\dd\nu}{2\pi\ii}\,
 \exp\!\left[J_+^{\rm pert}+J_-^{\rm pert}\right]
 \left(J_+^{(1)}+J_-^{(1)}\right)}
 {
 \int\frac{\dd\nu}{2\pi\ii}\,
 \exp\!\left[J_+^{\rm pert}+J_-^{\rm pert}\right]}\, .
\label{eq:TTI_grand_instanton_insertion}
\end{align}
Here the local grand potentials at the $+$ and $-$ poles
are evaluated at $\mu_+=\nu$ and $\mu_-=\mu-\nu$,
respectively.
The perturbative grand potential
$J_{\rm TTI}^{\rm pert}$ is the logarithm of the
denominator.

These are quantities evaluated in the grand canonical ensemble at fixed $\mu$; we next
transform to the fixed $N$ ensemble.

\subsection{Fixed rank and the small-squashing limit}
\label{subsec:TTI_fixed_rank}

The procedure is to assemble each membrane family
contribution at non-zero $b$, remove the auxiliary
refinement, perform the inverse Laplace transform
to fixed $N$, and then extract the small $b$ limit.
Terms proportional to $b^{-2}$ in the local grand
potentials can generate additional finite contributions
under this transform, as we illustrate at the universal
twist in section~\ref{subsec:TTI_universal_tests}.

We begin by writing the fixed $N$ partition function
as an inverse Laplace transform in the boundary
variable $\mu$:
\begin{align}
Z_{\rm TTI}(N)
 &=\int\frac{\dd\mu}{2\pi\ii}\,
       \ex^{J_{\rm TTI}(\mu)-N\mu}
 =\prod_{\sigma=\pm}\mathcal Z_\sigma(N)\, ,
 \nonumber\\
 \mathcal Z_\sigma(N)
 &=\int_{\Gamma_\sigma}\frac{\dd\mu_\sigma}{2\pi\ii}\,
       \ex^{J_\sigma(\mu_\sigma)-N\mu_\sigma}\, .
\label{eq:TTI_canonical_gluing}
\end{align}
The factorization follows by changing integration variables from
$(\mu,\nu)$ to $(\mu_+,\mu_-)$. At the perturbative level these are the two Airy gravitational
blocks. The perturbative squashed sphere partition function
takes the form
\begin{align}
Z_b^{\rm pert}(N,k,\Delta)
 &=\ex^{\mathcal A}\mathcal C^{-1/3}
 \operatorname{Ai}\!\left[
 \mathcal C^{-1/3}(N-\mathcal B)\right]\, ,
\label{eq:TTI_sphere_Airy_block}
\end{align}
with the coefficients in \eqref{eq:TTI_sphere_Airy_coefficients}.
The two factors in \eqref{eq:TTI_canonical_gluing} are obtained
by analytic continuation of the squashed sphere block, with
integration contours $\Gamma_\sigma$ and branches of
$\log\mathcal Z_\sigma$ chosen as in
\cite{BenettiGenolini:2026kjy}.
This choice of contours and branches selects perturbative saddles with opposite signs,
\begin{align}
 \mu_\sigma^*
 &=\sigma\sqrt{\frac{N-\mathcal B_\sigma}{\mathcal C_\sigma}}\, ,
 \label{eq:TTI_block_saddles}
\end{align}
such that the perturbative partition function $Z_{\rm TTI}(N)$ in the unrefined limit matches the field theory results of \cite{Bobev:2022jte}.

Expanding to first instanton order, the correction to the free energy
of a canonical block is obtained from
\begin{align}
\mathcal F_\sigma^{(1)}
 &=-\frac{1}{\mathcal Z_\sigma^{\rm pert}(N)}
 \int_{\Gamma_\sigma}\frac{\dd\mu_\sigma}{2\pi\ii}\,
 \ex^{J_\sigma^{\rm pert}(\mu_\sigma)-N\mu_\sigma}
 J_\sigma^{(1)}(\mu_\sigma)\, ,
 \nonumber\\
 \mathcal F_{\rm TTI}^{(1)}
 &=\mathcal F_+^{(1)}+\mathcal F_-^{(1)}\, .
\label{eq:TTI_canonical_membrane_insertion}
\end{align}
We define the block free energies by
$\mathcal F_\sigma=-\log\mathcal Z_\sigma$, with logarithm
branches fixed by the analytic continuation described above.
Since $\mathcal A_\sigma$ is independent of $\mu_\sigma$,
the factor $\ex^{\mathcal A_\sigma}$ cancels between the
numerator and denominator of
\eqref{eq:TTI_canonical_membrane_insertion}.
For a term of the form $P(\mu)\ex^{-a\mu}$ in $ J^{(1)}(\mu)$, where $P$ is a polynomial function, the corresponding
sphere block can be computed directly from the Airy function:
\begin{align}
 \mathcal F_b^{(1)}
 &=-\frac{1}{Z_b^{\rm pert}(N)}
\Big[
 P(-\partial_N)Z_b^{\rm pert}(N+a)
 \Big]\, .
 \label{eq:TTI_Airy_shift_rule}
\end{align}
No saddle
approximation has been made in this step. 
At fixed $N$, the Airy argument becomes large as
$b\to0$, and we extract the required small $b$
coefficients from its asymptotic expansion.
Its small $b$ expansion includes the change in the saddle
and Gaussian prefactor under $N\mapsto N+a$. These effects are missed by replacing the grand potential term
$P(\mu)\ex^{-a\mu}$ by its value
$P(\mu_*)\ex^{-a\mu_*}$ at the unshifted saddle. 
They can contribute at order $b^0$ in the small $b$ expansion
of the canonical free energy when $P(\mu)$ carries a
$b^{-2}$ prefactor, as in the examples below.

To describe the unrefined limit, write the single instanton
correction to the canonical sphere free energy at fixed
chemical potentials as
\begin{align}
\mathcal F_b^{(1)}(N,k,\Delta)
 &=\frac{f_{-1}^{(1)}(N,k,\Delta)}{b^2}
  +f_0^{(1)}(N,k,\Delta)+\cO(b^2)\, .
\label{eq:TTI_block_small_b_coefficients}
\end{align}
The local chemical potentials in
\eqref{eq:TTI_local_sphere_parameters} have the expansion
$\Delta_{\sigma,I}
 =\Delta_I+\sigma\epsilon(\mathfrak n_I-\Delta_I)+\cO(\epsilon^2)$.
Using $b_\sigma^2=\sigma\epsilon$, we insert these
expansions into the sum of canonical block corrections
in \eqref{eq:TTI_canonical_membrane_insertion}.
The $1/\epsilon$ terms cancel, leaving the gravitational
result for the unrefined TTI free energy correction:
\begin{align}
\lim_{\epsilon\to0}\mathcal F_{\rm TTI}^{(1)}
 &=2\left[f_0^{(1)}
   +\sum_I(\mathfrak n_I-\Delta_I)
      \partial_{\Delta_I}f_{-1}^{(1)}\right]\, .
\label{eq:TTI_general_finite_part}
\end{align}
The derivatives act on the chemical potentials $\Delta_I$
of the canonical block, and their combination in
\eqref{eq:TTI_general_finite_part} is tangent to the
constraint $\sum_I\Delta_I=2$.
Equations~\eqref{eq:TTI_local_membrane_grand_potential},
\eqref{eq:TTI_canonical_membrane_insertion} and
\eqref{eq:TTI_general_finite_part} 
give analytic expressions for
both the refined index and its unrefined limit.

We now relate these gravitational coefficients to
field theory. We use the real free energy
$F=-\operatorname{Re}\log Z=-\log|Z|$.
For the single instanton corrections, we therefore define
\begin{align}
\label{eq:TTI_real_instanton_corrections}
 F_{S_b^3,\rm np}^{(1)}
 &\equiv\operatorname{Re}\mathcal F_b^{(1)}\, ,\qquad
 F_{\rm TTI,np}^{(1)}
 \equiv\operatorname{Re}\!\left[
 \lim_{\epsilon\to0}\mathcal F_{\rm TTI}^{(1)}
 \right]\, .
\end{align}
Known field theory relations identify the coefficient
of $b^{-2}$ in the sphere free energy with
$\mathcal W/(2\pi)$, where
$\mathcal W=\operatorname{Im}\widetilde{\mathcal W}$
is the on-shell twisted superpotential at the selected
Bethe vacuum.\footnote{Equation~(5.1) of \cite{Bobev:2025ltz} equates the
leading SCI Cardy free energy to twice the leading
small-$b$ sphere free energy.
The Bethe vacuum Cardy expansion, equation~(4.3)
of \cite{Hosseini:2026lvj}, gives the SCI pole term
as $\mathcal W/(\pi\omega)$.
Comparing these coefficients with $\omega=b^2$,
for the same vacuum and compatible analytic
continuation, fixes the sphere coefficient to
$\mathcal W/(2\pi)$.
The finite $N$ Bethe relation includes the
non-perturbative part of the twisted superpotential;
the explicit large $N$ evaluation in
\cite{Bobev:2025ltz} retains its perturbative part.
A formal gravitational extension to the SCI,
including the issue of the different fixed point
spinor chiralities, is discussed in
Appendix~\ref{app:ABJM_SCI}.}
Our gravitational calculation of the pole coefficient
$f_{-1}^{(1)}$ in
\eqref{eq:TTI_block_small_b_coefficients} therefore predicts
$\mathcal W_{\rm np}^{(1)}
=2\pi\operatorname{Re}f_{-1}^{(1)}$.
For real chemical potentials $\Delta_I$, combining
\eqref{eq:TTI_block_small_b_coefficients}
and \eqref{eq:TTI_general_finite_part}
then gives
\begin{align}
 F_{S_b^3,\rm np}^{(1)}
 &=\frac{\mathcal W_{\rm np}^{(1)}}{2\pi b^2}
   +\frac12 F_{\rm TTI,np}^{(1)}(\Delta,\mathfrak n)
 -\frac1{2\pi}\sum_I(\mathfrak n_I-\Delta_I)
      \partial_{\Delta_I}\mathcal W_{\rm np}^{(1)}
   +\cO(b^2)\, .
 \label{eq:TTI_sphere_Bethe_dictionary}
\end{align}
For complex chemical potentials, we retain the analytic expression
\eqref{eq:TTI_general_finite_part} until taking the real part.
The last term vanishes when $\mathfrak n_I=\Delta_I$, for which
the local chemical potentials at the two poles coincide, as one sees 
from \eqref{eq:TTI_local_sphere_parameters}. The
finite sphere coefficient is then half the TTI correction.
This includes the ABJM universal values
$\Delta_I=\mathfrak n_I=\tfrac12$.
In the next subsection we evaluate the bulk coefficients and test
both predictions against field theory.

\subsection{The universal twist}
\label{subsec:TTI_universal_tests}
We now set $\Delta_I=\mathfrak n_I=\tfrac12$ and extract
the leading large $N$ exponential 
from the single M2-brane contribution associated
with full-stabilizer linear fillings.
 This determines the first instanton
sector used in the field theory comparison below.
As discussed in section 5 of
\cite{BenettiGenolini:2026kjy}, a globally defined
uplift to $D=11$ with these magnetic fluxes requires even $k$.
For odd $k$, we analytically continue the local block
expressions to these fluxes, allowing a comparison with
the field theory computation of \cite{Hosseini:2026lvj} for $k=1$.

To compute the M2-brane family
contributions, we introduce the auxiliary Nosaka weights
$\epsilon_A=\tfrac12(\xi,\eta,-\xi,-\eta)$ of
\eqref{eq:ABJM_round_auxiliary_weights}. 
For generic $\xi,\eta$, this auxiliary torus isolates the
coordinate fillings $W_{IJ}$ and is therefore sufficient
for the fixed point calculation at $\Delta_I=\tfrac12$.
We sum their contributions and remove $\xi,\eta$ at fixed
non-zero $b$. We then perform the inverse Laplace transform,
as in section~\ref{subsec:TTI_fixed_rank}, and take the
small $b$ limit at fixed $N$.

We start from the full-stabilizer family sum
$J_{\rm full}(k,\mu;b)$ in the first line of
\eqref{eq:ABJM_pure_squashing_family_sums},
keeping $b$ general.
For M2-branes whose covering planes are preserved by the
entire quotient group $\Z_k$, and hence lie in $\cM_k$,
the common M2-brane action is
\begin{align}
I_{\rm M2}&=\frac{16\mu}{kQ^2}\, .
\label{eq:TTI_massless_membrane_action}
\end{align}
At $k=1,2$, this sector contains the full
$\operatorname{Gr}(2,4)$ family. For $k>2$, it contains
$\{W_{12}\}\sqcup(\CP^1\times\CP^1)\sqcup\{W_{34}\}$.
For $k>2$, the generic orbit sector discussed in
section~\ref{subsubsec:ABJM_Round_kgt2} has action
$(k/h)I_{\rm M2}$, where $h=\gcd(k,2)$, and
enters at a higher exponential order than the one being considered in this subsection.

At $\Delta_I=\tfrac12$, the perturbative squashed sphere
partition function has Airy coefficients
\begin{align}
\mathcal C&=\frac{32}{\pi^2kQ^4}\, ,
 \qquad
 \mathcal B=\frac{k}{24}-\frac{2}{3k}+\frac{4}{kQ^2}\, .
\label{eq:TTI_massless_Airy_coefficients}
\end{align}
The positive sphere saddle, $\mu_*=\sqrt{(N-\mathcal B)/\mathcal C}$,
grows as $b^{-2}$, while the M2-brane action has the finite limit
\begin{align}
t&\equiv\lim_{b\to0}
 \left.I_{\rm M2}\right|_{\mu=\mu_*}
 =2\pi\sqrt{\frac{2\widehat N}{k}}\, ,
\label{eq:TTI_action_at_saddle}
\end{align}
where we have introduced the shifted rank $\widehat N$:
\begin{align}
\widehat N&\equiv N-\frac{k}{24}+\frac{2}{3k}\, .
\label{eq:TTI_rank_and_action}
\end{align}
The saddle identifies the instanton scale $t$;
the prefactors below are obtained from the Airy
transform \eqref{eq:TTI_Airy_shift_rule}.
We work with $\widehat N>0$ in the large $N$ regime
and extract the coefficients of the first exponential $\ex^{-t}$.

For $k=1,2$, the terms needed for this transform
are obtained by expanding the grand potential \eqref{eq:ABJM_sq_k1_closed_form}
at fixed $I_{\rm M2}$:
\begin{align}
 J_{\rm full}
 &=\ex^{-I_{\rm M2}}\bigg[
 -\frac{5k^2}{4\pi^2b^2}(I_{\rm M2}+1)
 -\frac{13k^2}{4\pi^2}I_{\rm M2}^2
 -\frac{5k^2}{\pi^2}(I_{\rm M2}+1)+18+\cO(b^2)
 \bigg]\, .
\label{k12eqjbbb}
\end{align}
The $b^{-2}$ term in \eqref{k12eqjbbb} leads to a singular
contribution to the free energy of each canonical sphere block. 
Given $b_+^2=\epsilon$ and $b_-^2=-\epsilon$, these singular
contributions cancel in the sum of their free energy corrections for the TTI,
as in
\eqref{eq:TTI_general_finite_part}. However, this term also generates
an additional finite contribution under the inverse Laplace transform.
To see this, we integrate the factor
$(I_{\rm M2}+1)\ex^{-I_{\rm M2}}$ with the perturbative weight
$\ex^{J^{\rm pert}-N\mu}$ and divide by the perturbative partition
function. Using \eqref{eq:TTI_Airy_shift_rule} and including the
overall minus sign in \eqref{eq:TTI_canonical_membrane_insertion},
the first term contributes
\begin{align}
\frac{\displaystyle\int\frac{\dd\mu}{2\pi\ii}\,
 \ex^{J^{\rm pert}-N\mu} \frac{5k^2}{4\pi^2b^2}(I_{\rm M2}+1)\ex^{-I_{\rm M2}}}
 {Z_b^{\rm pert}(N)}
 &=\ex^{-t}\left[
 \frac{5k^2}{4\pi^2b^2}(t+1)-20+\cO(b^2)
 \right]\, .
\label{eq:TTI_ensemble_finite_correction}
\end{align}
Combining the finite correction $-20\mskip1mu\ex^{-t}$ with the contribution
$-18\mskip1mu\ex^{-t}$ from the term $+18\mskip1mu\ex^{-I_{\rm M2}}$ in
\eqref{k12eqjbbb} gives $-38\mskip1mu\ex^{-t}$ in each canonical sphere
free energy.
Together with the remaining polynomial terms, this gives the
$k=1,2$ line of \eqref{eq:TTI_massless_sphere_checks} below.
At $k=4$, the analogous correction is $-4\mskip1mu\ex^{-t}$;
combining it with the grand potential constant $+6$ gives
the offset $-10\mskip1mu\ex^{-t}$ in the second line.

For $k=1,2,4$, the canonical sphere corrections are
\begin{align}
F_{S_b^3,\rm np}^{(1)}&=\ex^{-t}
 \begin{cases}
 \displaystyle
 \frac{5k^2}{4\pi^2b^2}(t+1)
 +\frac{13k^2}{4\pi^2}t^2
 +\frac{5k^2}{\pi^2}(t+1)-38+\cO(b^2)\, ,&k=1,2\, ,\\[6pt]
 \displaystyle
 \frac{4}{\pi^2b^2}(t+1)
 +\frac{4}{\pi^2}t^2
 +\frac{8}{\pi^2}(t+1)-10+\cO(b^2)\, ,&k=4\, .
 \end{cases}
\label{eq:TTI_massless_sphere_checks}
\end{align}
At $k=4$, the two isolated planes make no contribution at these
orders: their original determinants give
$Z^{\rm full}_{12,4}=Z^{\rm full}_{34,4}
 =\tfrac14\sin^2[\pi b^2/(1+b^2)]=\cO(b^4)$.
Using \eqref{eq:TTI_sphere_Bethe_dictionary}, we obtain
the corresponding corrections to the twisted superpotential,
\begin{align}
\label{eq:TTI_twisted_superpotential_checks}
 \mathcal W_{\rm np}^{(1)}
 &=\frac{k^2}{2\pi}(t+1)\ex^{-t}
 \begin{cases}5\, ,&\qquad k=1,2\, ,\\1\, ,&\qquad k=4\, ,\end{cases}
 \end{align}
and the 
TTI free energy:
\begin{align}\label{eq:TTI_index_checks}
 F_{\rm TTI,np}^{(1)}
 &=\ex^{-t}
 \begin{cases}
 \displaystyle
 \frac{k^2}{\pi^2}\Big[\frac{13}{2}t^2+10(t+1)\Big]-76\, ,
 &\qquad k=1,2\, ,\\[5pt]
 \displaystyle
 \frac{16}{\pi^2}\Big[\frac12t^2+t+1\Big]-20\, ,
 &\qquad k=4\, .
 \end{cases}
\end{align}
The result \eqref{eq:TTI_twisted_superpotential_checks} reproduces the complete single instanton
coefficient of the twisted superpotential reconstructed in
\cite{Hosseini:2026dkj}, while \eqref{eq:TTI_index_checks} reproduces the single instanton
polynomials of the TTI in \cite{Hosseini:2026lvj}
(see equation~(5.14) and Table~3),
including their constant terms.\footnote{In the field theory description, summing over the
$k$ symmetry-related Bethe vacua multiplies the contribution
of a single vacuum by $k$. This shifts the free energy by
$-\log k$ and leaves the instanton coefficients unchanged.}

The prescription also gives predictions at other values of $k$.
As an explicit example admitting a global uplift at the universal
twist, take $k=6$.
Here
$\widehat N=N-5/36$ and
$t=2\pi\sqrt{\widehat N/3}$. 
Both the mixed family and the isolated planes contribute
at this level. Expanding
\eqref{eq:ABJM_sq_full_isolated} and
\eqref{eq:ABJM_sq_mixed_coefficients} at small $b$ gives
the grand potential
\begin{align}\label{jttismallbbothfam}
	J_{\rm full} = &\Bigg[  \frac{1}{b^2}\left( -\frac{9}{\pi^2}(I_{\mathrm{M2}}+1)+\frac{\sqrt{3}}{\pi}\right)
 -\frac{9}{\pi^2}I_{\mathrm{M2}}^2
+\left(-\frac{18}{\pi^2}+\frac{6\sqrt{3}}{\pi}\right)I_{\mathrm{M2}}
-\frac{18}{\pi^2}+\frac{5\sqrt{3}}{\pi}+6 \nn\\
&+ \left(\rho_6-\frac{\sqrt{3}}{\pi}\right)
\left(
\frac{1}{b^2}+5+\frac{2\pi}{\sqrt{3}}
\right) +\cO(b^2) \Bigg] \ex^{-I_{\rm M2}} \, ,
\end{align}
where
\begin{align}
\rho_6&=\frac{\sqrt3}{2\pi}
 \left[2+\exp\!\left(-\frac9\pi
     \operatorname{Cl}_2\!\left(\frac{2\pi}{3}\right)\right)\right]\, ,
\label{eq:TTI_k6_constant}
\end{align}
and $\operatorname{Cl}_2(\theta)
 =\sum_{n\geq1}\frac{\sin(n\theta)}{n^2}$ is the Clausen function. 
The first line of \eqref{jttismallbbothfam} comes from the mixed family, whereas the last line is the contribution from isolated planes. 
Applying the same Airy transform and
\eqref{eq:TTI_sphere_Bethe_dictionary} gives
\begin{align}
\label{eq:TTI_k6_superpotential_prediction}
 \mathcal W_{\rm np}^{(1)}(N,6)
 &=\left[\frac{18}{\pi}(t+1)-2\pi\rho_6\right]\ex^{-t}\, ,\\
 F_{\rm TTI,np}^{(1)}(N,6)
 &=\ex^{-t}\Bigg[
 \frac{18}{\pi^2}t^2
 +\left(\frac{36}{\pi^2}-\frac{12\sqrt3}{\pi}\right)t
 +\frac{36}{\pi^2}-16\nonumber\\
 &\hspace{22mm}
 +2\rho_6\left(
 -5-\frac{2\pi}{\sqrt3}
 +\frac{4\pi^2}{9t}+\frac{8\pi^2}{9t^2}
 \right)\Bigg]\, .
\label{eq:TTI_k6_index_prediction}
\end{align}
The inverse powers of $t$ arise from the Airy transform of the
$\mu$-independent part of the leading $b^{-2}$ insertion.
At $k=6$, the generic orbit sector starts at $\ex^{-3t}$
and does not affect the displayed leading coefficient.
It would be interesting to test these predictions against field
theory, both at $k=6$ and at other levels.

\section{The \texorpdfstring{$V^{5,2}$}{V52} theory}
\label{sec:V52_Theory}

We next apply the M2-brane prescription to the $3d$ $\cN=2$
Chern--Simons-matter theory 
derived in \cite{Martelli:2009ga},
dual to M-theory on $EAdS_4\times V^{5,2}/\Z_k$, where
$\Z_k$ acts freely. The gauge group is
$U(N)_k\times U(N)_{-k}$, with two bifundamental and two
anti-bifundamental chiral multiplets, one adjoint chiral
multiplet at each node, and a cubic superpotential.\footnote{
At $k=1$, a related single node theory with the same
geometric branch of the moduli space was proposed in
\cite{Jafferis:2009th}.}

The perturbative M-theory partition function on the
three-sphere was computed in \cite{BenettiGenolini:2026kjy}.
Here we compute M2-instanton contributions arising from
linear holomorphic fillings. The Calabi--Yau cone over
$V^{5,2}$ is a non-toric quadric with an isolated singularity
at its apex. The linear fillings are smooth complex planes
passing through this singular point. We account for the
ambient singularity by constructing their fluctuation
character using an intrinsic virtual normal bundle as in \cite{BenettiGenolini:2026kjy}.

We test the resulting instanton contributions using
an exact finite $N$ relation to round sphere ABJM
at special squashing and a numerical comparison
with the round sphere matrix model in the planar
limit. These comparisons provide further checks
of our treatment of the singular cone.\footnote{It would be interesting to apply the
fixed-$N$, small-squashing analysis of
section~\ref{subsec:TTI_fixed_rank} to our instanton
contributions and compare the resulting predictions
with the recently obtained field theory results
for the on-shell twisted superpotential
\cite{Hosseini:2026kyx}.}

Our fixed point prescription accommodates general squashing
and Cartan flavour deformations. To illustrate, we explicitly evaluate the
family sums at vanishing flavour masses, using the Cartan
action as an auxiliary refinement to isolate the fillings.
Its parameters are removed after summing the fixed point
contributions, keeping the squashing fixed
(as in section~\ref{subsubsec:ABJM_pure_squashing}).
We first
describe the linear fillings and their quotient stabilizers,
before restricting the explicit computations to the sector
fully stabilized by~$\Z_k$.

\subsection{Linear fillings and their stabilizers}
\label{subsec:V52_geometry}

The homogeneous Sasaki--Einstein manifold
$V^{5,2}\cong SO(5)/SO(3)$ is non-toric and has a Calabi--Yau cone given by
\begin{align}
\label{eq:V52_cone}
 C(V^{5,2})
 =\left\{z_0^2+x_1y_1+x_2y_2=0\right\}
 \subset\C^5\, .
\end{align}
The Reeb circle acts by a common phase on $(z_0,x_1,y_1,x_2,y_2)$, each of which has 
R-charge $2/3$. The quotient of $V^{5,2}$ by the Reeb circle is the smooth projective
quadric $Q^3\subset\CP^4$
\begin{align}
\label{eq:V52_Q3}
	Q^3 = \bigg\{ [z_j] \in \CP^4 \mid \sum_j z_j^2 = 0 \bigg\} \cong \operatorname{Gr}_2^+(\R^5) \, .
\end{align}

The M-theory circle $U(1)_M$,
along which one may reduce to type IIA, is the diagonal
$U(1)$ in the maximal torus of $SO(5)$, acting with
charges $(0,1,-1,1,-1)$ on $(z_0,x_1,y_1,x_2,y_2)$.
The subgroup $\Z_k\subset U(1)_M$ is generated by
\begin{align}
\label{eq:V52_M_Action}
 (z_0,x_1,y_1,x_2,y_2)
 &\longmapsto
 (z_0,\omega x_1,\omega^{-1}y_1,
       \omega x_2,\omega^{-1}y_2)\, ,
 \qquad \omega=\ex^{2\pi\ii/k}\, .
\end{align}
This action is free on the $V^{5,2}$ link (a point fixed by a
non-trivial element would have all $x_i,y_i$ zero, and
the quadric equation would then force $z_0=0$). This action defines the $\Z_k$ quotient of \cite{Martelli:2009ga}.

For the sphere backgrounds, 
we take the auxiliary
twelve-dimensional filling to be
$W\cong M_4\times C(V^{5,2})/\Z_k$, with $M_4\cong\R^4$ and
the cone radially truncated to provide the internal
filling, as in \cite{BenettiGenolini:2026kjy}.
The M2-branes lie at the fixed point of $K$ in $M_4$.
We consider holomorphic fillings represented on the
covering cone $C(V^{5,2})$ by linear complex planes
$D\cong\C^2$. 
The space of linear covering planes is therefore
the Fano variety of lines in $Q^3$, which is
isomorphic to $\CP^3$:\footnote{
A line in $Q^3$ is the projectivization of a
two-dimensional subspace of $\C^5$ on which
the quadratic form vanishes identically.
Thus the Fano variety of lines is the
orthogonal Grassmannian $\operatorname{OGr}(2,5)$,
which is isomorphic to $\CP^3$ via the
spinor representation of $\operatorname{Spin}(5)$.
}
\begin{align}
\label{eq:V52_linear_family}
 F_1(Q^3)\cong\CP^3\, .
\end{align}
Planes in the same $\Z_k$ orbit describe the same
quotient configuration. As in the ABJM analysis of
section~\ref{subsec:Geometry_Semiclassical_Action},
let $\Gamma_D\subset\Z_k$ denote the setwise stabilizer
of $D$, with order $s$. The covering link is $S^3$,
and the membrane worldvolume is its free quotient
$S^3/\Gamma_D\cong S^3/\Z_s$.

Four coordinate planes will be useful:
\begin{align}
\label{eq:V52_coordinate_planes}
 D_{++}&=\operatorname{span}_{\C}
       \{e_{x_1},e_{x_2}\}\, ,
 &D_{+-}&=\operatorname{span}_{\C}
       \{e_{x_1},e_{y_2}\}\, ,
 \nonumber\\
 D_{-+}&=\operatorname{span}_{\C}
       \{e_{y_1},e_{x_2}\}\, ,
 &D_{--}&=\operatorname{span}_{\C}
       \{e_{y_1},e_{y_2}\}\, .
\end{align}
Each lies in $C(V^{5,2})$ and is preserved by the full $\Z_k$.
The quotient acts with equal charges on the two
coordinates of $D_{++}$ and $D_{--}$, and opposite
charges on those of $D_{+-}$ and $D_{-+}$.

The planes contained in $z_0=0$ correspond to lines
in the smooth quadric surface $Q^2\subset\CP^3$.
These planes form two disjoint families, each
parametrized by $\CP^1$.
We denote the family containing $D_{+-},D_{-+}$ by
$\CP^1_{\rm mix}$, and the family containing
$D_{++},D_{--}$ by $\CP^1_{\rm other}$.
The full-stabilizer locus is given by
\begin{align}
\label{eq:V52_full_stabilizer_locus}
 F_1(Q^3)^{\Z_k}
 =
 \begin{cases}
 \CP^3\, ,
     & k=1\, ,
 \\[2pt]
 \CP^1_{\rm mix}\sqcup\CP^1_{\rm other}\, ,
     & k=2\, ,
 \\[2pt]
 \{D_{++}\}\sqcup\CP^1_{\rm mix}\sqcup\{D_{--}\}\, ,
     & k>2\, .
 \end{cases}
\end{align}
For $k=2$, the generator in \eqref{eq:V52_M_Action} fixes $z_0$ and negates the
other four coordinates. Hence, an invariant plane lying in
the quadric must be contained in $z_0=0$,
giving precisely the two families above.
For $k>2$, the generator has distinct eigenvalues
$\omega,\omega^{-1},1$ on $D_{++}$, $D_{--}$ and
$\C e_{z_0}$, respectively. An invariant plane cannot
contain $e_{z_0}$ and must split between the first
two eigenspaces. It is therefore either $D_{++}$,
$D_{--}$, or the sum of one line in each eigenspace.
In the last case, the quadric requires the complex
bilinear pairing of the two lines to vanish.
Either line determines the other, leaving the single
family $\CP^1_{\rm mix}$.

There are also planes outside the full-stabilizer locus.
These have trivial stabilizer and worldvolume $S^3$,
with one exception. For even $k>2$, the planes within the family
$\CP^1_{\rm other}\setminus\{D_{++},D_{--}\}$
have stabilizer $\Z_2$ and worldvolume $\mathbb{RP}^3$.

For simplicity, we now restrict our explicit calculations
to M2-branes wrapping once the quotient links of planes
in the full-stabilizer locus
\eqref{eq:V52_full_stabilizer_locus}.
We evaluate this sector at vanishing flavour masses,
allowing arbitrary squashing of the three-sphere.
In this setting, the full-stabilizer configurations
have the smallest action among the linear fillings.\footnote{For $k>2$, these M2-branes wrap the reduction
circle $U(1)_M/\Z_k$ and give fundamental string
worldsheet instantons in type IIA. At $k=1,2$, the
larger families also contain transverse representatives
with a D2-brane interpretation.}
The sectors with smaller stabilizers contribute at
higher action. Their evaluation requires an analysis
of the corresponding covered limiting configurations,
which we will not pursue here.

\subsection{The action of \texorpdfstring{$K$}{K}
and the M2-brane action}
\label{subsec:V52_K_Action}

We use the external weights $b_1,b_2$ for the action of the Killing vector $K$ on $M_4$ and again define
the squashing parameter $b$ via \eqref{eq:ABJM_ExternalWeights},
with $b_1+b_2=2$ and $b_1b_2=4/Q^2$.
At vanishing flavour masses, the internal part of $K$ is proportional to the Reeb vector and
acts with common weight $-2/3$ on the five coordinates 
of $C(V^{5,2})$. It therefore preserves every linear filling,
independently of the squashing.

To compute the family contributions, we combine $K$ with
an auxiliary Cartan action parametrized by $u,v$.
The weights of this combined torus action $T$ on the two external
directions and the five internal coordinates, ordered as $(z_0,x_1,y_1,x_2,y_2)$, are
\begin{align}
\label{eq:V52_K_weights}
 \Bigl(
 b_1,b_2;\,
 -\frac23,
 -\frac23-u,-\frac23+u,
 -\frac23-v,-\frac23+v
 \Bigr)\, .
\end{align}
The quadric equation has weight $-4/3$ under this action, so the
holomorphic four-form on the smooth locus of $C(V^{5,2})$
has weight $-2$. Together with $b_1+b_2=2$, this
ensures that the holomorphic volume form on $W$
is uncharged under the combined action.
For generic $u,v$, the only linear fillings invariant under $T$
are the four coordinate planes given in
\eqref{eq:V52_coordinate_planes}.

All four planes pass through the cone apex and share
the same value of $\Phi^G_0$, hence the same flux
variable $\mu$ defined in \eqref{eq:mu_a_Def}.
On $D_{s_1s_2}$, where $s_1,s_2=\pm1$, the two
tangent weights are
$-(\frac23+s_1u)$ and $-(\frac23+s_2v)$.
Using the localization formula \eqref{eq:SM2_Generic}
and the full stabilizer $\Z_k$, we obtain the M2-brane action
\begin{align}
\label{eq:V52_refined_M2_action}
 I_{\rm M2}[D_{s_1s_2}]
 =\frac{4\mu}{
 kQ^2(\frac23+s_1u)(\frac23+s_2v)}\, .
\end{align}
Removing the auxiliary parameters gives the common
action throughout the full-stabilizer locus,
\begin{align}
\label{eq:V52_pure_squashing_action}
 I_{\rm M2}=\frac{9\mu}{kQ^2}\, .
\end{align}
At $b=1$, this reduces to $9\mu/(4k)$.\footnote{More generally, a linear configuration with stabilizer
order $s$ has action $9\mu/(sQ^2)$. Thus the smaller stabilizer
sectors discussed in the previous subsection, which we are not
considering here, have larger action than the full-stabilizer
sector $s=k$.}

In the family calculation below, we retain $u,v$ 
in both the M2-brane action and the fluctuation determinant, and
remove them only after summing the fixed point
contributions, with $b$ and $\mu$ held fixed.

\subsection{Fluctuations on the singular cone}
\label{subsec:V52_Fluctuations}

We now construct the fluctuation character, extending
the construction of section~\ref{subsec:Quantum_Brane_Families}
to the singular cone $C(V^{5,2})$. Following the treatment of the
bulk character in Appendix~E of \cite{BenettiGenolini:2026kjy},
we use the intrinsic virtual tangent bundle of the
hypersurface. Let $L_F$ denote the line associated with
the quadric equation, with weight $-4/3$ and zero
$\Z_k$ charge. Then
\begin{align}
\label{eq:V52_virtual_normal}
 [T_{C(V^{5,2})}^{\rm vir}]
 &=[T\C^5|_{C(V^{5,2})}]-[L_F]\, ,
 \nonumber\\
 [N_{D/{C(V^{5,2})}}^{\rm vir}]
 &=[T_{C(V^{5,2})}^{\rm vir}|_D]-[TD]
  =[N_{D/\C^5}]-[L_F|_D]\, .
\end{align}
The last expression contains the three ambient normal
directions minus the quadric constraint. It has virtual
complex rank two and remains well-defined at the apex.

Including the two external directions, we define
\begin{align}
\label{eq:V52_virtual_complex}
 V_D^{\rm vir}
 &=\bigl(\mathcal O_D\otimes T^{1,0}_0M_4\bigr)
   \oplus N_{D/{C(V^{5,2})}}^{\rm vir}\, ,
 \nonumber\\
 E_D^{\rm vir}
 &=V_D^{\rm vir}
   -K_D\otimes(V_D^{\rm vir})^*\, .
\end{align}
Here $K_D$ is the canonical bundle of the smooth
covering plane. For an ordinary rank four bundle,
the identity
$\Lambda^3V_D\cong(\det V_D)\otimes V_D^*$,
together with $\det V_D\cong K_D$, recovers the
smooth fluctuation complex \eqref{eq:ED_def}.
For the virtual bundle, we take the second line of
\eqref{eq:V52_virtual_complex} as the definition
of our extended fluctuation complex.\footnote{This definition is the rule of
section~\ref{subsec:Quantum_Brane_Families} applied to the virtual
normal class, $X_W(i_*V_D^{\rm vir}) = i_*E_D^{\rm vir}$, in the
notation of \eqref{eq:XW_def_main}.}

For the linear filling $D=D_{s_1s_2}$ and $j=0,\ldots,k-1$, 
for each quotient element $\omega^j$, define
\begin{align}
\label{eq:V52_character_variables}
 A_j=\omega^{-js_1}\ex^{-t(\frac23+s_1u)}\, ,
 \qquad B_j=\omega^{-js_2}\ex^{-t(\frac23+s_2v)}\, .
\end{align}
These are the characters of the two holomorphic
coordinate functions on $D$, and $A_jB_j$ is the
fibre character of $K_D$.
The three ambient normal directions in
\eqref{eq:V52_virtual_normal} have characters
$\ex^{2t/3}$, $\ex^{4t/3}A_j$ and
$\ex^{4t/3}B_j$.
Subtracting the quadric constraint gives
\begin{align}
\label{eq:V52_virtual_normal_character}
 \operatorname{ch}_j(N_{D/C(V^{5,2})}^{\rm vir})
 &=\ex^{2t/3}
   +\ex^{4t/3}(A_j+B_j-1)
 \nonumber\\
 &=\ex^{2t/3}+A_jB_j\ex^{4t/3}
   -\ex^{4t/3}(1-A_j)(1-B_j)\, .
\end{align}
Upon forming $E_D^{\rm vir}$ in
\eqref{eq:V52_virtual_complex} and dividing by the
Dolbeault denominator $(1-A_j)(1-B_j)$, the last
term contributes the finite pair
$-(\ex^{4t/3}-\ex^{-4t/3})$.
Applying the $\Z_k$ projection to $E_D^{\rm vir}$
then gives the full character:
\begin{align}
\label{eq:V52_projected_character}
 \cK^{\rm full}_{s_1s_2,k}(t)
 &=
 \frac1k\sum_{j=0}^{k-1}
\frac{ \Bigl[
 \sum_{\alpha=1}^{2}
 \bigl(\ex^{-tb_\alpha}-A_jB_j\ex^{tb_\alpha}\bigr)
 +(1-\ex^{-2t})
 \bigl(\ex^{2t/3}+A_jB_j\ex^{4t/3}\bigr)
 \Bigr]}{(1-A_j)(1-B_j)}
 \nonumber\\
 &\quad-\bigl(\ex^{4t/3}-\ex^{-4t/3}\bigr)\, .
\end{align}
In the first term,
the denominator accounts for holomorphic functions on
the plane, while the numerator contains the external
and internal fluctuations. As in
section~\ref{subsec:ABJM_Fluctuations}, the $\Z_k$
projection acts on both the fibre charges and the
holomorphic functions. 
The finite correction is neutral under $\Z_k$
and is therefore unchanged by the projection.
With the spectral convention of
\eqref{eq:1loop_def}, its zeta function
regularization contributes a factor $-1$
to each determinant. 

We denote the full determinant obtained by regularizing
\eqref{eq:V52_projected_character} by
$Z^{\rm full}_{s_1s_2,k}(b;u,v)$.
For fixed squashing and generic auxiliary parameters
$u,v$, all cohomological zero modes of the unrefined
$K$-spectrum acquire non-zero weights under $T$,
including the fermionic zero modes present at special
squashing values.
 The full determinants already include the Euler ratio
in \eqref{eq:Z1M}. We can therefore use them directly
in the fixed point sum, without  separate
tangent or obstruction Euler factors.
Combining them with the M2-brane actions
\eqref{eq:V52_refined_M2_action}, the full-stabilizer
linear contribution to the modified grand potential is then
\begin{align}
\label{eq:V52_full_family_sum}
 J_{\rm full}(k,\mu;b)
 =\lim_{u,v\to0}
 \sum_{s_1,s_2=\pm1}
 Z^{\rm full}_{s_1s_2,k}(b;u,v)\,
 \ex^{-I_{\rm M2}[D_{s_1s_2}]}\, .
\end{align}
The auxiliary limit is taken after summation, with
$b$ and $\mu$ held fixed.

\subsection{Family contributions on the squashed sphere}
\label{subsec:V52_squashed_results}

We now evaluate \eqref{eq:V52_full_family_sum} at vanishing
flavour masses. We use $f=(b^2-1)/(b^2+1)$, as in the
ABJM calculation.
All $\Z_k$-invariant linear fillings
then have the same instanton action \eqref{eq:V52_pure_squashing_action}, which here
we denote by
\begin{align}
I_k\equiv \frac{9\mu}{kQ^2}=\frac{9(1-f^2)\mu}{4k}\, .
\end{align}

We now identify the zero modes of the unrefined
$K$-spectrum by examining the bosonic and fermionic
mode weights at $u=v=0$.
Let $\ell\geq0$ denote the total degree of a holomorphic
monomial on the covering plane.
The relevant internal bosonic and external fermionic
towers in \eqref{eq:V52_projected_character} have the
following weights before the $\Z_k$ projection:
\begin{align}
\label{eq:V52_pure_squashing_mode_weights}
 \lambda^{\rm B}_{{\rm int},1}
 =\frac23(\ell-1)\, ,\qquad 
 \lambda^{\rm B}_{{\rm int},2}
 =\frac23\ell\, ,
\qquad
 \lambda^{\rm F}_{{\rm ext},\pm}
 &=\frac{1+2\ell}{3}\mp f\, .
\end{align}
For generic squashing, there are no fermionic
zero modes. 
The bosonic zero modes surviving the $\Z_k$ projection
are tangent to the components of the full-stabilizer locus
in \eqref{eq:V52_full_stabilizer_locus}:
there are
three complex zero modes on $\CP^3$ at $k=1$, one on each
of $\CP^1_{\rm mix}$ and $\CP^1_{\rm other}$ at $k=2$,
and one on $\CP^1_{\rm mix}$ at $k>2$.
The isolated planes $D_{++}$ and $D_{--}$ at $k>2$
have none. 

We now return to the refined calculation at
generic auxiliary parameters $u,v$ and evaluate
the fixed point sum \eqref{eq:V52_full_family_sum}.
For generic squashing, the poles cancel after
summation, giving a cubic prefactor in $I_1$
at $k=1$ and a linear prefactor in $I_k$
at $k\geq2$.
For $k=1$ and general squashing, evaluating the
fixed point sum as described in
Appendix~\ref{app:V52_determinants} gives
\begin{align}
\label{eq:V52_k1_general_squashing}
 &J_{\rm full}(1,\mu;b)
 ={}\frac{\ex^{-I_1}}{24\pi^3}
 \Biggl\{
 8\cos\!\left(\frac{3\pi f}{2}\right)I_1^3
 +48\left[
 \cos\!\left(\frac{3\pi f}{2}\right)
 +\frac{3\pi f}{2}\sin\!\left(\frac{3\pi f}{2}\right)
 \right]I_1^2
 \nonumber\\
 &+\Biggl[
 [108+8\pi^2(1-27f^2)]
 \cos\!\left(\frac{3\pi f}{2}\right)
 +252\pi f\sin\!\left(\frac{3\pi f}{2}\right)
 -\frac{15\pi^2(1-9f^2)}
        {\cos(3\pi f/2)}
 \Biggr](I_1+1)
 \nonumber\\
 &+12\pi^2(1-9f^2)
 \left[
 \cos\!\left(\frac{3\pi f}{2}\right)
 +2\pi f\sin\!\left(\frac{3\pi f}{2}\right)
 \right]
 \Biggr\}\, .
\end{align}
The prefactor is generically cubic in $\mu$.
The apparent singularities at $f=\pm1/3$ are removable as we elaborate on below.

For $k\geq2$, we obtain
\begin{align}
\label{eq:V52_kge2_general_squashing}
 J_{\rm full}(k,\mu;b)
 =\bigl[p_{k,1}(f)I_k+p_{k,0}(f)\bigr]\ex^{-I_k}\, .
\end{align}
The coefficient functions, including closed expressions
at $k=2$ and finite $k$ Barnes representations for
$k>2$, are given in Appendix~\ref{app:V52_determinants}.
The appendix also gives the pole cancellations and
coefficient extraction leading to
\eqref{eq:V52_k1_general_squashing}.
For $k\geq2$ and generic squashing, the term linear in
$\mu$ in the instanton prefactor comes entirely from
the $\CP^1$ components in
\eqref{eq:V52_full_stabilizer_locus}.
The mechanism is particularly simple.
The two fixed point determinants have opposite simple
poles in the auxiliary parameter. Their residues
cancel in the sum, while the first variation of the
instanton action multiplying these poles supplies the
term proportional to $I_k$.
At $k=2$, the families $\CP^1_{\rm mix}$ and
$\CP^1_{\rm other}$ give equal contributions in the
local phase convention used here.
At $k>2$, the linear term comes from
$\CP^1_{\rm mix}$, while the constant term also includes
the isolated planes $D_{++}$ and $D_{--}$.

On the round sphere, $b=1$, the explicit expressions for $k=1,2,3$ are:
\begin{align}
\label{eq:V52_round_low_levels}
 J_{\rm full}(1,\mu;1)
 &=\frac{\ex^{-9\mu/4}}{24\pi^3}
 \Bigl[
 \frac{729}{8}\mu^3+243\mu^2
 +\left(243-\frac{63\pi^2}{4}\right)\mu
 +108+5\pi^2
 \Bigr]\, ,
 \nonumber\\
 J_{\rm full}(2,\mu;1)
 &=\frac{3\ex^{-2G_{\rm C}/\pi}}{4\sqrt2\,\pi}
 (3\mu+4+\pi)\ex^{-9\mu/8}\, ,
 \nonumber\\
 J_{\rm full}(3,\mu;1)
 &=\left[
 \frac{3\mu+4}{8\pi}+\frac{2}{\sqrt3}
 \right]\ex^{-3\mu/4}\, ,
\end{align}
where
$G_{\rm C}=\operatorname{Cl}_2(\pi/2)$ is the Catalan constant.

The family sums simplify further at
$b^2=2$ and $b^2=1/2$, corresponding to $f=\pm1/3$.
In the unrefined spectrum, an external fermion of
monomial degree zero becomes a zero mode.
Its quotient charge is $-s_1-s_2$, so it survives
the $\Z_k$ projection at all four coordinate planes
$D_{s_1s_2}$ for $k=1,2$, and only at
$D_{+-}$ and $D_{-+}$ for $k>2$.
With auxiliary refinement, the surviving fermion
and a bosonic family mode have the same weight
$s_1u+s_2v$, which is non-zero for generic $u,v$.
Cancellations in the refined character then leave
two one-dimensional sums. The resulting full
determinant is 
\begin{align}
\label{eq:V52_special_squashing_determinant}
 Z^{\rm full}_{s_1s_2,k}
 =
 \frac{1}{
 4\sin\!\left(\frac{2\pi}{k(2+3s_1u)}\right)
  \sin\!\left(\frac{2\pi}{k(2+3s_2v)}\right)}\, .
\end{align}
The common instanton action is now $I_k=2\mu/k$.
Summing the four fixed point contributions before
removing the auxiliary parameters gives
\begin{align}
\label{eq:V52_special_squashing_results}
 J_{\rm full}(k,\mu;\sqrt2)
 &=J_{\rm full}(k,\mu;1/\sqrt2)
 =\begin{cases}
 \displaystyle
 \frac{4\mu^2+2\mu+1}{\pi^2}\ex^{-2\mu}\, ,
 &k=1\, ,
 \\[5pt]
 \displaystyle
 \frac{\ex^{-2\mu/k}}{\sin^2(\pi/k)}\, ,
 &k>1\, .
 \end{cases}
\end{align}
Thus the additional fermionic zero mode reduces the
cubic polynomial to a quadratic at $k=1$, and removes
the linear term at $k>1$.

These special squashing results admit an exact
field theory check. At fixed $N$ and vanishing
flavour masses, the exact $V^{5,2}$ partition function
at level $k$ and $b^2=2$ or $1/2$ equals the
round sphere ABJM partition function at level $2k$:
\begin{align}
\label{eq:V52_ABJM_special_squashing}
 Z_{V^{5,2}}(N,k;\sqrt2)
 &=Z_{V^{5,2}}(N,k;1/\sqrt2)
 =Z_{\rm ABJM}(N,2k;1)\, ,
\end{align}
as we show in Appendix~\ref{app:V52_matrix_identity}.
The known leading instanton contribution to the
round sphere ABJM grand potential at level $2k$
\cite{Hatsuda:2012dt,Hatsuda:2013gj} reproduces
\eqref{eq:V52_special_squashing_results} for
every $k$.
At $k=1$, this agreement includes the complete
quadratic prefactor multiplying $\ex^{-2\mu}$.

Returning to the round sphere, we can also take the
large $k$ limit with $x=\mu/k$ held fixed. We find
\begin{align}
\label{eq:V52_round_large_k}
 \lim_{k\to\infty}\frac{J_{\rm full}(k,kx;1)}{k^2}
 =
 \left[
 \frac{\frac92x+3+7\log2}{\Gamma(1/4)^4}
 +\frac4{9\pi^2}
 \right]\ex^{-9x/4}\, .
\end{align}
This gives the leading type IIA limit of the computed
full-stabilizer contribution. Its derivation from the
finite $k$ determinants is included in
Appendix~\ref{app:V52_determinants}.

After converting \eqref{eq:V52_round_large_k}
to fixed $N$ using a saddle point approximation
to the inverse Laplace transform, we compare it with a numerical
calculation of the round sphere field theory
matrix model in the planar limit.
The comparison supports the predicted instanton
action and both coefficients of the prefactor,
as detailed in
Appendix~\ref{app:V52_planar_numerics}.
The results at general squashing provide further
precise predictions for the leading instanton
contribution to the partition function of the
dual field theory.

\section{Discussion}
\label{sec:Discussion}

We have developed a precise localization formula for
M2-instanton contributions in supersymmetric M-theory
backgrounds. It reduces the calculation of the membrane
actions and fluctuation factors to fixed point data
on holomorphic fillings, making detailed instanton
computations tractable.
As in \cite{BenettiGenolini:2026kjy}, we first
compute at fixed chemical potential $\mu$,
conjugate to the M2-brane number $N$.
Partition functions at fixed $N$ may then be
obtained by an inverse Laplace transform.

We have applied our prescription to backgrounds dual
to ABJM theory on the three-sphere with arbitrary
squashing and flavour mass deformations, and to those
associated with its topologically twisted index.
For backgrounds dual to the $V^{5,2}$ theory,
we have evaluated the full-stabilizer linear contributions at arbitrary squashing and vanishing flavour masses.
The calculations reproduce detailed field theory
instanton data and give new predictions for deformed
backgrounds. Combined with the perturbative results, this remarkable
agreement provides quantitative evidence that the
auxiliary twelve-dimensional geometry captures both
perturbative M-theory corrections and exponentially
small contributions from wrapped membranes.
In particular, the topologically twisted index
calculations probe the non-perturbative quantum
description of supersymmetric $AdS_4$ black holes.

 Our gravitational calculations are complemented by
two field theory results for the $V^{5,2}$ theory,
providing highly non-trivial checks.
First, Appendix~\ref{app:V52_matrix_identity}
establishes an exact finite $N$ identity between
its partition function at level $k$ and
$b^2=2,\tfrac12$ and the round sphere partition
function of ABJM theory at level $2k$.
The known ABJM instanton coefficients then
reproduce our special squashing results for
every $k$.
Second, Appendix~\ref{app:V52_planar_numerics}
compares our large $k$ round sphere prediction
with a numerical field theory matrix model
calculation in the planar limit.
After converting the gravity prediction to fixed
$N$, the comparison supports the instanton action
and both coefficients of its prefactor.

The common rule \eqref{eq:XW_def_main} brings
the perturbative bulk fluctuations considered in
\cite{BenettiGenolini:2026kjy}\footnote{Independently, the $N$-independent contribution
to the perturbative ABJM free energy on the
round three-sphere has recently been reproduced
by a one-loop calculation in eleven-dimensional
supergravity \cite{Kurlyand:2026cyz}.} 
 and the M2-brane
fluctuations studied here into the same
holomorphic framework.
As in \cite{BenettiGenolini:2026kjy}, this
structure is motivated by twisted formulations
of eleven-dimensional supergravity
\cite{Costello:2016mgj,Costello:2018zrm,Raghavendran:2021qbh,Hahner:2023kts}
and protected M-theory indices
\cite{Nekrasov:2014nea}.
For the M2-brane sector, further motivation comes
from the cohomological fluctuation analysis in
two-derivative Freund--Rubin backgrounds
\cite{Kurlyand:2026yke,Kurlyand:2026gxi}.
Appendix~\ref{app:fluctuation_complex} shows
that our filling complex reproduces the graded
cohomological spectrum in the round ABJM example.
An important open problem is to derive
the common
rule \eqref{eq:XW_def_main} for bulk and M2-brane
fluctuations directly from M-theory and to fix
the normalization of the M2-brane contributions
in \eqref{eq:Z1M}.

Another open problem is to understand the full
moduli space of supersymmetric M2-brane configurations.
Our explicit calculations focus on linear fillings;
a complete treatment must incorporate additional
configurations and their degenerations.
A first-principles derivation of \eqref{eq:Z1M}
should also clarify the global geometric data
underlying the action, reduced determinant and
fermionic zero mode saturation.
The calculation in
Appendix~\ref{app:ABJM_b3_fixed_point_data}
shows how the local fixed point data for
the family of linear fillings
$\operatorname{Gr}(2,4)$ assemble into
characteristic classes whose integral
gives the family contribution.

Our results also motivate revisiting direct
M2-brane worldvolume calculations
\cite{Beccaria:2023ujc,Gautason:2025per,Kurlyand:2026yke}.
In particular, it would be valuable to derive
the Euler-weighted contributions of continuous
supersymmetric families, including the interactions
that saturate their fermionic zero modes and
produce the polynomial prefactors in $\mu$.
Such calculations could also clarify the
differences identified in the squashed sphere
comparison of
section~\ref{subsubsec:ABJM_pure_squashing}.

The relative phases between different membrane
sectors also require a global analysis.
In particular, the field theory comparisons in
Table~\ref{tab:Comparison_Round} select
$\ex^{\ii\vartheta_k}=(-1)^{k/h-1}$, with $h=\gcd(k,2)$
in \eqref{eq:ABJM_generic_orbit_refined_allk}.
Deriving these signs from the global definition
of the M2-brane amplitudes, including their
relation to the fermion Pfaffian and $C$-field
coupling, remains an open problem.

A further extension is a direct calculation of
M2-instanton corrections to the superconformal index (SCI).
Formally applying the SCI gluing rules to the
sphere blocks, including their M2-brane
contributions, gives the results presented
in Appendix~\ref{app:ABJM_SCI}, which we
expect to be correct.
As discussed in \cite{BenettiGenolini:2026kjy},
the relative signs in the SCI gluing reflect the chiralities of the
four-dimensional Killing spinor at the fixed points.
The SCI backgrounds involve both chiralities,
with opposite chiral components vanishing at
the two fixed points.
Understanding how the bulk and membrane fluctuation
complexes accommodate this structure is necessary
to give the formal sign rules a global
gravitational derivation.

We plan to report on some of these topics soon.

\section*{Acknowledgments} 
We thank Florian Gaar for discussions and collaboration at the initial stages of the project.
This work was supported by STFC grants ST/X000575/1 and ST/X000761/1, and SNSF Ambizione grant PZ00P2\_208666. 
JP is supported by a Dean's PhD studentship at Imperial College. 
Some of the ideas and calculations were
developed with the assistance of AI tools (ChatGPT and Claude). 

\appendix

\section{Comparison with the Freund--Rubin fluctuation complex}
\label{app:fluctuation_complex}

In this appendix we compare the equivariant Dolbeault complex on the holomorphic
filling, introduced in section~\ref{subsec:Quantum_Brane_Families},
with the cohomological fluctuation spectrum of an invariant
M2-brane in a two-derivative Freund--Rubin background derived
in \cite{Kurlyand:2026yke}. 
This comparison tests the graded cohomological
spectrum and its zero mode data. Their use in
the family contribution \eqref{eq:Z1M}, however,
is part of the localization proposal developed
in this paper. 
At the end of the appendix we compare the local determinants
and discuss how they enter the complete
instanton contributions.

Recall that $V_D=A\oplus N$, where
$A=\mathcal O_D\otimes T_p^{1,0}M_4$ describes the external
normal directions and $N=N^{1,0}_{D/Z_8}$ the internal ones.
Both have complex rank two. The virtual bundle
$E_D=\Lambda^1V_D-\Lambda^3V_D$ then decomposes as
\begin{align}
\label{eq:appendix_external_internal_complex}
 E_D=
 \underbrace{(A-A\otimes\det N)}_{\mathrm{RW}}
 +\underbrace{(N-\det A\otimes N)}_{\mathrm{ML}}\, .
\end{align}
The two terms have precisely the structure of the
Rozansky--Witten (RW) and McLean (ML) multiplets in
\cite{Kurlyand:2026yke}. The minus signs specify the grading
of the cohomological index, rather than a decomposition of
all fields in the quadratic worldvolume action.

For a concrete comparison consider the linear filling
$D=W_{12}\cong\C^2$ in the round ABJM background, with
$b=1$ and $\Delta_A=\tfrac12$. On the covering space its
link is $\Sigma=S^3$, with Reeb circle of period $2\pi$,
and $B=\Sigma/U(1)_r\cong\CP^1$. In the analysis of sections 4.2 and 4.3 of 
\cite{Kurlyand:2026yke}, the Reeb Fourier
mode $n\in\Z$ gives Dolbeault complexes on $B$ with
coefficients $\mathcal O(n)$ in the RW sector and
$N_{\CP^1/\CP^3}\otimes\mathcal O(n)
\cong\mathcal O(n+1)^{\oplus2}$ in the ML sector.
The corresponding indices and eigenvalues of the generator
$R=\mathcal Q^2$ are
\begin{align}
\label{eq:appendix_FR_indices}
 \chi_n(\mathcal O)&=n+1\, ,
 \qquad\quad\,  R_{\mathrm{RW}}=n+2\, ,\nonumber\\
 \chi_n(N\Sigma)&=2(n+2)\, ,
 \qquad R_{\mathrm{ML}}=n\, .
\end{align}
The RW multiplet has an additional multiplicity two from
its external doublet. In particular, these are signed
indices: negative Fourier modes may contribute through
$H^1$, rather than through holomorphic sections.
In the worldvolume calculation of
\cite{Kurlyand:2026yke}, supersymmetric cancellation
reduces the fluctuation determinant to the
cohomological spectrum encoded by
\eqref{eq:appendix_FR_indices}.
As explained in section~4.1 of that reference,
the quadratic kinetic operators can also have
additional zero modes in supersymmetric doublets.
These are treated by supersymmetric regularization
and do not contribute to the cohomological spectrum. 

We now compute the cohomological spectrum
using our Dolbeault complex on the filling
$D=W_{12}\cong\C^2$.
The holomorphic coordinate functions on $D$ have
weights $(\tfrac12,\tfrac12)$, while the fibre weights
of $A$ and $N$ are $(1,1)$ and
$(-\tfrac12,-\tfrac12)$, respectively.
The decomposition
\eqref{eq:appendix_external_internal_complex}
gives the covering character
\eqref{eq:ABJM_Unreduced_KD} at these weights:
\begin{align}
\label{eq:appendix_round_character}
 \cK^{\rm full}_D(t)=
 \frac{2\ex^{-t}+2\ex^{t/2}-2-2\ex^{-3t/2}}
 {(1-\ex^{-t/2})^2}\, .
\end{align}
The denominator counts holomorphic monomials on $\C^2$:
there are $\ell+1$ monomials of degree $\ell\geq0$,
each contributing weight $\ell/2$. The four terms in
the numerator give the RW and ML towers.

To compare these with the cohomological worldvolume
modes computed in \cite{Kurlyand:2026yke}, we use
holomorphic sections
$H^0(\CP^1,\mathcal O(\ell))$ for the bosonic towers
and Serre duality,
$H^1(\CP^1,\mathcal O(-\ell-2))
\cong H^0(\CP^1,\mathcal O(\ell))^*$,
for the fermionic towers. 
For example, the RW bosonic term $A$ gives
$n=\ell$ and $R=\ell+2$, whereas its negative graded
partner $A\otimes\det N$ corresponds to the RW cokernel
at $n=-\ell-2$. The other two towers follow in the same
way. The complete correspondence is
\begin{align}
\label{eq:appendix_round_towers}
 \begin{array}{c|cc|cc}
  &2\lambda^{\rm B}&2\lambda^{\rm F}
  &n_{\rm B}&n_{\rm F}\\ \hline
  \mathrm{RW}&\ell+2&\ell&\ell&-\ell-2\\
  \mathrm{ML}&\ell-1&\ell+3&\ell-1&-\ell-3
 \end{array}
 \qquad (\ell\geq0)\, .
\end{align}
Every entry has multiplicity $2(\ell+1)$. Here B and F
refer to the positive and negative grading of the filling
character, not to the sign of the corresponding eigenvalue:
$R=2\lambda^{\rm B}$ on the kernel towers and
$R=-2\lambda^{\rm F}$ on the cokernel towers.
The factor of two reflects our normalization of the filling
weights relative to the Reeb generator of \cite{Kurlyand:2026yke}.
Thus the filling complex reproduces the graded cohomological 
worldvolume spectrum and its multiplicities before
taking the $\Z_k$ quotient, after accounting for
Serre duality and the normalization of eigenvalues. 

The finite $\Z_k$ quotient provides a further check.
For the coordinate filling $W_{12}$, the respective quotient
charges of the RW B, RW F and both ML towers are
$-\ell$, $-\ell-2$ and $-\ell-1$. Thus the invariant
modes obey $\ell\equiv0$, $\ell+2\equiv0$ and
$\ell+1\equiv0\pmod{k}$, respectively. Using the Fourier
labels in \eqref{eq:appendix_round_towers}, these become
the conditions $n\equiv0\pmod{k}$ in the RW sector and
$n+2\equiv0\pmod{k}$ in the ML sector, in agreement with equations (4.33) and (4.41) of
\cite{Kurlyand:2026yke}.
The negative graded terms are dual to the corresponding
cokernel representations, which reverses their characters
but does not change the invariant multiplicities.

In particular, the cohomological zero modes are manifest.
The ML B tower at $\ell=1$ has four complex modes, giving
the infinitesimal deformations of the plane, while the RW F
tower at $\ell=0$ has two complex fermionic modes.
Both have quotient charge $-2$, so they survive for
$k=1,2$ and are projected out for $k>2$.
For $k=1,2$ these are precisely the tangent and obstruction
ranks of the Grassmannian family used in
section~\ref{subsubsec:ABJM_Round_k12}.

The preceding calculation establishes agreement of the
graded cohomological spectra, their quotient projections
and their zero modes. For the isolated filling $W_{12}$
at $k>2$, the spectral zeta function satisfies
$\zeta_{\cK^{\rm full}_{12,k}}(0)=0$, so the factor of two relating the
 eigenvalue conventions in
\eqref{eq:appendix_round_towers} leaves the regularized
determinant unchanged. Our local determinant agrees
with the factor computed in equation~(4.43) of
\cite{Kurlyand:2026yke}, reproducing the earlier
calculation of \cite{Beccaria:2023ujc}:
\begin{align}
 Z^{\rm full}_{12,k}
 &=\frac{1}{4\sin^2(2\pi/k)}\, .
\end{align}
This agreement provides a two-derivative check
of the fluctuation complex entering our
all-orders proposal \eqref{eq:Z1M}.

In the two-derivative worldvolume calculation,
the M2-brane action is given by
$I_{\rm M2}=2\pi\sqrt{2N/k}$.
Its comparison with the grand potential uses the
leading large $N$ saddle relation
$\mu_*\simeq\pi\sqrt{kN/2}$, giving
$I_{\rm M2}\simeq4\mu_*/k$.
Our localization formula yields the contribution
directly at fixed chemical potential $\mu$,
before the inverse Laplace transform to fixed $N$. 

A further distinction concerns the treatment of
membrane families and the counting of configurations.
The worldvolume derivation in
\cite{Kurlyand:2026yke} uses supersymmetric
cancellation in the quadratic action.
For configurations with genuine cohomological zero
modes, this calculation identifies the collective
coordinates and residual fermions, but does not
determine the family measure or the interactions
needed to saturate those fermions. These additional
ingredients are needed already at the two-derivative
level.

The local one-loop determinant at $W_{34}$ is the same as
that computed above for $W_{12}$, as follows by
exchanging $(z_1,z_2)$ with $(z_3,z_4)$.
Reference~\cite{Kurlyand:2026yke} suggests interpreting
an overall factor of four through the two
distinguished saddles and their opposite orientations.
By contrast, in our evaluation of \eqref{eq:Z1M}, the isolated
components $W_{12}$ and $W_{34}$ each supply one
quarter of the total, while the $\CP^1\times\CP^1$
component supplies the remaining half through the
Euler-weighted fixed point sum, as described in
section~\ref{subsubsec:ABJM_Round_kgt2}.
The mixed coordinate planes are also considered in
\cite{Kurlyand:2026yke}, where their residual
fermionic zero modes are identified, but the
corresponding family contribution is not evaluated.
In \eqref{eq:Z1M}, the fermionic Euler factor
represents the saturation of these modes, yielding
the non-zero $\CP^1\times\CP^1$ contribution.
Whether the additional orientation multiplicity
suggested in \cite{Kurlyand:2026yke} applies to
the same supersymmetric sector as our calculation
remains to be established.

\section{Refinement limits for M2-brane families}
\label{app:M2_family_localization}

The treatment of the family modes and the relation between
the full and reduced determinants were explained in
section~\ref{subsec:Quantum_Brane_Families}.
Here we give details of the refinement limit in
\eqref{eq:Z1M}. We derive a bound on the polynomial
dependence on the flux and express the finite limit as a
Taylor coefficient. We then illustrate the calculation
for ABJM on the round sphere  at $k=1,2$.

\subsection{Polynomial dependence on the flux}
\label{app:M2_family_collisions}

Consider a compact, smooth, connected family $\cM$ of
complex dimension $d$, with obstruction bundle of constant
complex rank $r$, and let $n=d-r\geq0$ be its virtual
complex dimension. Suppose that the unrefined M2-brane
action has the common value $\mu A_0$ on this family and
that the reduced determinant is independent of $\mu$.

We derive the polynomial dependence on $\mu$ directly
from the refinement limit in \eqref{eq:Z1M}.
Choose a generic direction in the refinement parameters,
\begin{align}
\label{eq:ABJM_family_limit_direction}
 \epsilon=\lambda\mskip1mu\widehat\epsilon\, ,
 \qquad \lambda\rightarrow0\, .
\end{align}
We restrict to contributing fixed points at which the
refinement lifts all the family modes. Fixed points with
an invariant fermionic obstruction mode and a regular reduced
determinant contribute zero, as explained in
section~\ref{subsec:Quantum_Brane_Families}.
Along this ray, the tangent and obstruction weights at
a fixed point $P$ are respectively $\lambda\tau_{P,a}$
and $\lambda\phi_{P,b}$. Consequently,
\begin{align}
 \frac{e_T(\operatorname{Obs}_P)}{e_T(T_P\cM)}
 =\lambda^{-n}C_P\, ,
 \qquad
 C_P=\frac{\prod_{b=1}^{r}\phi_{P,b}}
             {\prod_{a=1}^{d}\tau_{P,a}}\, .
\end{align}
Write the localized action as
$I_{{\rm M2},P}(\lambda)=\mu A_P(\lambda)$, with
$A_P(0)=A_0$. Assume that $A_P(\lambda)$ and
$Z^{\rm red}_{\text{1-loop},P}(\lambda)$ are analytic
near $\lambda=0$ and independent of $\mu$.
The refined fixed point sum then has the form
\begin{align}
 Z_\cM(\mu;\lambda)
 =\ex^{-\mu A_0}\lambda^{-n}
 \sum_P C_P Z^{\rm red}_{\text{1-loop},P}(\lambda)
 \ex^{-\mu[A_P(\lambda)-A_0]}\, .
\end{align}
For the limit $\lambda\to0$ to be finite, all negative
powers of $\lambda$ must cancel in the fixed point sum.
If these cancellations hold, the prefactor multiplying
$\ex^{-\mu A_0}$ in the finite limit is
\begin{align}
\label{eq:M2_family_Taylor_coefficient}
 P(\mu)=[\lambda^n]\sum_P
 C_P Z^{\rm red}_{\text{1-loop},P}(\lambda)
 \ex^{-\mu[A_P(\lambda)-A_0]}\, .
\end{align}
Here $[\lambda^n]$ denotes the coefficient of $\lambda^n$
in the Taylor expansion of the analytic numerator.
Since $A_P(\lambda)-A_0=O(\lambda)$, only powers of $\mu$
up to $\mu^n$ can contribute. Hence
\begin{align}
\label{eq:ABJM_family_coefficient_extraction}
 Z_\cM(\mu)=\ex^{-\mu A_0}P(\mu)\, ,
 \qquad \deg P\leq n\, .
\end{align}
Cancellation of the poles and independence of the limit
from the generic direction $\widehat\epsilon$ provide
consistency checks of the fixed point prescription
\eqref{eq:Z1M}.

\subsection{Example: round sphere ABJM}
\label{app:ABJM_round_collision_example}

For the round ABJM family at $k=1,2$, we have
$\cM=\operatorname{Gr}(2,4)$, $d=4$ and $r=2$, so $n=2$.
We now give the local data and explicit pole cancellations
leading to \eqref{eq:ABJM_round_F_collision}.

Use the auxiliary weights in
\eqref{eq:ABJM_round_auxiliary_weights} and set
$\xi=\lambda p_1$, $\eta=\lambda p_2$, with a fixed
direction satisfying $p_1p_2(p_1^2-p_2^2)\neq0$.
The physical round background is held fixed.
The planes $W_{13}$ and $W_{24}$ have vanishing obstruction
Euler factors and do not contribute, as explained in
section~\ref{subsubsec:ABJM_Round_k12}.
The remaining planes $W_{12},W_{14},W_{23},W_{34}$
correspond respectively to
$(\sigma,\tau)=(+,+),(+,-),(-,+),(-,-)$.
At such a plane $P$, the four tangent weights are
$\lambda\sigma p_1$, $\lambda\tau p_2$, and two copies
of $\lambda(\sigma p_1+\tau p_2)/2$.
Each of the two residual fermions has weight
$\lambda(\sigma p_1+\tau p_2)/2$.
All these modes survive the $\Z_k$ projection for $k=1,2$,
so their Euler ratio is $\lambda^{-2}C_P$.
Using the action and fluctuation lifts of
section~\ref{subsubsec:ABJM_Round_k12}, we obtain the
following local data in the notation of
Appendix~\ref{app:M2_family_collisions}:
\begin{align}
 C_P&=\frac{1}{\sigma\tau p_1p_2}\, ,
 \qquad
 A_P(\lambda)=
 \frac{4}{k(1+\sigma\lambda p_1)(1+\tau\lambda p_2)}\, ,
 \qquad A_0=\frac4k\, ,
 \nn\\
 Z^{\rm red}_{\text{1-loop},P}(\lambda)
 &=\frac{\lambda^2\sigma\tau p_1p_2}
 {4\sin\!\left(\frac{2\pi}{k(1+\sigma\lambda p_1)}\right)
    \sin\!\left(\frac{2\pi}{k(1+\tau\lambda p_2)}\right)}\, .
\end{align}
The reduced determinant is obtained by dividing the
full determinant in \eqref{eq:ABJM_round_Nosaka_determinant}
by the family Euler ratio $\lambda^{-2}C_P$.
This expression is regular at $\lambda=0$ and has
the expansion
\begin{align}
 Z^{\rm red}_{\text{1-loop},P}(\lambda)
 &=\frac{k^2}{16\pi^2}\biggl[
 1+(\sigma p_1+\tau p_2)\lambda
 +\Big(\sigma\tau p_1p_2
 +\frac{2\pi^2}{3k^2}(p_1^2+p_2^2)\Big)\lambda^2
 +\cO(\lambda^3)\biggr]\, .
\end{align}
The coefficients of the possible $\lambda^{-2}$ and
$\lambda^{-1}$ poles vanish, respectively:
\begin{align}
 \sum_{\sigma,\tau}
 C_P Z^{\rm red}_{\text{1-loop},P}(0)&=0\, ,
 \nn\\
 \sum_{\sigma,\tau}C_P\biggl[
 \left.\partial_\lambda
 Z^{\rm red}_{\text{1-loop},P}(\lambda)\right|_{\lambda=0}
 -\mu A'_P(0)Z^{\rm red}_{\text{1-loop},P}(0)
 \biggr]&=0\, .
\end{align}
Extracting the order $\lambda^2$ coefficient using
\eqref{eq:M2_family_Taylor_coefficient} gives
\begin{align}
 Z_\cM(\mu)=
 \frac{4\mu^2+k\mu+k^2/4}{\pi^2}\ex^{-4\mu/k}\, ,
 \qquad k=1,2\, .
\end{align}
This reproduces \eqref{eq:ABJM_round_F_collision}, with
all dependence on $(p_1,p_2)$ cancelling.
The virtual dimension $n=2$ explains the quadratic degree
bound, while the action and reduced determinant determine
the quadratic, linear and constant coefficients.

\section{ABJM instantons and the type IIA interpretation}
\label{app:ABJM_IIA}

We explain here how the linear M2-brane configurations
described in Section~\ref{subsec:Geometry_Semiclassical_Action}
are related to worldsheet and D2-brane instantons in type IIA.
The corresponding field theory conventions and results are
collected in Appendix~\ref{app:field_theory}.

\subsection{Circle actions and complex structures}

The relevant internal space is $S^7/\Z_k$. We first distinguish the Reeb circle $U(1)_r$, the M-theory
circle $U(1)_M$ used for reduction to type IIA, and the
internal action of $K$. Preservation by $K$ determines
which holomorphic fillings enter our localization
prescription, whereas their type IIA interpretation
depends on how $U(1)_M$ acts on the M2-brane.

On the covering sphere $S^7\subset\C^4$, the circle
$U(1)_M$ containing the $\Z_k$ quotient group
\eqref{eq:CG_ABJM_quotient} acts as
\begin{align}
 (z_1,z_2,z_3,z_4)
 \longmapsto
 (\ex^{\ii\psi}z_1,\ex^{\ii\psi}z_2,
  \ex^{-\ii\psi}z_3,\ex^{-\ii\psi}z_4)\, .
\end{align}
The effective reduction circle on $S^7/\Z_k$ is
$U(1)_M/\Z_k$.
By comparison, the Reeb circle $U(1)_r$ of the standard
Sasakian structure on the covering sphere acts with
weights $(1,1,1,1)$.

It is useful to distinguish the two quotients of the
covering sphere,
\begin{align}
 \CP^3_r\equiv S^7/U(1)_r\, ,
 \qquad
 \CP^3_{\rm IIA}\equiv S^7/U(1)_M\, .
\end{align}
Both are diffeomorphic to $\CP^3$, but their natural complex
structures are described using different coordinate choices.
The localization coordinates $z_I$ are adapted to $U(1)_r$,
while coordinates adapted to the type IIA reduction are
\begin{align}
 Z_1=z_1\, ,\qquad Z_2=z_2\, ,\qquad
 Z_3=\overline z_3\, ,\qquad Z_4=\overline z_4\, .
\end{align}
The circle $U(1)_M$ acts on the $Z_I$ by a common phase,
giving the usual projective complex structure on
$\CP^3_{\rm IIA}$.

\subsection{Worldsheet and D2-brane interpretations}

An M2-brane whose worldvolume contains the orbits of the
reduction circle $U(1)_M/\Z_k$ descends to a fundamental
string worldsheet in type IIA.
An M2-brane transverse to this circle has a D2-brane
interpretation. Holomorphy of the filling $D$ with respect to the localization
complex structure does not by itself distinguish these
possibilities.

The planes preserved by the entire $U(1)_M$ are $W_{12}$,
$W_{34}$, and the direct sums of one complex line in each.
For $k>2$, these form precisely the locus $\cM_k$ in
\eqref{eq:ABJM_full_stabilizer_locus}.
The circle is tangent to their links, and their images in
$\CP^3_{\rm IIA}$ are complex projective lines. An M2-brane wrapping such a quotient link once therefore
gives, in type IIA, a worldsheet instanton wrapping
$\CP^1$ once \cite{Beccaria:2023ujc}.

To describe a transverse M2-brane, consider a plane given
by \eqref{eq:Holomorphic_Plane_z12} with invertible $A$.
For $k>2$, such a plane belongs to the generic class
$D\notin\cM_k$.
In the coordinates adapted to the reduction, its equation
becomes
\begin{align}
 \begin{pmatrix}Z_3\\Z_4\end{pmatrix}
 =
 \overline A
 \begin{pmatrix}\overline Z_1\\\overline Z_2\end{pmatrix}\, .
\end{align}
The invertibility of $A$ ensures that the reduction circle
is nowhere tangent to its link.
Two points on this link project to the same point of
$\CP^3_{\rm IIA}$ precisely when they differ by the
antipodal map. Its image is therefore topologically
\begin{align}
 S^3/\{\pm1\}\cong\mathbb{RP}^3\, .
\end{align}
This identifies the topology of the projected D2-instanton cycle \cite{Drukker:2011zy};
its embedding in $\CP^3_{\rm IIA}$ depends on $A$.

\subsection{Lifts and wrapping numbers}
\label{app:ABJM_IIA_primitivity}

For the transverse M2-branes described above, the parity
of $k$ determines the degree of their projection onto
$\mathbb{RP}^3$. For the invertible graph above, the setwise stabilizer
under $U(1)_M$ on the covering space is $\{\pm1\}$.
Its intersection with $\Z_k$ has order $h=\gcd(k,2)$.
For even $k$, the antipodal element already belongs to
$\Z_k$. The M2-brane worldvolume is $\mathbb{RP}^3$,
and its projection onto the type IIA cycle is one-to-one.
For odd $k$, the M2-brane worldvolume is $S^3$, and its
projection onto the same cycle is two-to-one.

At the round superconformal point, the singly wrapped
$U(1)_M$-invariant M2-branes have instanton action
$I_{\rm M2}=4\mu/k$ and give unit worldsheet instantons.
The transverse linear M2-branes have instanton action
$I_{\rm M2}=4\mu/h$. For even $k$, they give singly wrapped
D2 instantons with $I_{\rm M2}=2\mu$, while for odd $k$
they give doubly wrapped D2 instantons with
$I_{\rm M2}=4\mu$.
 One might also expect
singly wrapped D2 instantons on the same $\mathbb{RP}^3$
in type IIA, but these are globally obstructed by the
non-trivial torsion RR two-form flux class on the cycle
\cite{deBoer:2001wca}.\footnote{The restricted M-theory circle bundle has first
Chern class $kt$, where $t$ generates
$H^2(\mathbb{RP}^3;\mathbb Z)\cong\mathbb Z_2$.
This class is non-trivial precisely for odd $k$.
Being torsion, it is invisible in de Rham cohomology.}

These examples explain why wrapping once in eleven
dimensions need not correspond to one unit of instanton
number after reduction. The degree of the projective curve
defining a holomorphic filling, the M2-brane wrapping number
in eleven dimensions, and the covering degree after type IIA
reduction are distinct notions.

The same distinction is relevant to the degeneration
described at the end of
Section~\ref{subsec:Geometry_Semiclassical_Action}.
For $k>2$, an M2-brane associated with a generic
orbit of linear fillings approaches a coordinate cycle as a cover of degree
$k/h$. Its limiting worldsheet interpretation therefore
has degree $k/h$, and must be distinguished from the
unit worldsheet instanton on the same support.

At the round superconformal point, for $k=1,2$, all linear planes belong to the same Grassmannian family.
Its $U(1)_M$-invariant members give
worldsheet instantons after reduction, while the transverse
members described above give D2 instantons.

\section{Field theory results and conventions}
\label{app:field_theory}

We collect field theory results and conventions
for the ABJM partition function on the three-sphere,
including mass and squashing deformations.
The geometric interpretation of the instanton
sectors, including their relation to M2-brane
wrapping numbers, is discussed in
Appendix~\ref{app:ABJM_IIA}.
As in the main text, supersymmetry-preserving mass deformations are encoded in four complexified parameters $\Delta_A$ constrained by $\sum_A \Delta_A = 2$, and the $U(1)\times U(1)$-preserving squashing parameter is $b>0$.

\subsection{ABJM on the round sphere}
\label{app:ABJM_round_field_theory}

Let $Z_{\rm ABJM}(N,k)$ be the partition function of ABJM theory on the unit round three-sphere,
in the normalization of
\cite{Kapustin:2009kz,Marino:2011eh}.
Its grand partition function $ \Xi(k,\mu)$ and modified grand
potential $ J(k,\mu)$ are related by (see section 3.2 of \cite{Hatsuda:2012dt}):
\begin{align}
\label{eq:ABJM_Xi}
 \Xi(k,\mu)
 &=\sum_{N=0}^{\infty}\ex^{\mu N}Z_{\rm ABJM}(N,k)
  =\sum_{n\in\Z}\exp\!\left[J(k,\mu+2\pi\ii n)\right]\, ,
 \nonumber\\
 J(k,\mu)&=J^{\rm pert}(k,\mu)+J^{\rm np}(k,\mu)\, .
\end{align}

The perturbative part of $J$ in the large $\mu$ expansion has the form
\begin{align}
\label{eq:ABJM_Jpert}
 J^{\rm pert}(k,\mu)=\frac{C(k)}{3}\mu^3+B(k)\mu+A(k)\, ,
\end{align}
where \cite{Marino:2011eh,Hanada:2012si,Hatsuda:2014vsa}
\begin{align}
\label{eq:ABJM_perturbative_coefficients}
 C(k)&=\frac{2}{k\pi^2}\, ,
 \qquad B(k)=\frac{k}{24}+\frac{1}{3k}\, ,\nonumber\\
 A(k)&=\frac{2\zeta(3)}{\pi^2k}
       \left(1-\frac{k^3}{16}\right)
       +\frac{k^2}{\pi^2}\int_0^{\infty}\dd x\,
        \frac{x\log(1-\ex^{-2x})}{\ex^{kx}-1}\, .
\end{align}
The integral representation is equation~(3.13) of
\cite{Hatsuda:2014vsa}.
The inverse Laplace transform gives the perturbative
canonical partition function
\begin{align}
\label{eq:ABJM_perturbative_Airy}
 Z^{\rm pert}_{\rm ABJM}(N,k)
 =\ex^{A(k)}C(k)^{-1/3}
   \operatorname{Ai}\!\left[C(k)^{-1/3}(N-B(k))\right]\, .
\end{align}

The non-perturbative expansion can be written as \cite{Hatsuda:2013gj}
\begin{align}
	J^{\rm np}(k,\mu) = \sum_{\substack{\ell, m = 0 \\ (\ell, m) \neq (0,0) }}^\infty f_{\ell,m}(k,\mu) \exp \left[ - \left( 2\ell + \frac{4m}{k} \right) \mu \right] \, .
\end{align}
The two sums take into account the contributions of the different types of IIA instantons described in Appendix \ref{app:ABJM_IIA}.
In particular, the terms with $\ell=0$ and $m=0$ are
\begin{align}
	J^{\rm WS}(k,\mu) &=\sum_{m=1}^\infty d_m(k)\ex^{-4m\mu/k} \, , \nn \\ 
	J^{\rm D2}(k,\mu) &= \sum_{\ell = 1}^\infty \left[a_\ell(k)\mu^2+b_\ell(k)\mu+c_\ell(k)\right] \ex^{-2\ell\mu} \, ,
\end{align}
respectively, which correspond, respectively, to the contributions of worldsheets and D2-brane instantons in type IIA (the latter sector is conventionally denoted $J^{\rm M2}$, or membrane instanton sector, in the Fermi gas literature). 
The mixed terms with both $\ell, m >0$ represent bound states and can be resummed by introducing a shifted effective chemical potential in the worldsheet series \cite{Hatsuda:2013gj}.

Both sectors have an eleven-dimensional M2-brane description,
but their instanton numbers need not coincide with
M2-brane wrapping numbers, as explained in Appendix~\ref{app:ABJM_IIA_primitivity}.
In particular, these sectors should not be identified
term by term at every integer $k$ with the
linear-filling M2-brane contributions $J_{\rm full}$
and $J_{\rm gen}$ from the full-stabilizer and
generic orbit sectors, defined in
\eqref{eq:ABJM_generic_fixed_point_sum} and
\eqref{eq:ABJM_generic_orbit_refined_allk}, respectively.
For reference, the first coefficients include
\begin{align}
\label{eq:ABJM_first_instanton_coefficients}
 d_1(k)=\frac1{\sin^2(2\pi/k)}\, ,
 \qquad
 a_1(k)=-\frac4{\pi^2k}\cos\!\left(\frac{\pi k}{2}\right)\, .
\end{align}
The remaining membrane instanton coefficients $b_1,c_1$
and several higher orders are listed in Appendix B of
\cite{Hatsuda:2013gj}.
The complete coefficient system is organized by
ordinary and refined topological strings on local
$\CP^1\times\CP^1$
\cite{Hatsuda:2013oxa,Hatsuda:2015gca}.
At fixed instanton number, the topological string
expressions retain the full dependence on $k$.
The Fermi gas WKB expansion gives the D2-instanton
coefficients as small $k$ series, while the
worldsheet contributions $\ex^{-4m\mu/k}$ are
non-perturbative in $k$ and are therefore invisible
at every finite WKB order
\cite{Hatsuda:2013gj,Nosaka:2015iiw}.

Individual coefficients can have poles at integer
$k$. Terms with coincident instanton actions must
then be combined before taking the limit; this is
the pole cancellation mechanism of
\cite{Hatsuda:2012dt,Hatsuda:2013gj}. 
The first orders in $J^{\rm np}(k,\mu)$ for $k =3,4,5,6$ are collected in table~\ref{tab:Comparison_Round}.

\subsection{The ABJM two-mass locus}
\label{app:ABJM_mass_field_theory}

On the round sphere Nosaka locus, the Fermi gas
description determines the perturbative Airy
coefficients and gives WKB information about the
non-perturbative terms \cite{Nosaka:2015iiw}.
In the mass normalization of \cite{Nosaka:2024gle},
write $\alpha_\sigma=1+\sigma\ii\mathfrak m_1/\pi$
and $\beta_\tau=1+\tau\ii\mathfrak m_2/\pi$,
with $\sigma,\tau=\pm1$.
Thus $\xi=\ii\mathfrak m_1/\pi$ and
$\eta=\ii\mathfrak m_2/\pi$ in
\eqref{eq:ABJM_round_Nosaka_weights}.
For equal ranks, the four first worldsheet terms
are conjectured to be (see equation~(5.12) of
\cite{Nosaka:2024gle}):
\begin{align}
\label{eq:ABJM_mass_WS1}
 J_{\rm WS}^{(1)}(k,\mu;\mathfrak m_1,\mathfrak m_2)
 \overset{\mathrm{conj.}}{=}
 \sum_{\sigma,\tau=\pm1}
 \frac{\exp\!\left[-\frac{4\mu}
                       {k\alpha_\sigma\beta_\tau}\right]}
      {4\sin\!\left(\frac{2\pi}{k\alpha_\sigma}\right)
        \sin\!\left(\frac{2\pi}{k\beta_\tau}\right)}\, .
\end{align}
This expression is supported by high precision
finite $N$ field theory partition functions.
At equal masses, the opposite sign terms disagree
with the leading 't~Hooft calculation in
\cite{Gautason:2023igo}, as discussed in
\cite{Nosaka:2024gle}.
Away from poles, its zero-mass limit gives
$d_1(k)\ex^{-4\mu/k}$.

The M2-instanton sector $J^{\rm M2}$ also splits
under the mass deformation. The Fermi gas WKB analysis
gives the scales $\exp[-2\ell\mu/\alpha_\sigma]$ and
$\exp[-2\ell\mu/\beta_\tau]$, which reduce to
$\exp[-2\ell\mu]$ at zero mass
\cite{Nosaka:2015iiw}.
Their exact finite $k$ prefactors are not determined
by that analysis.

These asymptotic expansions apply in the mass
chamber connected to the undeformed theory;
their breakdown at large real masses is studied
in \cite{Nosaka:2024gle}.

\subsection{ABJM with squashing}
\label{app:ABJM_squashed_field_theory}

For pure squashing, a useful exact result is
the massless theory at $k=1$ and $b^2=3$, or
equivalently $b^2=\tfrac13$.
The mirror theory Fermi gas is related to the
quantized mirror curve of local $\mathbb{CP}^2$
\cite{Hatsuda:2016uqa}.
Its leading non-perturbative term is (see equation~(3.22) of
\cite{Hatsuda:2016uqa}):
\begin{align}
\label{eq:ABJM_squashed_Jnp_b3}
 J^{\rm np}_{b^2=3}(1,\mu)
 =\left[-\frac{45\mu^2}{8\pi^2}
        -\frac{9\mu}{4\pi^2}
        +\frac38-\frac{3}{4\pi^2}\right]\ex^{-3\mu}
   +\cO(\mu^2\ex^{-6\mu})\, .
\end{align}
Higher instanton coefficients at this point are
also given in \cite{Hatsuda:2016uqa}.

An exact continuous mass-squashing relation is
available for arbitrary $N,k$ (see 
equation~(3.2) of \cite{Chester:2021gdw}).
In the three-mass conventions of that reference,
define $m_\pm=m_2\pm m_1$; these masses differ
in normalization from $\mathfrak m_i$ above.
Then
\begin{align}
\label{eq:ABJM_mass_squashing_identity}
 Z_b\!\Big(N,k;m_1,m_2,\frac{\ii}{2}(b-b^{-1})\Big)
 =Z_{\rm round}\!\left(N,k;b^{-1}m_+,b\,m_-\right)\, ,
\end{align}
where the round sphere function on the right is
parametrized by its two mass combinations.
This identity includes non-perturbative terms.
Setting $m_1=m_2=0$ gives the undeformed round sphere
partition function, but requires a non-zero imaginary
third mass when $b\neq1$. It therefore does not
describe pure squashing, as discussed in the main text.
These give complementary checks: \eqref{eq:ABJM_squashed_Jnp_b3} fixes the complete leading term at a special value of pure squashing, while \eqref{eq:ABJM_mass_squashing_identity} constrains the combined mass and squashing dependence. Our comparison with the latter identity concerns the single instanton contributions computed in section~\ref{subsubsec:ABJM_squashed_Nosaka}.

For another tuned mass-squashing locus at $k=1$,
\cite{Kubo:2024qhq} provides a Fermi gas construction,
perturbative Airy coefficients and exact small $N$
partition functions.

\section{Pure-squashing results for ABJM}
\label{app:Squashed}

We give the complete polynomial prefactors for
the pure-squashing family sums of
section~\ref{subsubsec:ABJM_pure_squashing},
together with details of their evaluation.
For $k=1$ at $b^2=3$, we also express the family
contribution in terms of characteristic classes
on $\operatorname{Gr}(2,4)$.

We use the projected characters and spectral
conventions of
section~\ref{subsec:ABJM_Fluctuations}.
To evaluate the projected double zeta function
\eqref{eq:ABJM_projected_double_zeta} for
stabilizer order $s$, write the monomial degrees as
\begin{align}
 m=r_I+sM\, ,\qquad n=r_J+sN\, ,
 \qquad 0\leq r_I,r_J<s\, ,\qquad M,N\geq0\, .
\end{align}
The decomposition into residue classes gives
\begin{align}
\label{eq:ABJM_sq_residue_Barnes}
 \zeta^{IJ}_{2,s}(z,x;c)
 =
 \sum_{\substack{0\leq r_I,r_J<s\\
 c-r_I\gamma_I-r_J\gamma_J\equiv0\ ({\rm mod}\ s)}}
 \zeta_2\!\left(
 z,x+r_I\Delta_I+r_J\Delta_J
 \mid s\Delta_I,s\Delta_J
 \right)\, .
\end{align}
Here $\zeta_2$ is the ordinary Barnes double
zeta function with the displayed periods.
The residue sum implements the stabilizer
projection, so there is no additional factor
$1/s$.
Substitution into
\eqref{eq:ABJM_generic_Barnes_determinant}
gives the full determinants at generic weights.

For the pure-squashing family calculation,
we use the Nosaka parameters $\xi,\eta$ in
\eqref{eq:ABJM_round_Nosaka_weights} as auxiliary
weights and sum the fixed point contributions
before taking $\xi,\eta\to0$.
Using the squashing parameter $f$ defined in
\eqref{eq:ABJM_squashing_f}, we write
\begin{align}
 \lim_{\xi,\eta\to0}F_s^{(b)}(\mu;\xi,\eta)
 =\ex^{-\alpha_s\mu}P_s(\mu;f)\, ,
 \qquad
 \alpha_s=\frac{4(1-f^2)}{s}\, .
\end{align}
Here $\mu\alpha_s$ is the M2-brane action at
zero refinement, and $P_s(\mu;f)$ is the
complete polynomial prefactor.
The corresponding sector contributions are
specified by
\eqref{eq:ABJM_pure_squashing_family_sums}.

\subsection{Stabilizer orders one and two}
\label{app:ABJM_sq_low_stabilizers}

For $s=1,2$, all four internal tangent modes
survive the stabilizer projection.
At generic squashing, the unrefined spectrum
has no residual fermionic zero modes.
Write $\Delta_A=\tfrac12+\epsilon_A$, with the
Nosaka refinement
\eqref{eq:ABJM_round_auxiliary_weights}.
Separating the tangent Euler factor gives
\begin{align}
 Z^{\rm full}_{IJ,s}
 =
 \frac{Z^{\rm red}_{IJ,s}}
 {
  \prod_{i\in\{I,J\}}
  \prod_{A\notin\{I,J\}}(\epsilon_i-\epsilon_A)}\, .
\end{align}
The reduced determinant is regular near
$\epsilon=0$.

At zero refinement, the spectrum can be
organized by the total monomial degree
$\ell=m+n$.
For $s=1$, all degrees survive, while for $s=2$
the projection retains even degrees in the
external sector and odd degrees in the internal
sector.
Omitting the four internal degree one tangent
modes, the resulting Hurwitz zeta sums give
the same value
\begin{align}
\label{eq:ABJM_sq_quartic_normalization}
 D_s(f)
 \equiv
 \left.Z^{\rm red}_{IJ,s}\right|_{\epsilon=0}
 =
 -\frac{s^4}{64\pi^4}
 \sin^2\!\left(\frac{2\pi f}{s}\right)\, ,
 \qquad s=1,2\, .
\end{align}

The tangent Euler denominator is homogeneous
of degree four in the auxiliary parameters.
The finite family sum is therefore extracted
from the fourth order expansion of
$Z^{\rm red}_{IJ,s}\ex^{-I_{\rm M2}[W_{IJ}]}$,
as in \eqref{eq:M2_family_Taylor_coefficient}.
In particular, the coefficient of $\mu^4$
depends only on $D_s(f)$ and the linear
variation of the action,
\begin{align}
 I_{\rm M2}[W_{IJ}]
 =
 \alpha_s\mu
 \bigl[1-2(\epsilon_I+\epsilon_J)
       +\cO(\epsilon^2)\bigr]\, .
\end{align}
Using the fixed point identity for $\operatorname{Gr}(2,4)$
\begin{align}
 \sum_{I<J}
 \frac{(\epsilon_I+\epsilon_J)^4}
 {
  \prod_{i\in\{I,J\}}
  \prod_{A\notin\{I,J\}}(\epsilon_i-\epsilon_A)}
 =2\, ,
\end{align}
we obtain
\begin{align}
 [\mu^4]P_s(\mu;f)
 &=
 \frac{(2\alpha_s)^4}{4!}\,2D_s(f)
 =
 \frac43\alpha_s^4D_s(f)\, ,
\end{align}
where $[\mu^4]$ extracts the coefficient of
$\mu^4$.
This reproduces
\eqref{eq:ABJM_pure_squashing_Ps4}.

The lower powers of $\mu$ also involve
higher-order terms in the action and reduced
determinant.
Evaluating the complete fourth order coefficient
in the auxiliary parameters gives the following
common expression for $s=1,2$:
\begin{align}
\label{eq:ABJM_sq_k1_closed_form}
 &\lim_{\xi,\eta\to0}F_s^{(b)}(\mu;\xi,\eta)
 =
 -\frac{\ex^{-\alpha_s\mu}}{4\pi^4}
 \Biggl\{
 -\frac{64\pi^4}{s^4}
 D_s(f)\cP_4^{(s)}(\mu,f)
 +\pi f\sin\!\left(\frac{4\pi f}{s}\right)
       \cP_3^{(s)}(\mu,f)\nn\\
 &\qquad\qquad+\pi^2\cP_2^{(s)}(\mu,f)
 +\frac{20\pi^3}{3}f(1-4f^2)
 [4(1-f^2)\mu+s]
 \cot\!\left(\frac{2\pi f}{s}\right)
 \Biggr\}\, .
\end{align}
The degree $a$ polynomials appearing here are
\begin{align}
\label{laterintext}
 &\cP_4^{(s)}(\mu,f)
 =
 \frac{64}{3}(1-f^2)^4\mu^4
 +\frac{160s}{3}(1-f^2)^3\mu^3
 \nn\\
 &\quad
 +\frac{8}{3}(1-f^2)^2
 [21s^2+\pi^2(2-108f^2)]\mu^2
 +\frac{4s}{9}(1-f^2)
 [66s^2+\pi^2(25-702f^2)]\mu
 \nn\\
 &\quad
 +\frac{22s^4}{3}
 +s^2\pi^2\left(\frac{43}{9}-92f^2\right)
 +12\pi^4f^2(9f^2-1)\, ,
 \nn\\[4pt]
 &\cP_3^{(s)}(\mu,f)
 =
 -64(1-f^2)^3\mu^3
 -112s(1-f^2)^2\mu^2
 \nn\\
 &\quad
 +8(1-f^2)
 [\pi^2(18f^2-1)-9s^2]\mu
 +s(72\pi^2f^2-8\pi^2-18s^2)\, ,
 \nn\\[4pt]
 &\cP_2^{(s)}(\mu,f)
 =
 16(1-f^2)^2(14f^2-1)\mu^2
 +\frac{4s}{3}(1-f^2)(168f^2-13)\mu
 \nn\\
 &\quad
 +s^2\left(56f^2-\frac{13}{3}\right)
 +8\pi^2f^2(2-11f^2)\, .
\end{align}

For $k=1,2$, setting $s=k$ in
\eqref{eq:ABJM_sq_k1_closed_form} gives
$J_{\rm lin}(k,\mu;b)$.
In particular, $s=k=1$ gives
\eqref{eq:Intro_Jlin_1}, where the stabilizer
label is suppressed by writing
$\cP_a\equiv\cP_a^{(1)}$.
The same local expressions enter the proposed
generic orbit contribution through
\eqref{eq:ABJM_pure_squashing_family_sums}.

Special squashing values are approached after
assembling the complete prefactor $P_s(\mu;f)$.
The limits $f\to0$ for $s=1,2$ reproduce the
quadratic round sphere results
\eqref{eq:ABJM_round_F_collision}, while
$f\to\pm\tfrac12$ for $s=1$ gives
\eqref{eq:ABJM_squashed_b3_result}.
For $s=2$, the points $f=\pm\tfrac12$ remain
quartic.
The geometric explanation of these degree reductions
was given in
section~\ref{subsubsec:ABJM_pure_squashing}.
The next subsection illustrates this mechanism
explicitly at $k=1$ and $b^2=3$.

\subsection{The special value \texorpdfstring{$k = 1$, $b^2 = 3$}{k=1, b squared=3}}
\label{app:ABJM_b3_fixed_point_data}

The result \eqref{eq:ABJM_squashed_b3_result} follows directly by taking $f \to \tfrac12$ in the compact expression \eqref{eq:ABJM_sq_k1_closed_form}. Here we give the fixed point calculation and its interpretation as a characteristic class integral over the membrane family. We use the candidate obstruction bundle \eqref{eq:ABJM_b3_fermion_bundle} and the family prescription \eqref{eq:Z1M}.

At $k = 1$ and $b^2 = 3$, the external weights are $(b_1,b_2) = (\tfrac32,\tfrac12)$ and the bosonic family is $\cM = \operatorname{Gr}(2,4)$. Introduce generic auxiliary parameters $\epsilon_A $ with
\begin{align}
 \Delta_A = \frac12+\epsilon_A \, , \qquad \sum_A\epsilon_A = 0 \, .
\end{align}
At each coordinate plane $W_{IJ}$, the four tangent weights are $\epsilon_i-\epsilon_A$, with $i \in \{I,J\}$ and $A \notin \{I,J\}$, and the two residual fermionic weights are $2\epsilon_I+\epsilon_J$ and $\epsilon_I+2\epsilon_J$. The full determinant therefore includes their Euler ratio as in \eqref{ObsIJ}. The sum to be evaluated is
\begin{align}
 J_{\rm lin}(1,\mu;\sqrt3)
 = \lim_{\epsilon\to0}\sum_{I<J}
 \frac{(2\epsilon_I+\epsilon_J)(\epsilon_I+2\epsilon_J)}
 {
\prod_{i\in\{I,J\}}\prod_{A\notin\{I,J\}}(\epsilon_i-\epsilon_A)}
 Z^{\rm red}_{IJ,1}(\epsilon)\,
 \ex^{-I_{\rm M2}[W_{IJ}]} \, .
\end{align}
The Euler ratio is homogeneous of degree $-2$ in the auxiliary parameters. Consequently, the finite limit requires the action weight and reduced determinant through quadratic order; the lower order poles cancel after summing all six terms.

At $k=1$ and $b^2=3$, the reduced determinant
$Z^{\rm red}_{IJ,1}$ omits the four bosonic
tangent modes and the two residual fermionic
modes. We denote its common value at zero
auxiliary parameters by $D_{b^2=3}$.
The function $D_1(f)$ in
\eqref{eq:ABJM_sq_quartic_normalization} omits
only the tangent modes.
At zero refinement, the two additional fermionic
weights are both $\tfrac12-f$, so
\begin{align}
\label{eq:ABJM_sq_wall_normalization}
D_{b^2=3}
 &\equiv Z^{\rm red}_{IJ,1}(0)
 =\lim_{f\to1/2}
   \frac{D_1(f)}{(\tfrac12-f)^2}
 =-\frac{1}{16\pi^2}\, .
\end{align}
The sign follows the spectral branch used for
the full determinant. 

Since $Q^2 = 16/3$, the localized action and its exponential are
\begin{align}
\label{eq:ABJM_sq_b3_action_expansion}
\begin{aligned}
 I_{\rm M2}[W_{IJ}]
 &= \frac{3\mu}{(1+2\epsilon_I)(1+2\epsilon_J)} \, ,\\
 \ex^{-I_{\rm M2}[W_{IJ}]}
 &= \ex^{-3\mu}\Bigl[1+6\mu(\epsilon_I+\epsilon_J)
 +(18\mu^2-12\mu)(\epsilon_I+\epsilon_J)^2
 +12\mu\epsilon_I\epsilon_J+\cO(\epsilon^3)\Bigr] \, .
\end{aligned}
\end{align}
To obtain the reduced determinant, remove the same six modes from the full character:
\begin{align}
\label{eq:ABJM_sq_b3_reduced_character}
\begin{aligned}
 \cK^{\rm red}_{IJ,1}(t;\epsilon)
 = {}&\cK^{\rm full}_{IJ,1}(t;\epsilon)
 -\sum_{A\notin\{I,J\}}\sum_{i\in\{I,J\}}
       \ex^{t(\epsilon_A-\epsilon_i)}
 +\ex^{-t(2\epsilon_I+\epsilon_J)}
  +\ex^{-t(\epsilon_I+2\epsilon_J)} \, .
\end{aligned}
\end{align}
The bosonic terms are subtracted, whereas the fermionic terms, which enter the full index with a minus sign, 
are added back. 
We expand the spectral zeta function up to 
quadratic order in the auxiliary parameters,
grouping monomials by their total degree
$\ell=m+n$.
At zero refinement, the remaining eigenvalues
lie in the towers $(\ell-1)/2$, $(\ell+1)/2$
and $(\ell+3)/2$, with the zero weights omitted.
Evaluating the resulting one-dimensional zeta
sums, differentiating at zero and exponentiating,
as in \eqref{eq:1loop_def}, gives 
\begin{align}
\label{eq:ABJM_sq_b3_reduced_expansion}
\begin{aligned}
 Z^{\rm red}_{IJ,1}(\epsilon)
 =D_{b^2=3}\Biggl[{}&1+2(\epsilon_I+\epsilon_J)
 -2\pi^2(\epsilon_I+\epsilon_J)^2\\
 &+\left(4+\frac{4\pi^2}{3}\right)\epsilon_I\epsilon_J
 +\frac{4\pi^2}{3}\sum_A\epsilon_A^2
 +\cO(\epsilon^3)\Biggr] \, .
\end{aligned}
\end{align}
Multiplying the two expansions, the product of their linear terms cancels $-12\mu(\epsilon_I+\epsilon_J)^2$ in the action weight. The terms quadratic in the auxiliary parameters are thus
\begin{align}
  \label{eq:App_ABJM_Dwall}
D_{b^2=3}\ex^{-3\mu}\left[
 (18\mu^2-2\pi^2)(\epsilon_I+\epsilon_J)^2
 +\left(12\mu+4+\frac{4\pi^2}{3}\right)\epsilon_I\epsilon_J
 +\frac{4\pi^2}{3}\sum_A\epsilon_A^2
 \right] \, .
\end{align}

We now rewrite the fixed point sum as an
integral over the membrane family
$\cM=\operatorname{Gr}(2,4)$.
At the order needed for the finite refinement
limit, this amounts to identifying the weight
polynomials in \eqref{eq:App_ABJM_Dwall}
with the fixed point restrictions of equivariant
characteristic classes on $\cM$. 
For the candidate obstruction bundle in
\eqref{eq:ABJM_b3_fermion_bundle}, the ordinary
Euler class is 
\begin{align}
\label{eq:ABJM_b3_global_classes}
 e(\operatorname{Obs}) = 2c_1(\cS^*)^2+c_2(\cS^*) \, .
\end{align}
The rank two Euler class supplies two units of complex degree, leaving two for the action weight and reduced determinant on the four-dimensional complex Grassmannian. As discussed in the main text, this is the geometric reason that the prefactor is quadratic in $\mu$.

The term proportional to $\sum_A\epsilon_A^2$ in \eqref{eq:App_ABJM_Dwall} is independent of the plane and therefore factors out of the equivariant family integral. Its contribution vanishes because $\int_{\cM}e_T(\operatorname{Obs}) = 0$ by degree. The same degree argument explains the cancellation of the lower order poles in the fixed point sum.
In the remaining quadratic terms we recognise the equivariant restrictions at $W_{IJ}$ of the characteristic classes of $\cS^*$, namely $c_1(\cS^*) = \epsilon_I+\epsilon_J$ and $c_2(\cS^*) = \epsilon_I\epsilon_J$, so that
\begin{align}
\label{eq:ABJM_b3_geometric_family_integral}
 J_{\rm lin}(1,\mu;\sqrt3)
 = {}&-\frac{\ex^{-3\mu}}{16\pi^2}
 \int_{\operatorname{Gr}(2,4)}\bigl(2c_1(\cS^*)^2+c_2(\cS^*)\bigr)\nn\\
 &\qquad\times\Biggl[(18\mu^2-2\pi^2)c_1(\cS^*)^2
 +\left(12\mu+4+\frac{4\pi^2}{3}\right)c_2(\cS^*)\Biggr] \, .
\end{align}
The required integrals of Chern classes over the
bosonic moduli space $\cM=\operatorname{Gr}(2,4)$ are 
\begin{align}
\label{eq:ABJM_b3_Grassmannian_intersections}
\begin{aligned}
 &\int_{\operatorname{Gr}(2,4)}c_1(\cS^*)^4 = 2 \, , \qquad \int_{\operatorname{Gr}(2,4)}c_1(\cS^*)^2c_2(\cS^*)
 = \int_{\operatorname{Gr}(2,4)}c_2(\cS^*)^2 = 1 \, .
\end{aligned}
\end{align}
They imply $\int_{\cM}e(\operatorname{Obs})c_1(\cS^*)^2 = 5$ and $\int_{\cM}e(\operatorname{Obs})c_2(\cS^*) = 3$, which give all three coefficients in \eqref{eq:ABJM_squashed_b3_result}. Thus the fixed point sum and the characteristic class integral evaluate the same family prescription, reproducing the field theory coefficient \eqref{eq:ABJM_squashed_Jnp_b3}. Exchanging the two external directions gives the identical result at $b^2 = \tfrac13$.

\subsection{Higher stabilizer orders}
\label{app:ABJM_sq_higher_stabilizers}

For $s=k>2$, the full-stabilizer locus consists
of two isolated planes and the mixed family
$\CP^1\times\CP^1$.
We now derive the complete polynomial prefactor
\begin{align}\label{eq:ABJM_sq_higher_k_decomposition}
 P_k(\mu;f)
 &=
 2Z^{\rm full}_{{\rm iso},k}(f)
 +D_k(f)\cP_{\rm mix}^{(k)}(\mu,f)\, ,
\end{align}
where
\begin{align}
 \cP_{\rm mix}^{(k)}(\mu,f)
 &=
 4\alpha_k^2\mu^2+B_k(f)\mu+C_k(f)\, ,
 \nn\\
 D_k(f)
 &=
 -\frac{k^2}{16\pi^2}
 \frac{\sin^2(2\pi f/k)}{\sin^2(2\pi/k)}\, .
 \label{eq:app_sq_higher_k}
\end{align}
Here $\alpha_k=4(1-f^2)/k$, and the coefficients
$B_k(f)$ and $C_k(f)$ are given explicitly in
\eqref{eq:ABJM_sq_mixed_coefficients} below.
At generic squashing, $D_k(f)$ is the common
value at zero refinement of the mixed family
determinant with its two complex tangent modes
omitted.
The mixed contribution in
\eqref{eq:ABJM_Squashed_FullStabilizer} is
$P_k^{\rm mix}=D_k\cP_{\rm mix}^{(k)}$.

For the isolated planes, the external bosonic
weights are $kn/2+1\pm f$, with multiplicity
$kn+1$ for $n\geq0$.
The external fermionic weights are
$kn/2-1\pm f$, with multiplicity $kn-1$ for
$n\geq1$.
The internal bosonic and fermionic weights are
$kn/2-1$ and $kn/2+1$, respectively, each with
multiplicity $2kn$ for $n\geq1$.
With the spectral convention used in the
projected zeta functions, these multiplicities
give
\begin{align}
\label{eq:codex_sq_isolated}
 Z^{\rm full}_{{\rm iso},k}(f)
 &=
 \frac{1}{1-f^2}
 \prod_{n\geq1}^{\rm reg}
 \frac{
 [(kn/2-1)^2-f^2]^{kn-1}
 (kn/2+1)^{2kn}}
 {
 [(kn/2+1)^2-f^2]^{kn+1}
 (kn/2-1)^{2kn}}\, .
\end{align}
The prefactor $1/(1-f^2)$ comes from the
$n=0$ external bosons.
This product evaluates to
\eqref{eq:ABJM_sq_full_isolated}.
Its comparison with \cite{Gautason:2025per}
is discussed in
section~\ref{subsec:ABJM_Squashed_Sphere}.

To evaluate the mixed family contribution,
we work at generic squashing and specialize
the Nosaka refinement to $\xi=w$, $\eta=-w$.
For $w\neq0$, this refinement lifts both
tangent directions of $\CP^1\times\CP^1$.
The tangent Euler factors are $w^2$ at
$W_{14},W_{23}$ and $-w^2$ at $W_{13},W_{24}$.
With $k$ and $f$ held fixed, define the
regular factors
\begin{align}
 D_{14,k}(w)
 &=
 w^2 Z^{\rm full}_{14,k}
       \bigl(b,\Delta(w,-w)\bigr)\, ,
 \nn\\
 D_{13,k}(w)
 &=
 -w^2 Z^{\rm full}_{13,k}
       \bigl(b,\Delta(w,-w)\bigr)\, .
\end{align}
They satisfy
\begin{align}
 D_{13,k}(-w)&=D_{13,k}(w)\, ,
 \qquad
 D_{14,k}(0)=D_{13,k}(0)=D_k(f)\, .
\end{align}
The actions at $W_{14},W_{23}$ are
$\alpha_k\mu/(1\pm w)^2$, while those at
$W_{13},W_{24}$ are
$\alpha_k\mu/(1-w^2)$.
Their assembled contribution is therefore
\begin{align}
 \ex^{-\alpha_k\mu}P_k^{\rm mix}(\mu;f)
 &=
 \lim_{w\to0}\frac{1}{w^2}
 \Bigl[
 D_{14,k}(w)
 \ex^{-\alpha_k\mu/(1+w)^2}
 \nn\\
 &\qquad
 +D_{14,k}(-w)
 \ex^{-\alpha_k\mu/(1-w)^2}
 -2D_{13,k}(w)
 \ex^{-\alpha_k\mu/(1-w^2)}
 \Bigr]\, .
 \label{eq:ABJM_sq_mixed_direct_limit}
\end{align}
The constant and linear terms in the numerator
cancel.
Expanding up to quadratic order in $w$ gives
\begin{align}
 P_k^{\rm mix}(\mu;f)
 &=
 4\alpha_k^2\mu^2D_k
 +4\alpha_k\mu\bigl[D_{14,k}'(0)-D_k\bigr]
 +D_{14,k}''(0)-D_{13,k}''(0)\, ,
\end{align}
where primes denote derivatives with respect
to $w$.
Comparison with \eqref{eq:app_sq_higher_k}
identifies
\begin{align}
 \frac{B_k(f)}{4\alpha_k}
 &=
 \frac{D_{14,k}'(0)}{D_k(f)}-1\, ,
 \qquad
 C_k(f)
 =
 \frac{D_{14,k}''(0)-D_{13,k}''(0)}
      {D_k(f)}\, .
\end{align}

The required derivatives can be evaluated
directly from the projected spectrum.
At $W_{14}$,
\eqref{eq:ABJM_ordinary_Nosaka_character_continuation}
gives
\begin{align}
 \cK^{\rm full}_{14,k}(t;w,f)
 &=
 2\frac{
 \ex^{-t}+\ex^{-[k(1+w)/2-1]t}}
 {1-\ex^{-k(1+w)t/2}}
 \nn\\
 &\quad
 +2[\cosh(ft)-1](\ex^{-t}-\ex^{-wt})
 \frac{1+\ex^{-k(1+w)t/2}}
 {(1-\ex^{-(1+w)t})
  (1-\ex^{-k(1+w)t/2})}\, .
\end{align}
Omitting the two tangent modes amounts to
subtracting $2\ex^{-wt}$ from this character.
Zeta function regularization then gives
$D_{14,k}(w)$.
At $W_{13}$, the projected character reduces
to one-dimensional towers.
The Hurwitz identities
\eqref{eq:ABJM_Hurwitz_Cool} give
\begin{align}
 D_{13,k}(w)
 &=
 -\frac{w^2}{4\sin^2(2\pi/k)}
 \frac{
 \sin\!\left(\frac{2\pi f}{k(1+w)}\right)
 \sin\!\left(\frac{2\pi f}{k(1-w)}\right)}
 {
 \sin\!\left(\frac{2\pi w}{k(1+w)}\right)
 \sin\!\left(\frac{2\pi w}{k(1-w)}\right)}\, .
\end{align}
Its limit at $w=0$ reproduces $D_k(f)$.
These regular factors omit the two bosonic
tangent modes.
Upon continuation to $f=0$, they still contain
the residual fermionic factors and are therefore
not the reduced determinants of the round family.

Evaluating the derivatives gives closed
trigonometric expressions for $B_k$ and $C_k$.
Define
\begin{align}
\label{eq:ABJM_sq_mixed_coefficient_arguments}
 h&=\gcd(k,2)\, ,
 \qquad
 \theta_j=\pi\left(1-\frac{2j}{k}\right)\, ,
 \qquad
 g(z)=z\cot z\, ,
\end{align}
with $g(0)=1$ understood by continuity.
Then
\begin{align}
 \frac{B_k(f)}{4\alpha_k}
 ={}&
 \frac{4\pi}{k}\cot\!\left(\frac{2\pi}{k}\right)
 -1
 +\left(2+\frac{4h}{k}\right)
 \left[1-g\!\left(\frac{2\pi f}{k}\right)\right]
 \nn\\
 &-\frac4k
 \sum_{j=1}^{\lfloor(k-1)/2\rfloor}
 \left[
 g\!\left(\theta_j+\frac{2\pi f}{k}\right)
 +g\!\left(\theta_j-\frac{2\pi f}{k}\right)
 -2g(\theta_j)
 \right]\, ,
 \nn\\[4pt]
 C_k(f)
 ={}&
 \left(\frac{B_k(f)}{4\alpha_k}\right)^2
 -2f\,\partial_f\!\left(
       \frac{B_k(f)}{4\alpha_k}\right)
 +3
 -\frac{16\pi f}{k}
  \cot\!\left(\frac{2\pi f}{k}\right)
 \nn\\
 &+\frac{8\pi^2}{k^2}
 \left[
 \frac{1}{\sin^2(2\pi/k)}
 +\frac{2f^2-1}{\sin^2(2\pi f/k)}
 \right]\, .
 \label{eq:ABJM_sq_mixed_coefficients}
\end{align}
These expressions hold for $0<|f|<1$.
The round limit is taken in the complete
mixed family prefactor:
\begin{align}
 \lim_{f\to0}
 D_k(f)\cP_{\rm mix}^{(k)}(\mu,f)
 &=
 \frac{1}{2\sin^2(2\pi/k)}\, .
\end{align}
The two isolated planes supply the same total,
recovering \eqref{eq:ABJM_round_full_result}.

\section{The $V^{5,2}$ determinants and field theory checks}
\label{app:V52_determinants}

In this appendix we derive the family sums and
coefficient functions presented in
section~\ref{subsec:V52_squashed_results},
including the large $k$ limit for the round sphere.
For the determinant calculations, we use the
virtual complex \eqref{eq:V52_virtual_complex}
and the regularization specified in
section~\ref{subsec:V52_Fluctuations}.

The final two subsections provide field theory
checks of these gravity results.
We derive the exact matrix model identity
\eqref{eq:V52_ABJM_special_squashing} at special
squashing, and then compare the round large $k$
prediction \eqref{eq:V52_round_large_k},
after conversion to fixed $N$, with a numerical
calculation of the planar matrix model of \cite{Martelli:2011qj}.

\subsection{The $\CP^3$ family at $k=1$}

At $k=1$, the four coordinate planes
$D_{s_1s_2}$ in \eqref{eq:V52_coordinate_planes}
are the fixed points of the auxiliary Cartan action
on the $\CP^3$ family.
At each fixed point, write
$x=s_1u$ and $y=s_2v$, so that the four contributions
are obtained by evaluating the same function at
$(x,y)=(\pm u,\pm v)$.
The three bosonic zero modes are lifted by the
auxiliary Cartan action, with weights $x$, $y$ and $x+y$.
For generic squashing there are no fermionic zero
modes, so \eqref{eq:M2_family_recombination} gives
\begin{align}
\label{eq:V52_app_k1_full_reduced}
 Z^{\rm full}(x,y;f)
 =\frac{Z^{\rm red}(x,y;f)}{xy(x+y)}\, .
\end{align}

In the family sum \eqref{eq:V52_full_family_sum},
the numerator
$Z^{\rm red}(x,y;f)\ex^{-I_{\rm M2}(x,y)}$
is symmetric in $x,y$ and analytic near the origin
for $f\neq\pm1/3$.
We therefore expand it in symmetric homogeneous
polynomials in $x,y$.
Let $\mathscr L(P)$ denote the sum of
$P(x,y)/[xy(x+y)]$ over $x=\pm u$, $y=\pm v$.
Direct summation gives
\begin{align}
\label{eq:V52_app_k1_collision_identities}
 \mathscr L(1)
 &=\mathscr L(x+y)
 =\mathscr L((x+y)^2)
 =\mathscr L(xy)=0\, ,
 \nonumber\\
 \mathscr L((x+y)^3)&=8\, ,
 \qquad
 \mathscr L(xy(x+y))=4\, .
\end{align}
When $u,v$ are scaled together towards zero, the
denominator scales cubically. The constant, linear
and quadratic terms in the numerator could therefore
produce cubic, quadratic and simple poles, respectively.
The first line shows that all these poles cancel
after summing the four fixed point contributions.
Only the cubic terms give a finite contribution;
higher degree terms vanish in the limit.
Since every symmetric cubic is a linear combination
of $(x+y)^3$ and $xy(x+y)$, the second line in \eqref{eq:V52_app_k1_collision_identities} determines
the finite result and shows that it is independent
of the generic direction of the auxiliary limit.

We can perform this coefficient extraction using a
single formal variable $H$, by making the replacements
$ x+y\longmapsto2H$, and $xy\longmapsto2H^2$. These send $(x+y)^3$ and $xy(x+y)$ to $8H^3$ and
$4H^3$, respectively, reproducing
\eqref{eq:V52_app_k1_collision_identities}.
The finite family sum is therefore
\begin{align}
\label{eq:V52_app_k1_coefficient_rule}
 J_{\rm full}(1,\mu;b)
 =[H^3]\left[
 Z^{\rm red}(H;f)\ex^{-I_{\rm M2}(H)}
 \right]\, ,
\end{align}
where $[H^3]$ denotes the coefficient of $H^3$ in
the Taylor expansion of the enclosed expression
about $H=0$. Consequently we need the reduced determinant only
up to cubic order. We write
\begin{align}
 Z^{\rm red}(H;f)
 =R_0(f)\exp\!\left[
 L_1H+L_2H^2+L_3H^3
 \right]+\cO(H^4)\, .
\end{align}
Here $R_0(f)=Z^{\rm red}(0,0;f)$ is the reduced
determinant with the auxiliary parameters removed,
while $L_1,L_2,L_3$ describe its dependence on the
auxiliary weights:
\begin{align}
\label{eq:V52_app_k1_reduced_expansion}
 R_0(f)&=\frac{2\cos(3\pi f/2)}{27\pi^3}\, ,
 \qquad
 L_1=9+9\pi f\tan\!\left(\frac{3\pi f}{2}\right)\, ,
 \nonumber\\
 L_2&=\frac98\left[
 -6-18\pi f\tan\!\left(\frac{3\pi f}{2}\right)
 +\frac{4\pi^2}{3}
 -\frac{\pi^2}{2}(27f^2+5)
  \sec^2\!\left(\frac{3\pi f}{2}\right)
 \right]\, ,
 \nonumber\\
 L_3&=\frac98\left[
 -2\pi^2+18\pi f\tan\!\left(\frac{3\pi f}{2}\right)
 +\pi^2(27f^2+15)\sec^2\!\left(\frac{3\pi f}{2}\right)
 \right.
 \nonumber\\
 &\hspace{5em}\left.
 +\frac{9\pi^3f}{2}(5+3f^2)
  \sec^2\!\left(\frac{3\pi f}{2}\right)
  \tan\!\left(\frac{3\pi f}{2}\right)
 \right]\, .
\end{align}
For example, for $f\ne\pm1/3$, the factor
$R_0(f)=\exp[\partial_s\zeta^{\rm red}_0(s;f)|_{s=0}]$
is obtained from the reduced spectral zeta function
at zero auxiliary parameters,
\begin{align}
\label{eq:V52_app_k1_zeta_constant}
\zeta^{\rm red}_0(s;f)
 &=\left(\frac32\right)^s\Bigl[
 6\zeta_{\rm R}(s)-1+2^{-s}
 +\ex^{-\ii\pi s}(1-2^{-s})
 \nonumber\\
 &\hspace{3em}
 -\zeta_{H}(s,\tfrac12+\tfrac32f)
  -\zeta_{H}(s,\tfrac12-\tfrac32f)
 \Bigr]\, ,
\end{align}
after omitting the three tangent modes. The higher
coefficients follow by expanding the monomial weights
and using the Hurwitz reflection identities. The factor $\cos(3\pi f/2)$ changes sign when $|f|$
crosses $1/3$, reflecting the fact that one of the lowest external fermionic weights in
\eqref{eq:V52_pure_squashing_mode_weights} crosses through zero.

Applying the same replacements to the localized action
\eqref{eq:V52_refined_M2_action} gives
\begin{align}
\label{eq:V52_app_k1_action_expansion}
 I_{\rm M2}(H)
 =I_1\Big(1-3H+\frac92H^2\Big)+\cO(H^4)\, .
\end{align}
Extracting the cubic coefficient of
$Z^{\rm red}(H;f)\ex^{-I_{\rm M2}(H)}$ gives
\eqref{eq:V52_k1_general_squashing}.
In particular, the apparent singularities of the
final expression are removable because as $f\to\pm1/3$,
$\pi^2(1-9f^2)/\cos(3\pi f/2)\to4\pi$.

\subsection{The two families at $k=2$}

At $k=2$, the full-stabilizer locus consists of
$\CP^1_{\rm mix}$ and $\CP^1_{\rm other}$, as in
\eqref{eq:V52_full_stabilizer_locus}.
We continue to write $x=s_1u$ and $y=s_2v$.
Since $+1=-1$ modulo two, the quotient projection
is independent of the signs $s_1,s_2$, and the four
fixed point determinants are obtained from the same
function of $x,y$. 

For generic squashing, each fixed point has one
bosonic zero mode and no fermionic zero modes.
The auxiliary weight of this tangent mode is
$w=x+y$. Writing the full determinant in terms
of the reduced one, using
\eqref{eq:M2_family_recombination}, gives
\begin{align}
\label{eq:V52_app_k2_expansion}
 Z^{\rm full}(x,y;f)
 =\frac{R_2(f)}{w}
 \bigl[1+L_2^{(2)}(f)w+\cO(x^2,xy,y^2)\bigr]\, .
\end{align}
Here $R_2(f)$ is the reduced determinant at $x=y=0$,
while $L_2^{(2)}(f)$ is the coefficient of $w$ in
its logarithmic expansion.
Applying the projection in
\eqref{eq:V52_projected_character} and regularizing
the remaining spectrum gives
\begin{align}\label{R2L2appF2exp}
 R_2(f)&=\frac{E(q_+)E(q_-)}{6\pi}\, ,
 \qquad \qquad q_\pm=\frac34(1\pm f)\, ,
 \nonumber\\
 L_2^{(2)}(f)
 &=\frac94+\frac{9\pi}{16}
 \frac{1+3f^2+4f\sin(3\pi f/2)}
      {\cos(3\pi f/2)}\, ,
\end{align}
where \begin{align}
E(q)&=(2\pi)^{2q-1}
 \frac{\Gamma(1-q)}{\Gamma(q)}
 \left[\frac{G(1-q)}{G(q)}\right]^2\, ,
\end{align} and
$G$ is the Barnes $G$ function.
For $0<q<1$, an equivalent expression is
\begin{align}
\label{eq:V52_app_k2_E_clausen}
 \log E(q)
 =(2q-1)\log(2\sin\pi q)
 +\frac{\operatorname{Cl}_2(2\pi q)}{\pi}\, .
\end{align}
The Barnes expression for $E(q)$ gives the continuation
specified by the determinant regularization in
section~\ref{subsec:V52_Fluctuations}.
In particular, $R_2(f)$ changes sign when $|f|$
crosses $1/3$, where an external fermionic weight
passes through zero.

Each $\CP^1$ component contains a pair of opposite
fixed planes: $D_{+-},D_{-+}$ for $\CP^1_{\rm mix}$,
and $D_{++},D_{--}$ for $\CP^1_{\rm other}$.
Within either pair, $(x,y)$ changes to $(-x,-y)$,
so the two tangent weights are $w$ and $-w$.
The localized action
\eqref{eq:V52_refined_M2_action} has expansion
\begin{align}
 I_{\rm M2}(x,y)
 =I_2\left[1-\frac32w+\cO(x^2,xy,y^2)\right]\, .
\end{align}
The simple poles in
\eqref{eq:V52_app_k2_expansion} therefore cancel
between the two fixed points.
The finite terms come from the linear correction
to the reduced determinant and the linear correction
to the action. Each pair contributes
\begin{align}
 2R_2(f)\left[L_2^{(2)}(f)+\frac32I_2\right]
 \ex^{-I_2}\, .
\end{align}
The two families give equal contributions, so comparison
with \eqref{eq:V52_kge2_general_squashing} gives
\begin{align}
\label{eq:V52_app_k2_coefficients}
 p_{2,1}(f)=6R_2(f)\, ,
 \qquad
 p_{2,0}(f)=4R_2(f)L_2^{(2)}(f)\, .
\end{align}

At the round point, $q_\pm=3/4$, and
\eqref{eq:V52_app_k2_E_clausen} gives
$R_2(0)=\ex^{-2G_{\rm C}/\pi}/(3\sqrt2\,\pi)$,
using $\operatorname{Cl}_2(3\pi/2)=-G_{\rm C}$.
Together with $L_2^{(2)}(0)=9(4+\pi)/16$,
this reproduces the $k=2$ expression in
\eqref{eq:V52_round_low_levels}.
As $f\to\pm1/3$, we have $R_2\to0$ and
$R_2L_2^{(2)}\to1/4$.
The term proportional to $I_2$ consequently vanishes,
while the constant term tends to one, giving the
$k=2$ case of \eqref{eq:V52_special_squashing_results}.

\subsection{The mixed family and isolated planes at $k>2$}

For $k>2$, the full-stabilizer locus consists of
$\CP^1_{\rm mix}$ and the isolated planes
$D_{++},D_{--}$, as in
\eqref{eq:V52_full_stabilizer_locus}.
We first evaluate the contribution from the mixed
family, whose two auxiliary fixed points are
$D_{+-}$ and $D_{-+}$, and then add the isolated
contributions.

For the mixed family, choose the auxiliary parameters
$u=m/2$ and $v=-m/2$.
The tangent weights at $D_{+-}$ and $D_{-+}$ are
then $m$ and $-m$.
At $D_{+-}$, the two coordinate functions have
common weight $a=2/3+m/2$ and opposite quotient
charges. A monomial with powers $p,q\geq0$ therefore
has weight $a(p+q)$ and survives the $\Z_k$ projection
precisely when $p-q=0$ modulo $k$.
The invariant function character is thus
\begin{align}
\label{eq:V52_app_mixed_function_character}
 H_k(z)
 =\sum_{\substack{p,q\geq0\\p-q=0\ ({\rm mod}\ k)}}
 z^{p+q}
 =\frac{1+z^k}{(1-z^2)(1-z^k)}\, .
\end{align}
Here $z$ counts total monomial degree.
Setting $z=\ex^{-ta}$ shows that the invariant
spectrum consists of two sums with spacings
$2a$ and $ka$, one shifted relative to the other
by $ka$. 
To evaluate the regularized products associated with
these two sums, using \eqref{eq:1loop_def}, define
\begin{align}
\label{eq:V52_app_Barnes_functions}
 \Gamma_2^\star(z\mid\alpha,\beta)
 &=\exp\!\Bigg[
 \left.\partial_s
 \sum_{n,l\geq0}(z+n\alpha+l\beta)^{-s}
 \right|_{s=0}\Bigg]\, ,
 \nonumber\\
 \mathcal G_{k,a}(z)
 &=\Gamma_2^\star(z\mid2a,ka)
   \Gamma_2^\star(z+ka\mid2a,ka)\, .
\end{align}
The two factors in $\mathcal G_{k,a}$ correspond to
the unshifted sum and the sum shifted by $ka$,
respectively.
The defining sum is initially convergent for
$\operatorname{Re}s>2$ and positive arguments.
We analytically continue to $s=0$ and to the
required arguments using the same logarithms of
spectral weights as in \eqref{eq:1loop_def}.

The finite pair of terms outside the quotient sum
in \eqref{eq:V52_projected_character} contributes
a common factor $-1$ to each determinant, as
explained in section~\ref{subsec:V52_Fluctuations}.
We keep this factor separate and denote the
regularized product of the remaining towers by
$\mathcal D_k(m;f)$:
\begin{align}
\label{eq:V52_app_mixed_determinant}
 \mathcal D_k(m;f)
 =
 \frac{
 \mathcal G_{k,a}(1+f)\mathcal G_{k,a}(1-f)
 \mathcal G_{k,a}(-\tfrac23)\mathcal G_{k,a}(m)}
 {\mathcal G_{k,a}(2a-1-f)\mathcal G_{k,a}(2a-1+f)
  \mathcal G_{k,a}(2a+\tfrac23)
  \mathcal G_{k,a}(\tfrac43)}\, .
\end{align}
The four factors in the numerator represent the bosonic
towers, and the four in the denominator represent
their fermionic partners.
Restoring this factor of $-1$, the full determinants
at $D_{+-}$ and $D_{-+}$ are
$-\mathcal D_k(m;f)$ and $-\mathcal D_k(-m;f)$,
respectively.

To evaluate the sum of the two fixed point contributions
in the auxiliary limit $m\to0$, we need only the pole
and constant terms of $\mathcal D_k(m;f)$. Define
\begin{align}
\label{eq:V52_app_mixed_residues}
 \mathcal R_k(f)
 =\lim_{m\to0}2m\mathcal D_k(m;f)\, ,\qquad
 \mathcal S_k(f)
 &=\left.\frac{\dd}{\dd m}
       [2m\mathcal D_k(m;f)]\right|_{m=0}\, ,
\end{align}
so that
\begin{align}
 \mathcal D_k(m;f)
 =\frac{\mathcal R_k(f)}{2m}
  +\frac{\mathcal S_k(f)}2+\cO(m)\, .
\end{align}
The localized action
\eqref{eq:V52_refined_M2_action} gives
\begin{align}
 I_{\rm M2}(m)
 =I_k\left[1-\frac32m+O(m^2)\right]\, ,
\end{align}
with $I_{\rm M2}(-m)$ at the opposite fixed point.
The poles cancel when the two contributions are
summed. Including the finite terms from both the
determinant and the action, the mixed family gives
\begin{align}
 &-\lim_{m\to0}\Bigl[
 \mathcal D_k(m;f)\ex^{-I_{\rm M2}(m)}
 +\mathcal D_k(-m;f)\ex^{-I_{\rm M2}(-m)}
 \Bigr]
 \nonumber\\
 &\qquad\qquad\qquad\qquad\qquad\qquad=
 -\left[\mathcal S_k(f)
       +\frac32I_k\mathcal R_k(f)\right]\ex^{-I_k}\, .
\end{align}

The isolated planes $D_{++},D_{--}$ have no zero
modes, so their determinants can be evaluated
directly at $u=v=0$.
At total monomial degree $\ell$, the multiplicity
is $\ell+1$. 
In the numerator of the quotient sum in
\eqref{eq:V52_projected_character}, terms without
$A_jB_j$ have zero fibre charge under $\Z_k$,
while those proportional to $A_jB_j$ carry the
quotient charge of $K_D$.
The projection therefore retains monomial degrees
$\ell=0$ modulo $k$ and $\ell+2=0$ modulo $k$,
respectively.
Each tower therefore reduces to a weighted Hurwitz
zeta sum. Evaluating these sums with
\eqref{eq:1loop_def} and using reflection identities
gives a common determinant for the two planes.

For these isolated contributions, put
$q_\pm=3(1\pm f)/(2k)$.
Using the same convention as for $\mathcal D_k$,
we write the full determinant at either isolated
plane as $-\mathcal D_{\rm iso}(k;f)$, where
\begin{align}
\label{eq:V52_app_isolated_determinant}
 \mathcal D_{\rm iso}(k;f)
 ={}&-\frac{
 [2\sin(\pi q_+)]^{kq_+-1}
 [2\sin(\pi q_-)]^{kq_--1}}
 {[2\sin(\pi/k)]^2[2\sin(2\pi/k)]}
 \nonumber\\
 &\times\exp\!\left\{
 \frac{k}{2\pi}\left[
 \sum_{\sigma=\pm}\operatorname{Cl}_2(2\pi q_\sigma)
 -\operatorname{Cl}_2(2\pi/k)
 -\operatorname{Cl}_2(4\pi/k)
 \right]\right\}\, .
\end{align}
For $k>2$ and $|f|<1$, all sine bases in this
expression are positive.
The  minus sign comes from the single
negative bosonic weight $-2/3$.

Adding the two isolated contributions to the mixed
family and comparing with
\eqref{eq:V52_kge2_general_squashing}, we obtain
\begin{align}
\label{eq:V52_app_kge2_coefficients}
 p_{k,1}(f)&=-\frac32\mathcal R_k(f)\, ,
 \nonumber\\
 p_{k,0}(f)&=-2\mathcal D_{\rm iso}(k;f)
             -\mathcal S_k(f)\, ,
 \qquad k>2\, .
\end{align}
This also makes explicit why the isolated planes
contribute only to the constant term in the
instanton prefactor.

At the round point and $k=3$, the Barnes and
Clausen expressions reduce to
\begin{align}
\label{eq:V52_app_k3_round_data}
 \mathcal R_3(0)&=-\frac1{3\pi}\, ,
\qquad
 \mathcal S_3(0)=-\frac1{2\pi}-\frac2{3\sqrt3}\, ,
 \qquad
 \mathcal D_{\rm iso}(3;0)=-\frac2{3\sqrt3}\, .
\end{align}
Substituting these values reproduces the $k=3$
expression in \eqref{eq:V52_round_low_levels}.

At $f=\pm1/3$, the additional external fermion
has the same auxiliary weight as the tangent
boson of the mixed family.
Their contributions cancel in the character,
removing the pole in $\mathcal D_k$.
The resulting values are
\begin{align}
 \mathcal R_k&=0\, ,
 &\mathcal S_k&=-\frac1{2\sin^2(\pi/k)}\, ,
 &\mathcal D_{\rm iso}&=-\frac1{4\sin^2(\pi/k)}\, .
\end{align}
These give \eqref{eq:V52_special_squashing_results}.

\subsection{The round sphere, large $k$ limit}

At the round point $f=0$, the Barnes functions
defined in \eqref{eq:V52_app_Barnes_functions}
have the large $k$ expansion
\begin{align}
 \mathcal G_{k,a}(z)
 =C_{k,a}
 \frac{\Gamma(z/(2a))}{\sqrt{2\pi}}
 \left(\frac{k}{4\pi}\right)^{\frac12-\frac{z}{2a}}
 [1+o(1)]\, ,
\end{align}
where $C_{k,a}$ is independent of $z$ and cancels
from the determinant ratio
\eqref{eq:V52_app_mixed_determinant}.

To obtain the mixed family contribution, we insert
this expansion into $2m\mathcal D_k(m;0)$.
Its value and first derivative at $m=0$ determine
$\mathcal R_k(0)$ and $\mathcal S_k(0)$ through
\eqref{eq:V52_app_mixed_residues}.
After removing the simple pole, the expansion can
be differentiated at $m=0$.
Gamma and digamma identities give the first two
limits below:
\begin{align}
 \frac{\mathcal R_k(0)}{k^2}
 &\to-\frac4{3\Gamma(1/4)^4}\, ,
 \qquad
 \frac{\mathcal S_k(0)}{k^2}
 \to-\frac{3+7\log2}{\Gamma(1/4)^4}\, ,
 \nonumber\\
 \frac{\mathcal D_{\rm iso}(k;0)}{k^2}
 &\to-\frac2{9\pi^2}\, .
\end{align}
The isolated plane limit in the second line follows
directly from \eqref{eq:V52_app_isolated_determinant},
using
$\operatorname{Cl}_2(\vartheta)
=\vartheta(1-\log\vartheta)+\cO(\vartheta^3)$
as $\vartheta\to0^+$.

Finally, substituting these limits into the coefficient
relations \eqref{eq:V52_app_kge2_coefficients}
and then into \eqref{eq:V52_kge2_general_squashing},
with $x=\mu/k$ held fixed and $I_k=9x/4$,
gives \eqref{eq:V52_round_large_k}.

\subsection{The matrix model identity at special squashing}
\label{app:V52_matrix_identity}

Here we derive the exact field theory identity
\eqref{eq:V52_ABJM_special_squashing} from the
localized field theory matrix model integral at $b^2=2$.
We consider the $3d$ $\cN=2$
$U(N)_k\times U(N)_{-k}$ Chern--Simons-matter
theory proposed in \cite{Martelli:2009ga}.
Its matter content consists of two bifundamental
and two anti-bifundamental chiral multiplets,
with one adjoint chiral multiplet at each gauge
node and a cubic superpotential.

The round sphere matrix model integral and its matter
contributions are given in Eqs.~(4.6)--(4.9) of
\cite{Martelli:2011qj}, with $n=2$ in their
notation.
At vanishing flavour masses, all six chiral
multiplets have R-charge $2/3$.
We denote the two sets of Cartan eigenvalues by
$\lambda_i,\widetilde\lambda_i$ and write
$\lambda_{ij}=\lambda_i-\lambda_j$, with analogous
notation for the second node.

On the squashed sphere, a chiral multiplet contributes
$s_b(\ii Q/6-z)$, 
where $s_b$ is the double sine
function and $2\pi z$ is the appropriate
Cartan eigenvalue difference.
We use the standard localization normalization and
double sine conventions of equation~(2.1) and Appendix~A
of \cite{Bobev:2025ltz}.
At $b^2=2$, we have $Q/6=1/(2b)$, and the
double sine identity gives
\begin{align}
\label{eq:V52_matrix_chiral_pair}
 s_b\!\left(\frac{\ii Q}{6}+z\right)
 s_b\!\left(\frac{\ii Q}{6}-z\right)
 =\frac{1}{2\cosh(\pi z/b)}\, .
\end{align}

For each of the two $U(N)$ gauge factors, the
off-diagonal adjoint contributions pair with opposite
eigenvalue differences.
Combining them with the vector multiplet contribution
gives, for each $i<j$,
\begin{align}
 \frac{
 4\sinh(b\lambda_{ij}/2)\sinh(\lambda_{ij}/(2b))}
 {2\cosh(\lambda_{ij}/(2b))}
 =4\sinh^2\!\left(\frac{\lambda_{ij}}{2b}\right)\, ,
\end{align}
where we used $b^2=2$.
Each adjoint also has $N$ diagonal components,
each contributing $s_b(\ii Q/6)$.
Setting $z=0$ in \eqref{eq:V52_matrix_chiral_pair}
gives $s_b(\ii Q/6)^2=1/2$.
The diagonal components of the two adjoint multiplets
therefore contribute $s_b(\ii Q/6)^{2N}=2^{-N}$.
The two bifundamental pairs give
\begin{align}
 \prod_{i,j=1}^N
 \left[
 2\cosh\!\left(
 \frac{\lambda_i-\widetilde\lambda_j}{2b}
 \right)
 \right]^{-2}\, .
\end{align}

Now rescale
$\lambda_i=b x_i$ and
$\widetilde\lambda_i=b\widetilde x_i$.
The integration measure acquires the Jacobian
$b^{2N}=2^N$, which cancels the diagonal adjoint
factor exactly.
The remaining hyperbolic factors coincide with
those in the ABJM matrix integral on the round
sphere,
while the Chern--Simons factor
becomes
\begin{align}
 \exp\!\left[
 \frac{\ii k}{4\pi}
 \sum_i(\lambda_i^2-\widetilde\lambda_i^2)
 \right]
 =
 \exp\!\left[
 \frac{\ii(2k)}{4\pi}
 \sum_i(x_i^2-\widetilde x_i^2)
 \right]\, .
\end{align}
Thus the integral is precisely the round sphere
ABJM partition function at level $2k$.
The symmetry under $b\leftrightarrow b^{-1}$
gives the same result at $b^2=1/2$, proving
\eqref{eq:V52_ABJM_special_squashing}.
The identity holds at every finite $N$ and provides
the field theory check of
\eqref{eq:V52_special_squashing_results}.

\subsection{A numerical check in the planar limit}
\label{app:V52_planar_numerics}

Here we test our round sphere, large $k$ result
\eqref{eq:V52_round_large_k} using the field theory
matrix model discussed in
Appendix~\ref{app:V52_matrix_identity}.
At vanishing flavour masses, let
$Z_{V^{5,2}}(N,k;1)$ denote the exact partition
function on the round sphere.
Its canonical free energy and planar limit are
\begin{align}
 F(N,k)
 &=-\log Z_{V^{5,2}}(N,k;1)\, ,
 \nonumber\\
 \Phi_0(\lambda)
 &=\lim_{\substack{N,k\to\infty\\N/k=\lambda}}
   \frac{F(N,k)}{N^2}\, ,
\end{align}
where $\lambda=N/k$ is the 't Hooft coupling.
The numerical observable that we will use to make the field theory/gravity comparison will
be the derivative
$H_0(\lambda)=\Phi_0'(\lambda)$.
Both $\Phi_0$ and $H_0$ include the exponentially
small corrections at strong coupling that we
wish to test.

We solve the finite $N$ saddle equations in
equation~(4.10) of \cite{Martelli:2011qj}, with $n=2$,
on the conjugation-symmetric branch
$\widetilde\lambda_i=\overline{\lambda_i}$.
Denote the action evaluated at this saddle by
$F_N^{\rm saddle}(\lambda)$.
Writing $\lambda_i=x_i+\ii y_i$, we compute
\begin{align}
\label{eq:V52_numerical_H}
 H_N(\lambda)
 &\equiv
 \frac{\mathrm d \ \ }{\mathrm d\lambda}
 \left(\frac{F_N^{\rm saddle}(\lambda)}{N^2}\right)
 =-\frac{1}{\pi\lambda^2N}
   \sum_{i=1}^N x_i y_i\, .
\end{align}
We assume that this saddle controls the planar
limit over the range of couplings considered below.
Its action then gives the leading $N^2$
contribution to $F(N,k)$, so $H_N$ approaches
$H_0$.\footnote{At fixed $x=\mu/k$, the full-stabilizer
membranes have finite action $9x/4$ and reduce
to fundamental string worldsheet instantons.
The smaller-stabilizer linear configurations
have action $9kx/(4s)$, with $s\leq2$, and are
exponentially suppressed as $k\to\infty$;
they therefore do not contribute to the planar
comparison below.}
We extract this limit by extrapolating the
finite $N$ results.

We separate the planar free energy as
$\Phi_0=\Phi_{0,\rm pert}+\Phi_{0,\rm np}$.
To extract the instanton contribution, we subtract
the derivative of the perturbative part, obtained
from the perturbative grand potential:
\begin{align}
\label{eq:V52_numerical_planar_pert}
 \Phi_{0,\rm pert}(\lambda)
 &=\frac{16\pi}{27}
   \frac{(\lambda-\frac1{12})^{3/2}}{\lambda^2}
   +\frac{\gamma}{\lambda^2}\, .
\end{align}
The shift $1/12$ is the planar limit of the
perturbative Airy coefficient $\mathcal B/k$,
while $\gamma$ comes from the constant term
in the grand potential.\footnote{
The planar shift is appropriate to the freely
acting quotient used here. The deformed cone has
two isolated $\C^4/\Z_k$ singularities
\cite{Martelli:2009ga}. Each contributes the
anomalous M2 charge $-(k-1/k)/24$
\cite{Bergman:2009zh}, giving
$\mathcal B=k/12+O(k^{-1})$ in the sector
without additional torsion contributions
of order $k$.}

The gravity result \eqref{eq:V52_round_large_k}
gives the instanton contribution to the grand
potential in the large $k$ limit with
$x=\mu/k$ held fixed.
To obtain a prediction for the canonical planar free
energy, we evaluate the inverse Laplace transform
to fixed $N$ by a saddle point approximation.
The perturbative saddle is
\begin{align}
\label{eq:V52_numerical_planar_saddle}
 x_*(\lambda)
 \equiv\frac{\mu_*}{k}
 =\frac{8\pi}{9}
  \sqrt{\lambda-\frac1{12}}\, .
\end{align}
At first instanton order, the displacement of the
saddle does not change the stationary perturbative
action. Evaluating the membrane grand potential
at $\mu=\mu_*$ therefore gives a gravity prediction
for the first instanton correction to the field
theory planar free energy $\Phi_0$ defined above.
We denote this prediction by
$\delta\Phi_0^{(1)}$, and its derivative by
$\delta H_0^{(1)}$:
\begin{align}
\label{eq:V52_numerical_planar_prediction}
 \delta\Phi_0^{(1)}(\lambda)
 &=-\frac{P(x_*)}{\lambda^2}
   \ex^{-9x_*/4}\, ,
   \qquad\qquad
 P(x)
 =\frac{\frac92x+3+7\log2}{\Gamma(1/4)^4}
   +\frac{4}{9\pi^2}\, ,
 \nonumber\\
 \delta H_0^{(1)}(\lambda)
 &=\frac{\mathrm d}{\mathrm d\lambda}
   \delta\Phi_0^{(1)}(\lambda)\, .
\end{align}

We now compare this gravity prediction with
the non-perturbative part of $H_0$ extracted
numerically from the field theory matrix model.
To eliminate the unknown perturbative constant
$\gamma$ from this comparison, put
$H_{\rm pert}=\Phi_{0,\rm pert}'$ and, for a
fixed reference coupling $\lambda_0$, define
\begin{align}
\label{eq:V52_numerical_subtraction}
 \mathcal D(\lambda;\lambda_0)
 &=
 \lambda^3
 \bigl[H_0(\lambda)-H_{\rm pert}(\lambda)\bigr]
 -\lambda_0^3
 \bigl[H_0(\lambda_0)-H_{\rm pert}(\lambda_0)\bigr]\, .
\end{align}
The $\gamma$-dependent part of
$\lambda^3H_{\rm pert}(\lambda)$ is the constant
$-2\gamma$, which cancels between the two
couplings.
Inserting the extrapolated matrix model values
of $H_0$ into
\eqref{eq:V52_numerical_subtraction} gives
$\mathcal D_{\rm num}$.
Its evaluation therefore requires no choice
of $\gamma$.

The gravity prediction for this quantity at
leading instanton order is
\begin{align}
\label{eq:V52_numerical_subtraction_prediction}
 \mathcal D^{(1)}(\lambda;\lambda_0)
 &=
 \lambda^3\delta H_0^{(1)}(\lambda)
 -\lambda_0^3\delta H_0^{(1)}(\lambda_0)\, .
\end{align}
The instanton action and both coefficients
of $P(x)$ in \eqref{eq:V52_numerical_planar_prediction} are fixed by the membrane calculation;
none is fitted to the numerical data.

On the field theory side, we solve the saddle
equations by Newton iteration.\footnote{The numerical code used to produce
Table~\ref{tab:V52_planar_numerics} was written
and executed by ChatGPT (OpenAI), following the
saddle equations and extrapolation procedure
described in this subsection.}
We extrapolate
$H_N=H_0+\sum_{r=1}^4 c_rN^{-r}$
using five ranks $N=2^jN_0$, $j=0,\ldots,4$.
At each coupling, the levels $k=N/\lambda$
are integers.
To assess sensitivity to the extrapolation,
we repeat the calculation using a cubic
polynomial in $1/N$ and the first four ranks.
The maximum absolute saddle equation residual
is below $2\times10^{-10}$.
Table~\ref{tab:V52_planar_numerics} gives the
results with reference coupling $\lambda_0=6$,
for which we use $N_0=60$.
The values $\mathcal D_{\rm num}$ are obtained
from the full matrix model saddle, whereas
the gravity prediction $\mathcal D^{(1)}$
retains only the leading instanton contribution.

\begin{table}[ht]
\centering
\small
\begin{tabular}{|c|c|c|c|c|}
\hline
$\lambda$ & $N_0$
& $\mathcal D_{\rm num}$
& $\mathcal D^{(1)}$
& $\left|\mathcal D_{\rm num}/
             \mathcal D^{(1)}-1\right|$
\\
\hline
$1.4$ & $63$
& $7.15484\times10^{-4}$
& $7.16706\times10^{-4}$
& $1.71\times10^{-3}$
\\
\hline
$1.8$ & $63$
& $2.97949\times10^{-4}$
& $2.98122\times10^{-4}$
& $5.82\times10^{-4}$
\\
\hline
$2.2$ & $66$
& $1.35475\times10^{-4}$
& $1.35505\times10^{-4}$
& $2.21\times10^{-4}$
\\
\hline
$2.5$ & $60$
& $7.82739\times10^{-5}$
& $7.82825\times10^{-5}$
& $1.11\times10^{-4}$
\\
\hline
$3.0$ & $60$
& $3.33151\times10^{-5}$
& $3.33162\times10^{-5}$
& $3.20\times10^{-5}$
\\
\hline
\end{tabular}
\caption{
Field theory matrix model values
$\mathcal D_{\rm num}$ compared with the gravity
prediction $\mathcal D^{(1)}$ in
\eqref{eq:V52_numerical_subtraction_prediction}.
The reference coupling is $\lambda_0=6$, and
$N_0$ is the smallest rank used at each coupling.
Changing from the four rank to the five rank
extrapolation changes $\mathcal D_{\rm num}$
by approximately $2\times10^{-9}$.
}
\label{tab:V52_planar_numerics}
\end{table}

The matrix model data support the gravity
prediction for the instanton action and both
coefficients of $P(x)$.
The relative difference decreases over the
range of couplings shown.
At $\lambda=3$, the absolute difference is
approximately $10^{-9}$, comparable to the
change under extrapolation.
These data therefore test the leading
instanton contribution but do not resolve
the next instanton sector.

\section{A formal extension to the ABJM superconformal index}
\label{app:ABJM_SCI}

We extend the gravitational block calculation of the TTI in
section~\ref{sec:ABJM_TTI} to the ABJM superconformal index
(SCI). The ingredients are the same perturbative block for
the squashed three-sphere $S^3_b$ obtained in
\cite{BenettiGenolini:2026kjy} and the linear filling
M2-brane contributions computed in this paper; the change is simply in the
relative signs and branches used to glue the two blocks
\cite{Hristov:2022lcw,BenettiGenolini:2026cyc}.
We conjecture that this gluing applies to the sphere blocks,
including the M2-brane contributions, at general
squashing and flavour chemical potentials $\Delta_I$. 
We test this proposal in the Cardy limit of small angular
potential, $\omega\to0$, reproducing the instanton
coefficients determined by field theory at Chern--Simons
levels $k=1,2,4$
\cite{Hosseini:2026dkj,Hosseini:2026lvj}.

For the TTI, the external filling has topology
$M_4\simeq\R^2\times S^2$, with one fixed point of $K$
over each pole of $S^2$.
The holomorphic description of the twelve-dimensional
filling $W$ allows us to use the fluctuation complexes of
\cite{BenettiGenolini:2026kjy} and the present paper
at both external fixed points. For the SCI, the four-dimensional
Killing spinor instead contains both chiralities, with opposite
chiral components vanishing at the two fixed points
\cite{BenettiGenolini:2026kjy}. A corresponding global
fluctuation complex has yet to be constructed. We therefore
formulate the SCI extension directly in terms of gravitational
blocks, carrying over the sphere normalization and local phase
conventions.
Throughout this appendix, $\mathcal I_{\rm SCI}$ denotes the
selected two block saddle contribution. Its global integration
cycle and phase assignment, and its place in the full sum over
SCI saddles, remain to be determined.\footnote{For a related
localization derivation of holomorphic and anti-holomorphic
factorization for asymptotically flat black hole indices,
see \cite{Bobev:2026ije}.}

\subsection{Block parameters and perturbative gluing}
\label{app:SCI_perturbative_blocks}

The SCI is a supersymmetric partition function on
$S^2\times S^1$, with a complex angular potential $\omega$
for rotations of $S^2$ and chemical potentials $\chi_I$
encoding the R-symmetry and flavour fugacities.
The block associated with each pole is obtained from
the partition function on the squashed three-sphere
$S^3_b$, evaluated at local parameters
$(b_\sigma,\Delta_{\sigma,I})$ specified below.
Background magnetic fluxes $\mathfrak n_I$ through $S^2$
are allowed, with supersymmetry imposing
\begin{align}
\label{eq:SCI_constraints}
 \sum_I\chi_I = 2(1+\omega) \, , \qquad
 \sum_I\mathfrak n_I = 0 \, .
\end{align}
The ordinary SCI has $\mathfrak n_I=0$; non-zero flavour fluxes give its generalized version. 
We express the contribution at each pole, labelled by
$\sigma=\pm$, using the squashed sphere functions
evaluated at
\begin{align}
\label{eq:SCI_local_sphere_parameters}
 b_\sigma^2 = \omega \, , \qquad
 \Delta_{\sigma,I}
 = \frac{\chi_I+\sigma\omega\mathfrak n_I}{1+\omega} \, ,
 \qquad \sum_I\Delta_{\sigma,I} = 2 \, .
\end{align}
For comparison, the TTI refinement was denoted by
$\epsilon$ in section~\ref{sec:ABJM_TTI}.
There, the fluxes obey $\sum_I\mathfrak n_I=2$,
implementing the topological twist, and
\eqref{eq:TTI_local_sphere_parameters} assigns
opposite local sphere rotations,
$b_\sigma^2=\sigma\epsilon$, with chemical potential
normalization containing $1+\sigma\epsilon$.
For the SCI, the different spinor
chirality assignments instead give the same effective
rotation $b_\sigma^2=\omega$ at both poles, and hence the
common denominator $1+\omega$. The magnetic shifts
$\sigma\omega\mathfrak n_I$ still have opposite signs.

In the proposed SCI gluing, we use the same local
$S^3_b$ functions as in the TTI calculation, evaluated at the parameters in
\eqref{eq:SCI_local_sphere_parameters}.
In particular, the perturbative grand potential
at each pole is
\begin{align}
\label{eq:SCI_local_perturbative_grand_potential}
 J_\sigma^{\rm pert}(\mu_\sigma)
 &= \frac{\mathcal C_\sigma}{3}\mu_\sigma^3
    +\mathcal B_\sigma\mu_\sigma+\mathcal A_\sigma \, ,\nn\\
 \mathcal C_\sigma
 &= \frac{2\omega^2}
 {\pi^2k\prod_I(\chi_I+\sigma\omega\mathfrak n_I)} \, ,\nn\\
 \mathcal B_\sigma
 &= \mathcal B(k,\sqrt{\omega},\Delta_\sigma) \, , \qquad
 \mathcal A_\sigma
 = \mathcal A_{\rm ABJM}(k,\sqrt{\omega},\Delta_\sigma) \, .
\end{align}
Here $\mathcal B$ is given in
\eqref{eq:TTI_sphere_Airy_coefficients}, and
$\mathcal A_{\rm ABJM}$ is the sphere one-loop function
of \cite{BenettiGenolini:2026kjy}. The cubic and linear
terms reproduce the perturbative SCI gravitational block
data of \cite{Hristov:2022lcw,BenettiGenolini:2026cyc}.
We extend the same gluing to $\mathcal A_{\rm ABJM}$ and,
below, to the M2-brane contributions.

For the TTI, the continued contours select the opposite
saddle branches in \eqref{eq:TTI_block_saddles}.
For the SCI, we instead choose
$\mu_\sigma^*=+\sqrt{(N-\mathcal B_\sigma)/\mathcal C_\sigma}$
at both poles, initially in a chamber where the Airy
functions decay. The corresponding canonical
perturbative blocks are
\begin{align}
\label{eq:SCI_perturbative_Airy_blocks}
 \mathcal Z_\sigma^{\rm pert}(N)
 &= \int_{\Gamma_\sigma^{\rm SCI}}\frac{\dd\mu_\sigma}{2\pi\ii}\,
    \ex^{J_\sigma^{\rm pert}(\mu_\sigma)-N\mu_\sigma}\nn\\
 &= \ex^{\mathcal A_\sigma}\mathcal C_\sigma^{-1/3}
    \operatorname{Ai}\!\left[
       \mathcal C_\sigma^{-1/3}(N-\mathcal B_\sigma)
    \right] \, .
\end{align}
The blocks are then analytically continued in $\omega$
and the chemical potentials from this chamber.
The perturbative contribution of the selected saddle is
\begin{align}
\label{eq:SCI_perturbative_gluing}
 \mathcal I_{\rm SCI}^{\rm pert}(N)
 = \mathcal Z_+^{\rm pert}(N)\mathcal Z_-^{\rm pert}(N) \, .
\end{align}
As for the TTI, this is a product of analytic blocks;
no complex conjugation is implicit. Relations of the
schematic form $\mathcal I_{\rm SCI}\sim|Z_{S_b^3}|^2$
require the corresponding parameter and conjugation
conventions \cite{Bobev:2026lvl}.

The field theory relation of \cite{Bobev:2026lvl} applies
to saddle contributions to all algebraic orders in the
Cardy expansion, subject to the appropriate contour
assumptions. It omits terms that are non-perturbative in the angular
potential, schematically $\ex^{-c/|\omega|}$ with $c>0$.
These are distinct from the M2-brane terms
$\ex^{-c'\sqrt N}$ retained here. The expressions at finite
$\omega$ below are the proposed gravitational block
continuation beyond this established Cardy relation.

\subsection{M2-brane corrections}
\label{app:SCI_membrane_blocks}

For the TTI, we first computed the M2-brane contribution
to each local grand potential, then transformed to fixed
$N$ and summed the two canonical free energy corrections,
as in \eqref{eq:TTI_canonical_membrane_insertion}.
We follow exactly these steps for the SCI, using instead
the local parameters \eqref{eq:SCI_local_sphere_parameters}
and the Ai contours in
\eqref{eq:SCI_perturbative_Airy_blocks}. The change in
gluing does not introduce an additional sign in the
local M2-brane determinant.

Let $J_{{\rm lin},\sigma}(\mu_\sigma)$ denote the
contribution of the linear fillings at these sphere
parameters, with $\mu_\sigma$ the inverse Laplace variable
of the block. For example, at generic weights its singly
wrapped full-stabilizer part is
\begin{align}
\label{eq:SCI_full_stabilizer_insertion}
 J_{{\rm full},\sigma}(\mu_\sigma)
 = \sum_{I<J}
 Z^{\rm full}_{IJ,k}(\sqrt{\omega},\Delta_\sigma)
 \exp\!\left[
 -\frac{4\omega\mu_\sigma}
 {k(\chi_I+\sigma\omega\mathfrak n_I)
    (\chi_J+\sigma\omega\mathfrak n_J)}
 \right] \, .
\end{align}
Compared with the TTI expression
\eqref{eq:TTI_local_membrane_grand_potential}, the action
contains $\omega$ in place of $\sigma\epsilon$, while the
flux shifts retain their pole dependence. The determinants
are the same projected Barnes zeta expressions
\eqref{eq:ABJM_projected_double_zeta} and
\eqref{eq:ABJM_generic_Barnes_determinant}. When families
occur, their auxiliary fixed point contributions are
combined before taking the refinement limit, as in
section~\ref{subsec:Quantum_Brane_Families}. The additional
generic orbit term at $k>2$ may be included using
\eqref{eq:ABJM_generic_orbit_refined_allk}, retaining its
separate assumptions about covered maps, quotient measure
and phase.

The corrected canonical block and its gluing are
\begin{align}
\label{eq:SCI_corrected_block_gluing}
 \mathcal Z_\sigma(N)
 &= \int_{\Gamma_\sigma^{\rm SCI}}\frac{\dd\mu_\sigma}{2\pi\ii}\,
 \ex^{J_\sigma^{\rm pert}(\mu_\sigma)-N\mu_\sigma}
 \left[1+J_{{\rm lin},\sigma}(\mu_\sigma)+\cdots\right] \, ,\nn\\
 \mathcal I_{\rm SCI}(N)
 &= \mathcal Z_+(N)\mathcal Z_-(N) \, .
\end{align}
The ellipsis denotes sectors beyond the single M2-brane
insertions retained here. Writing
$\mathcal F_{\rm SCI}=-\log\mathcal I_{\rm SCI}$ gives
\begin{align}
\label{eq:SCI_linear_membrane_correction}
 \mathcal F_{\rm SCI,lin}^{(1)}
 = -\sum_{\sigma = \pm}
 \frac{1}{\mathcal Z_\sigma^{\rm pert}(N)}
 \int_{\Gamma_\sigma^{\rm SCI}}\frac{\dd\mu_\sigma}{2\pi\ii}\,
 \ex^{J_\sigma^{\rm pert}(\mu_\sigma)-N\mu_\sigma}
 J_{{\rm lin},\sigma}(\mu_\sigma) \, .
\end{align}
This has the same form as
\eqref{eq:TTI_canonical_membrane_insertion}, with the SCI
block assignments. Polynomial insertions multiplying
instanton exponentials are evaluated by the same Airy
shift rule \eqref{eq:TTI_Airy_shift_rule}; no saddle
approximation is required for this transform. In particular,
the $\mu_\sigma$-independent factor
$\ex^{\mathcal A_\sigma}$ cancels from each normalized
insertion.

\subsection{The symmetric limit and field theory comparison}
\label{app:SCI_symmetric_limit}

We now make a non-trivial check
against the leading field theory instanton coefficients
in the Cardy expansion. We specialize 
to vanishing magnetic fluxes and equal
chemical potentials:\footnote{The restriction to even $k$ for the
gravitational universal TTI twist, discussed in
section~\ref{subsec:TTI_universal_tests}, does not
apply to this zero flux SCI background.}
\begin{align}
\label{eq:SCI_symmetric_parameters}
 \mathfrak n_I = 0 \, , \qquad
 \chi_I = \frac{1+\omega}{2} \, .
\end{align}
Both poles then have the same normalized sphere parameters,
$\Delta_{\sigma,I}=\tfrac12$ and $b_\sigma^2=\omega$.
The two blocks therefore give identical M2-brane
corrections to the canonical free energy, and their
combined contribution is
\begin{align}
\label{eq:SCI_symmetric_membrane_relation}
 \mathcal F_{\rm SCI,lin}^{(1)}
 = 2\left.\mathcal F_b^{(1)}\right|_{b^2 = \omega,\,\Delta_I = 1/2} \, .
\end{align}
Here $\mathcal F_b^{(1)}$ is the analytic canonical sphere
correction, before taking a real part. For the full-stabilizer contribution considered below,
the normalized Airy expression is real for positive
real $b$ in the chosen spectral convention. Thus the real coefficients displayed
in \eqref{eq:TTI_massless_sphere_checks} also determine
the analytic coefficients on this branch, which we then
continue in $\omega$.

The limit $\omega\to0$ is the Cardy limit. 
We extract the leading exponential from the
full-stabilizer sector using the small-$b$ coefficients
$f_{-1}^{(1)}$ and $f_0^{(1)}$ of the canonical sphere
correction, defined in
\eqref{eq:TTI_block_small_b_coefficients}.
For the
unrefined TTI at the universal twist, $b_\pm^2=\pm\epsilon$ makes the pole
terms cancel; for the SCI, $b_\pm^2=\omega$ makes them
add. The finite terms add in both cases, giving
\begin{align}
\label{eq:SCI_symmetric_Cardy_structure}
 \mathcal F_{\rm SCI,np}^{(1)}
 = \frac{2f_{-1}^{(1)}}{\omega}
 +2f_0^{(1)}+\cO(\omega) \, .
\end{align}
We use the shifted rank and action of \eqref{eq:TTI_rank_and_action} and \eqref{eq:TTI_action_at_saddle},
\begin{align}
 \widehat N = N-\frac{k}{24}+\frac{2}{3k} \, , \qquad
 t = 2\pi\sqrt{\frac{2\widehat N}{k}} \, ,
\end{align}
with $\widehat N>0$ in the large $N$ regime.
The small $\omega$ expansion is taken at fixed $N$,
after the Airy transform, just as for the TTI.

We retain the complex SCI free energy here.
For comparison with field theory, we approach
$\omega=0$ along the positive imaginary axis,
$\omega\to\ii0^+$.
On this ray, the pole term is imaginary and
contributes to the phase of the index, while
the real finite term gives the universal twist
TTI correction.

The definitions and normalization in
section~\ref{subsec:TTI_fixed_rank} give
$2f_{-1}^{(1)}=\mathcal W_{\rm np}^{(1)}/\pi$
and $2f_0^{(1)}=F_{\rm TTI,np}^{(1)}$,
where the TTI is evaluated at the universal twist
$\Delta_I=\mathfrak n_I=\tfrac12$.
Here $\mathcal W_{\rm np}^{(1)}$ and
$F_{\rm TTI,np}^{(1)}$ are our gravitational
predictions for the on-shell twisted superpotential
and the real unrefined TTI correction, respectively.
Their appearance in
\eqref{eq:SCI_symmetric_Cardy_structure} reproduces
the field theory Cardy structure at the selected
Bethe vacuum, up to a convention-dependent finite
imaginary phase \cite{Hosseini:2026lvj}.

The coefficient of $b^{-2}$ and the finite term in
\eqref{eq:TTI_massless_sphere_checks} give
$f_{-1}^{(1)}$ and $f_0^{(1)}$, respectively, since
the sphere correction is real on the branch used here.
Inserting these coefficients into
\eqref{eq:SCI_symmetric_Cardy_structure} gives the explicit SCI corrections at $k=1,2,4$:
\begin{align}
\label{eq:SCI_symmetric_leading_instantons}
 \mathcal F_{\rm SCI,np}^{(1)}
 &=2\ex^{-t}
 \begin{cases}
 \displaystyle
 \frac{5k^2}{4\pi^2\omega}(t+1)
 +\frac{13k^2}{4\pi^2}t^2
 +\frac{5k^2}{\pi^2}(t+1)-38+\cO(\omega)
 \, ,&k=1,2\, ,\\[6pt]
 \displaystyle
 \frac{4}{\pi^2\omega}(t+1)
 +\frac{4}{\pi^2}t^2
 +\frac{8}{\pi^2}(t+1)-10+\cO(\omega)
 \, ,&k=4\, .
 \end{cases}
\end{align}
The pole coefficient and the real finite term agree
with the field theory results
\cite{Hosseini:2026dkj,Hosseini:2026lvj}, including
the instanton prefactors. This provides a non-trivial
consistency check of the SCI gluing, using the same
twisted superpotential and TTI data tested in
section~\ref{subsec:TTI_universal_tests}.

At $k=6$, the same calculation gives the prediction
\begin{align}
\label{eq:SCI_symmetric_k6}
 \mathcal F_{\rm SCI,np}^{(1)}(N,6)
 &= \ex^{-t}\Biggl[
 \frac{1}{\omega}
 \left(\frac{18}{\pi^2}(t+1)-2\rho_6\right)
 +\frac{18}{\pi^2}t^2
 +\left(\frac{36}{\pi^2}-\frac{12\sqrt3}{\pi}\right)t
 \nonumber\\
 &\qquad
 +\frac{36}{\pi^2}-16
 +2\rho_6\left(
 -5-\frac{2\pi}{\sqrt3}
 +\frac{4\pi^2}{9t}+\frac{8\pi^2}{9t^2}
 \right)
 +\cO(\omega)
 \Biggr] \, ,
\end{align}
where $\rho_6$ is given in \eqref{eq:TTI_k6_constant},
$\widehat N=N-5/36$ and
$t=2\pi\sqrt{\widehat N/3}$.
The pole and finite terms follow from
\eqref{eq:TTI_k6_superpotential_prediction} and
\eqref{eq:TTI_k6_index_prediction}, respectively.
At any $k>2$, the generic orbit sector has action
$(k/h)t$, with $h=\gcd(k,2)$, and hence does not affect
the leading exponential given here. At $k=6$ it
starts at $\ex^{-3t}$.

\bibliographystyle{utphys} 
\bibliography{references}

\end{document}